\pdfoutput=1
\documentclass{aa}
\usepackage{multirow}
\usepackage{tablefootnote}
\usepackage{txfonts}
\usepackage{natbib}
\usepackage{xcolor, tikz}
\usepackage{graphicx}

\bibpunct{(}{)}{;}{a}{}{,}
\defcitealias{Anders2019}{A19}

\begin{document}

   \title{Photo-astrometric distances, extinctions, and astrophysical parameters for {\it Gaia} EDR3 stars brighter than $G=18.5$}

   \author{F. Anders\inst{1,2}, A. Khalatyan\inst{2}, A. B. A. Queiroz\inst{2,3,4}, C. Chiappini\inst{2,3}, \\
   J. Ardèvol\inst{1}, L. Casamiquela\inst{5}, F. Figueras\inst{1}, Ó. Jiménez-Arranz \inst{1}, C. Jordi\inst{1}, M. Monguió\inst{1}, M. Romero-Gómez\inst{1}, \\
   D. Altamirano\inst{1}, T. Antoja\inst{1}, R. Assaad\inst{6}, T. Cantat-Gaudin\inst{7}, A. Castro-Ginard\inst{8}, H. Enke\inst{2}, L. Girardi\inst{9}, G. Guiglion\inst{2}, \\ 
   S. Khan\inst{10}, X. Luri\inst{1}, A. Miglio\inst{11,12,13}, I. Minchev\inst{2}, 
   P. Ramos\inst{14}, B. X. Santiago\inst{15,3}, M. Steinmetz\inst{2}
   }

   \authorrunning{Anders, Khalatyan, et al.}
   \titlerunning{{\tt StarHorse} parameters for {\it Gaia} EDR3 stars}

    \institute{Institut de Ci\`encies del Cosmos, Universitat de Barcelona (IEEC-UB), Mart\'i i Franqu\`es 1, 08028 Barcelona, Spain\\
              \email{fanders@icc.ub.edu, akhalatyan@aip.de}
	\and{Leibniz-Institut f\"ur Astrophysik Potsdam (AIP), An der Sternwarte 16, 14482 Potsdam, Germany}
	\and{Laborat\'orio Interinstitucional de e-Astronomia - LIneA, Rua Gal. Jos\'e Cristino 77, Rio de Janeiro, RJ - 20921-400, Brazil}
    \and{Institut f\"{u}r Physik und Astronomie, Universit\"{a}t Potsdam, Haus 28 Karl-Liebknecht-Str. 24/25, D-14476 Golm, Germany}
    \and{Laboratoire d’Astrophysique de Bordeaux, Univ. Bordeaux, CNRS, B18N, allée Geoffroy Saint-Hilaire, 33615 Pessac, France}
    \and{Department of Physics, University of Surrey, Guildford GU2 7XH, United Kingdom}
    \and{Max-Planck-Institut für Astronomie, Königstuhl 17, 69117 Heidelberg, Germany}
    \and{Leiden Observatory, Leiden University, Niels Bohrweg 2, 2333 CA Leiden, The Netherlands}
    \and{Osservatorio Astronomico di Padova, INAF, Vicolo dell'Osservatorio 5, 35122 Padova, Italy}
    \and{Institute of Physics, Laboratory of Astrophysics, École Polytechnique Fédérale de Lausanne (EPFL), Observatoire de Sauverny, 1290 Versoix, Switzerland}
    \and{Dipartimento di Fisica e Astronomia, Universit\`{a} di Bologna, Via Gobetti 93/2, 40129 Bologna, Italy}
    \and{INAF - Osservatorio di Astrofisica e Scienza dello Spazio di Bologna, Via Gobetti 93/3, I-40129 Bologna, Italy}
    \and{School of Physics and Astronomy, University of Birmingham, Edgbaston B15 2TT, UK}
    \and{Observatoire astronomique de Strasbourg, Université de Strasbourg, CNRS, 11 rue de l’Université, 67000 Strasbourg, France}
    \and{Instituto de F\'\i sica, Universidade Federal do Rio Grande do Sul, Caixa Postal 15051, Porto Alegre, RS - 91501-970, Brazil}
    }
   \date{Received 5.10.2021; accepted 8.11.2021}

  \abstract
  {We present a catalogue of 362 million stellar parameters, distances, and extinctions derived from {\it Gaia}'s Early Data Release (EDR3) cross-matched with the photometric catalogues of Pan-STARRS1, SkyMapper, 2MASS, and AllWISE. The higher precision of the {\it Gaia} EDR3 data, combined with the broad wavelength coverage of the additional photometric surveys and the new stellar-density priors of the {\tt StarHorse} code, allows us to substantially improve the accuracy and precision over previous photo-astrometric stellar-parameter estimates. At magnitude $G=14$ (17), our typical precisions amount to 3\% (15\%) in distance, 0.13 mag (0.15 mag) in $V$-band extinction, and 140 K (180 K) in effective temperature. Our results are validated by comparisons with open clusters, as well as with asteroseismic and spectroscopic measurements, indicating systematic errors smaller than the nominal uncertainties for the vast majority of objects. We also provide distance- and extinction-corrected colour-magnitude diagrams, extinction maps, and extensive stellar density maps that reveal detailed substructures in the Milky Way and beyond. The new density maps now probe a much greater volume, extending to regions beyond the Galactic bar and to Local Group galaxies, with a larger total number density.
  We publish our results through an ADQL query interface ({\tt gaia.aip.de}) as well as via tables containing approximations of the full posterior distributions. Our multi-wavelength approach and the deep magnitude limit render our results useful also beyond the next {\it Gaia} release, DR3.
  }

   \maketitle


\section{Introduction}

Since its launch in 2013 the European Space Agency's flagship mission {\it Gaia} \citep{GaiaCollaboration2016} has revolutionised Galactic astronomy and its neighbouring fields \citep{Brown2021}. 
The precision and accuracy of our knowledge of the Solar System (e.g. \citealt{Spoto2018, BailerJones2018a, PortegiesZwart2021}), stellar astrophysics (e.g. \citealt{Jao2018, Lanzafame2019, Mowlavi2021}), the immediate solar vicinity (e.g. \citealt{Klioner2021, Smart2021, Reyle2021}), open star clusters \citep[e.g.][]{Cantat-Gaudin2018, Cantat-Gaudin2020a, Castro-Ginard2020}, distant regions of the Milky Way (e.g. \citealt{Ramos2021, GaiaCollaboration2021Antoja, Zari2021}) and the Local Group (e.g. \citealt{GaiaCollaboration2018Helmi, Antoja2020, Luri2021}), the Galactic potential (e.g. \citealt{Crosta2020, Cunningham2020, Hattori2021}), and even the Hubble constant (e.g. \citealt{Breuval2020, Riess2021, Baumgardt2021}) are constantly increasing thanks to {\it Gaia}. 
In the context of Galactic archaeology, {\it Gaia} has also enabled a completely new line of precision studies tracing the past accretion events of the Milky Way \citep{Helmi2020, Kruijssen2020, Pfeffer2021}, often via the combination of complete phase-space information with detailed chemistry from spectroscopic surveys (e.g. \citealt{Aguado2021, Gudin2021, Limberg2021a, Limberg2021, Montalban2021, Naidu2021, Shank2021}).

The latest {\it Gaia} data release, Early Data Release 3 ({\it Gaia} EDR3; \citealt{GaiaCollaboration2021}), covers the first 34 months of observations with positions and photometry for $1.8\cdot10^9$ sources \citep{Riello2021}, proper motions and parallaxes for $1.5\cdot10^9$ sources \citep{Lindegren2021a}, and radial velocities for $7\cdot10^6$ sources \citep{Seabroke2021, GaiaCollaboration2021}. With respect to Data Release 2 ({\it Gaia} DR2; \citealt{GaiaCollaboration2018}), the proper motions are by a factor of 2 more precise, and parallax uncertainties are reduced by 20\% (see \citealt{Fabricius2021} for details).

\begin{table*}
\centering
\caption{Summary of the calibrations and data curation applied to the astrometric and photometric data for this work.}
\begin{tabular}{c|c|l|l}
Parameter & Parameter regime & Calibration choice & Reference \\
\hline
 $\varpi^{\rm cal}$  &  &   use {\tt zpt.py} calibration & \citet{Lindegren2021}  \\
\hline
 $\sigma_{\varpi}^{\rm cal}$  &  & ${\tt uwu} (G) \cdot {\tt parallax\_error}$    &  Fit to \citet{Fabricius2021} Fig. 19a\&b  \\
\hline
 $G$  &  ${\tt astrometric\_params\_solved}=95$     &  colour-dependent correction    &  \citet{GaiaCollaboration2021} App. A \\
\hline
 $ g_{\rm PS1}$   &  ${\tt g\_mean\_psf\_mag}>14$  &   ${\tt g\_mean\_psf\_mag} - 0.020 $  &   \\
 $ r_{\rm PS1}$   &  ${\tt r\_mean\_psf\_mag}>15$  &   ${\tt r\_mean\_psf\_mag} - 0.033 $  &   \\
 $ i_{\rm PS1}$   &  ${\tt i\_mean\_psf\_mag}>15$ &   ${\tt i\_mean\_psf\_mag} - 0.024 $  & \citet{Scolnic2015}  \\
 $ z_{\rm PS1}$   &  ${\tt z\_mean\_psf\_mag}>14$  &   ${\tt z\_mean\_psf\_mag} - 0.028 $  &   \\
 $ y_{\rm PS1}$   &  ${\tt y\_mean\_psf\_mag}>13$  &   ${\tt y\_mean\_psf\_mag} - 0.011 $  &   \\
\hline
 $ g_{\rm SM}, r_{\rm SM}$   &   & $E(B-V)$ dependent zero-point shifts  & \citet{Huang2020}  \\
\hline
 $ u_{\rm SM}, v_{\rm SM}$   &   & not used in inference  & \citet{Schlafly2016}  \\
\hline
\multirow{4}{*}{$\sigma_{\rm mag}$}  & {\it Gaia} EDR3  & $\max\{{\tt \sigma_{\rm mag, source}}, 0.02 {\rm mag}\} $  &    \\
  & 2MASS, AllWISE  & $\max\{{\tt \sigma_{\rm mag, source}}, 0.03 {\rm mag}\} $  &    \\
  & Pan-STARRS1 DR1  & $\max\{{\tt \sigma_{\rm mag, source}}, 0.04 {\rm mag}\} $  &    \\
  & SkyMapper DR2  & $\max\{{\tt \sigma_{\rm mag, source}}, 0.05 {\rm mag}\} $  &    \\
\end{tabular}
\label{calibtable}
\end{table*}

In a previous work (\citealt{Anders2019}, hereafter \citetalias{Anders2019}) based on {\it Gaia} DR2, our group derived Bayesian stellar parameters, distances, and extinctions for 265 million stars brighter than $G=18$ with the {\tt StarHorse} code \citep{Santiago2016, Queiroz2018}. The combination of precise {\it Gaia} DR2 parallaxes and optical photometry with the multi-wavelength photometry of Pan-STARRS1 \citep{Chambers2016}, 2MASS \citep{Cutri2003}, and AllWISE \citep{Cutri2013} substantially improved the accuracy of the extinction and effective temperature estimates provided with only {\it Gaia} DR2 \citep{Andrae2018}. A selection of the most reliable in- and output data, a sample of 137 million stars, allowed \citetalias{Anders2019} to detect the imprint of the Galactic bar both in the stellar density distribution and in proper motion maps (further studied with APOGEE spectroscopy in \citealt{Queiroz2021}).

The results of our {\it Gaia} DR2 {\tt StarHorse} run presented in \citetalias{Anders2019} have been used in a wide variety of science cases, including exoplanetary research \citep{Sozzetti2021}, interstellar extinction \citep{Leike2020}, runaway stars from supernova remnants \citep{Lux2021}, X-ray transients \citep{Lamer2021}, ,$\gamma$-ray astronomy \citep{Steppa2020}, ,the Galactic escape speed curve \citep{Monari2018}, the three-dimensionl phase-space structure of the Milky Way disc \citep{Carrillo2019}, and spectroscopic survey simulations \citep{Chiappini2019}.

Anticipating a significant improvement thanks to the new {\it Gaia} data, we update our analysis using the new EDR3 data in this paper, addressing some of the known caveats of our previous data release and reducing the uncertainties of the main output parameters by a factor of 2. In a parallel effort we will be publishing {\tt StarHorse} results for spectroscopic surveys combined with {\it Gaia} ($\approx 6$ million stars) in Queiroz et al. (2022, in prep.).

This paper is structured as follows: 
Section \ref{data} presents the input data and Sect. \ref{method} our method. In particular, Sect. \ref{updates} describes the updates to our code with respect to previous applications, and Sect. \ref{flags} explains how we flagged the new {\tt StarHorse} results for {\it Gaia} EDR3. We then present some first astrophysical results in Sect. \ref{results}, mainly focussing on colour-magnitude diagrams (CMDs), extinction maps, and stellar density maps. The stellar density maps demonstrate the emergence of substructure beyond the detection of the Galactic bar, for example when focussing on metal-poor stars, the Magellanic Clouds, or the outer Milky Way halo. The precision and accuracy of the {\tt StarHorse} EDR3 parameters are discussed in Sect. \ref{uncertainties}, providing comparisons to Galactic open clusters \citep[OCs][]{Cantat-Gaudin2020, Dias2021}, asteroseismically derived parameters for giant stars \citep{Miglio2021}, and spectroscopic stellar parameters from the GALAH survey \citep{Buder2021}. We also make comparisons to previous results obtained from {\it Gaia} DR2 and EDR3 in Sect. \ref{comp}. Finally, we conclude the paper with a summary and a brief outlook to the near future in Sect. \ref{conclusions}.

\section{Data}\label{data}

As input for {\tt StarHorse}, we use the {\it Gaia} EDR3 data cross-matched with 2MASS, AllWISE, Pan-STARRS1, and SkyMapper \citep{Onken2019}, in the sense that all available good photometric measurements are used in the inference. The calibrations used in this paper are summarised in Table \ref{calibtable}.

From {\it Gaia} EDR3 we use the parallaxes and three-band photometry, together with their associated uncertainties. We recalibrate the parallaxes following the recommendations of \citet{Lindegren2021} who assessed the variations in the parallax zero point as a function of sky position, magnitude, and colour\footnote{code available at \url{https://gitlab.com/icc-ub/public/gaiadr3_zeropoint/-/tree/master/}}. Furthermore, we inflate the corresponding parallax uncertainties by a magnitude-dependent factor, following \citet[][see also \citealt{El-Badry2021}]{Fabricius2021}. In particular, we fit the inflation factor to their worst-case scenario of Fig. 19 in \citet{Fabricius2021}  (a crowded Large Magellanic Cloud field) to make sure that our parallax uncertainties are not underestimated. A further discussion of the fidelity of the {\it Gaia} parallaxes in our context is available in \citet{Rybizki2021} and in our Sect. \ref{gaiaflag}.

\begin{figure*}
\centering
\includegraphics[width=.8\textwidth]{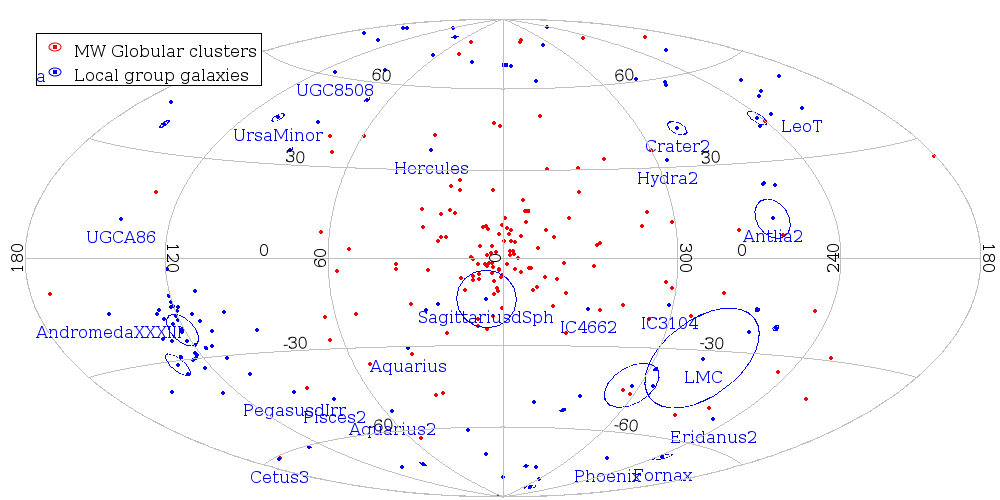}
\includegraphics[width=.85\textwidth]{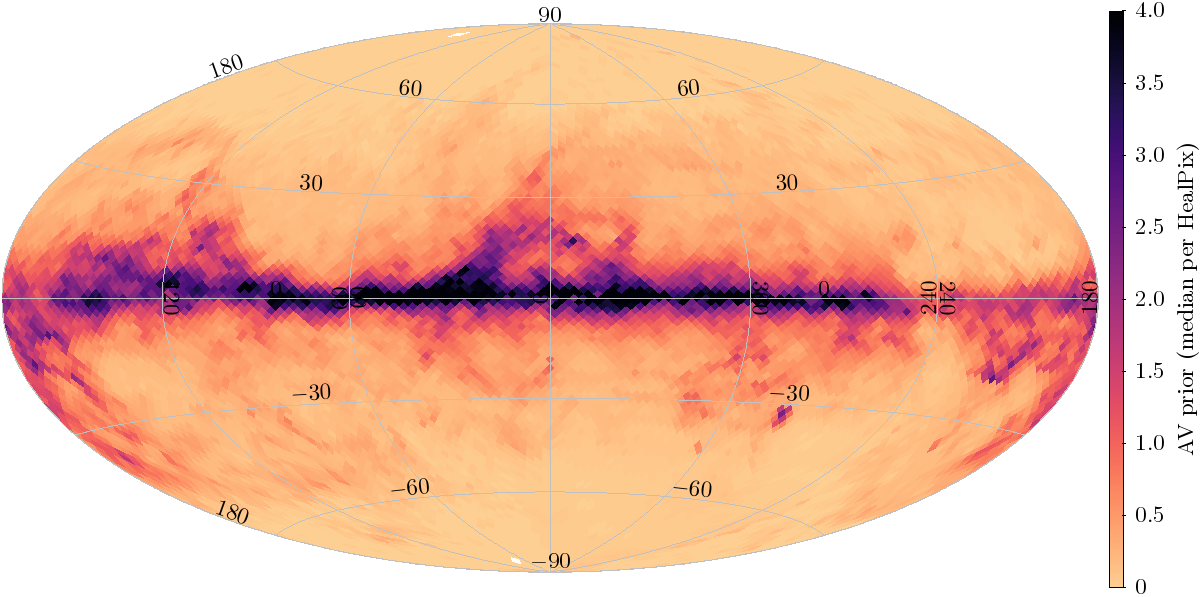}
\caption{Newly implemented priors in the {\tt StarHorse} code. Top panel: Sky distribution (in Galactic coordinates) of extragalactic and globular-cluster priors added in the new {\tt StarHorse} version. The angular extents (5 effective radii) of each of the Local Group priors are shown as circles, highlighting the most prominent objects: the Magellanic Clouds, the Sgr dSph, and Andromeda. Bottom panel: Median prior $V$-band extinction per HealPix. The extinction prior is calculated individually for each star from the three-dimensional extinction maps of either \citet{Green2019} or \citet{Drimmel2003}.}
  \label{extragal}
\end{figure*}

\begin{figure}
\centering
\includegraphics[width=.49\textwidth]{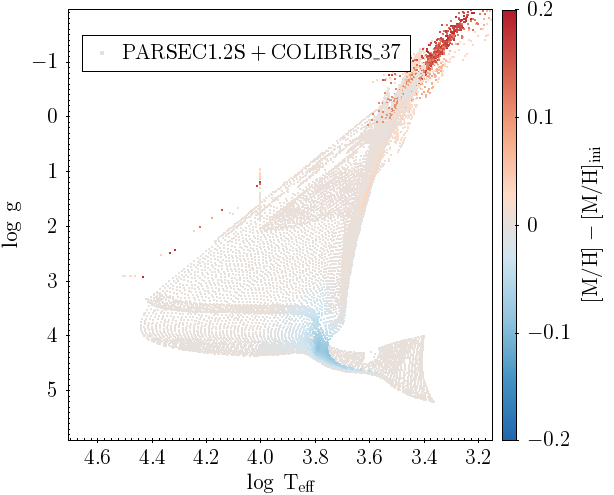}
\includegraphics[width=.49\textwidth]{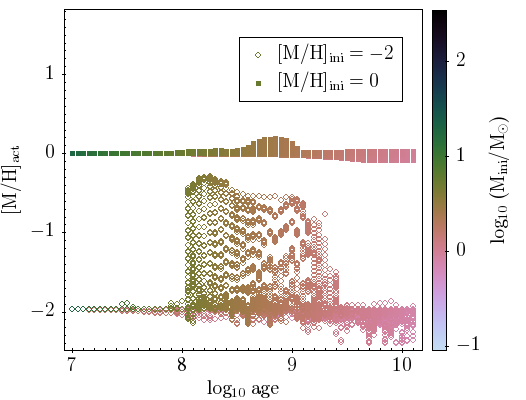}
\caption{Stellar evolution effects on surface metallicity in the PARSEC 1.2S + COLIBRI S37 stellar models. Top: Kiel diagram colour-coded by the difference between the surface metallicity and the initial metallicity. The evolution effects of diffusion and dredge-up are clearly visible. Bottom: Age dependence of the surface metallicity, for two initial metallicities, colour-coded by stellar mass.}
  \label{metallicity_effect}
\end{figure}

Regarding the {\it Gaia} photometry, we use the precise EDR3 magnitudes \citep{Riello2021} without any posterior correction other than the $G$-band correction advertised in Appendix A of \citet{GaiaCollaboration2021}\footnote{\url{https://github.com/agabrown/gaiaedr3-6p-gband-correction/blob/main/GCorrectionCode.ipynb}}, since they show much lower systematics  ($\lesssim 0.01$ mag; e.g. \citealt{Fabricius2021, Niu2021}) than the previous DR2 photometry (see e.g. \citealt{Maiz2018}).
For the BP/RP photometry, we follow the recommendation of \citet{Fabricius2021} and do not use magnitudes {\tt phot\_bp\_mean\_mag}$>20.5$ or {\tt phot\_rp\_mean\_mag}$>20$ in the inference.

The {\it Gaia} EDR3 cross-match to the large-area photometric surveys 2MASS, AllWISE, Pan-STARRS1 DR1, and SkyMapper DR2 is documented in \citet{Marrese2021}. The main novelty with respect to \citetalias{Anders2019} is the inclusion of SkyMapper data. From the SkyMapper DR2 data we only use the $griz$ bands (with zero points recalibrated following \citealt{Huang2020}) and refrain from using the $u$ and $v$ bands, because our default extinction law \citep{Schlafly2016} should not be extrapolated to the ultraviolet.

For Pan-STARRS1, we apply the zero-point corrections recommended by \citet{Scolnic2015}, and do not use magnitudes brighter than the saturation limit. With respect to \citetalias{Anders2019}, we apply more restrictive filters to the 2MASS and AllWISE photometry: only magnitudes with corresponding photometric quality flags 'A' or 'B' are accepted. The minimum photometric uncertainties used in the inference (reflecting also the systematic uncertainties of the passbands and the bolometric corrections) are given in Table \ref{calibtable}. We also note that for $\lesssim0.5$\% of the {\it Gaia EDR3} sources, the {\it Gaia} cross-match with 2MASS returns in multiple matches. In these cases (pointing towards possible confusion), we do not use any 2MASS photometry.

\begin{figure*}
\centering
    \begin{tikzpicture}
       \node[anchor=south west,inner sep=0] (image) at (0,0) {\includegraphics[width=0.9\textwidth]{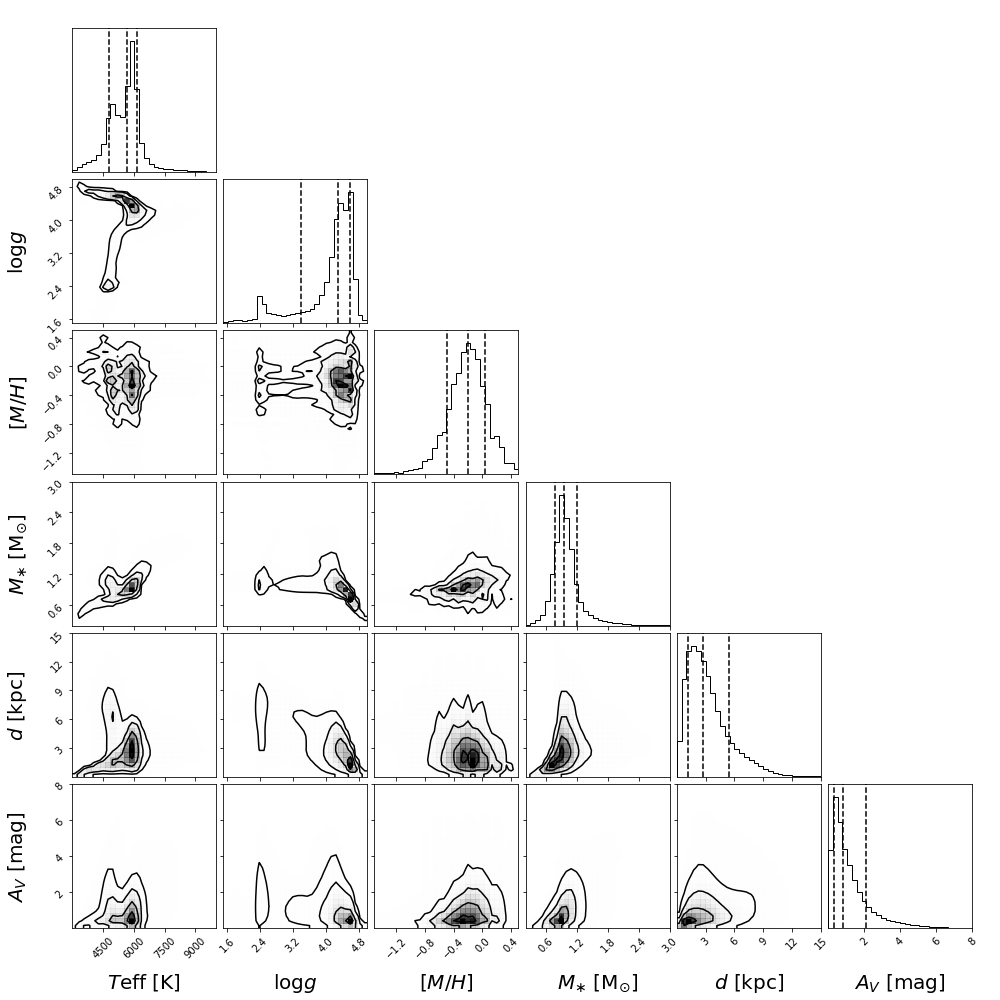}};
       \begin{scope}[x={(image.south east)},y={(image.north west)}]
       \node[anchor=south west,inner sep=0] (image) at (0.045,0.25) {\includegraphics[width=0.9\textwidth]{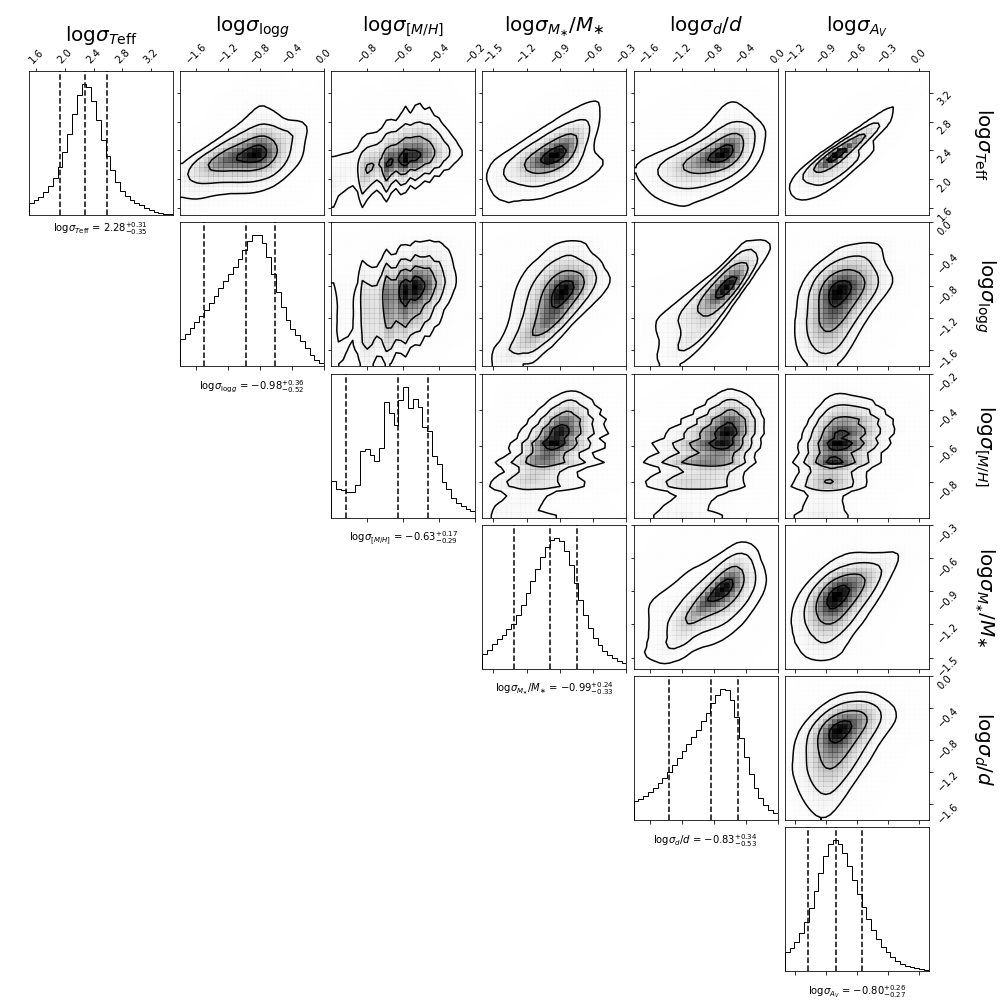}};
       \end{scope}
    \end{tikzpicture}
 	\caption{{\tt corner} plots showing the correlations and distributions of {\tt StarHorse} median posterior output values $T_{\rm eff}, \log g,$ [M/H], $M_{\ast}, d, $ and $A_V$ (lower-left panels), and their corresponding uncertainties (in logarithmic scale; top-right panels) for all stars in our catalogue. The dashed vertical lines in the diagonal panels show the 16th, 50th, and 84th percentiles of each parameter.}
 	\label{corner}
 \end{figure*}

\section{Method: The {\tt StarHorse} code}\label{method}

{\tt StarHorse} \citep{Queiroz2018} is an isochrone-fitting code tailored to derive distances $d$, extinctions (at $\lambda=542$ nm) $A_V$, ages $\tau$, masses $m_{\ast}$, effective temperatures $T_{\mathrm{eff}}$, metallicities [M/H], and surface gravities $\log g$ for field stars. In the absence of spectroscopic input data, it takes as input only the measured parallax $\varpi$ and a set of observed magnitudes $m_{\lambda}$ to estimate how close a stellar model is to the observed data. 

{\tt StarHorse} also includes priors about the geometry, metallicity and age characteristics of the main Galactic components. The priors adopted here are very similar to those in \citet{Queiroz2018}, \citetalias{Anders2019}, and \citet{Queiroz2020}: a \citet{Chabrier2003} initial mass function; exponential spatial density profiles for thin and thick discs; a spherical halo and a triaxial bulge/bar component, as well as broad Gaussian distributions for the age and metallicity distribution priors. The normalisation of each Galactic component, as well as the solar position, were taken from \citet{Bland-Hawthorn2016}. 

\subsection{Code updates and improvements}\label{updates}

With respect to \citetalias{Anders2019} and \citet{Queiroz2020}, we have implemented some changes that help to improve the performance of {\tt StarHorse} in the context of {\it Gaia} EDR3.

\subsubsection{A more informative interstellar extinction prior}\label{extprior}

One of the drawbacks of the \citetalias{Anders2019} {\it Gaia} DR2 run was the a priori limit in interstellar extinction to $A_V < 4$ mag for sources with low signal-to-noise parallax measurements ({\tt parallax\_over\_error} $<5$). This resulted in poor convergence or biased results for distant obscured objects in the Galactic plane. For the present EDR3 run we therefore update our previously uninformative top-hat $A_V$ prior to a prior that takes into account our knowledge of Galactic interstellar extinction. 

For the region of the sky covered by Pan-STARRS1, we use the large-scale three-dimensional extinction map of \citet{Green2019}. For the missing part (1/4) of the sky, we use the 2MASS-derived three-dimensional extinction model by \citet{Drimmel2003}. To get the range of possible distances for each star (needed to query the extinction maps), we invert the EDR3 zero-point-corrected parallax measurements to estimate a prior extinction value range for each star (using a maximum of $A_{V_{prior}}=10$ mag, considering that our sample is limited by $G<18.5$). The extinction prior is then defined as a very broad Gaussian distribution around the central value (with $\sigma_{AV,prior} = \max\{0.2, 0.33\cdot A_{V_{prior}}\}$). 

\subsubsection{Extragalactic priors}\label{extraprior}

In \citetalias{Anders2019} we saw that the stellar populations of the Magellanic Clouds, the Sagittarius (Sgr dSph) galaxy, and a number of relatively nearby globular clusters, whose stellar densities were not accounted for by our priors until now, left a spurious imprint on the posterior Galactic density distribution inferred with {\tt StarHorse}.

For the new runs we therefore included new extragalactic and globular cluster priors in the calculation of the global prior. For the extragalactic resolved stellar population we use the updated list (October 2019) of Local Group Galaxies from \citet{McConnachie2012}, which comprises sky positions, distances, foreground extinctions, apparent dimensions, central densities, metallicities, masses, and other basic quantities for the Local Group. We manually added M31 to this list, and curated the list for the most prominent objects on the sky: the Magellanic Clouds and the Sgr dSph galaxy. For all but these objects we estimate the mass of the external galaxy by inverting the mass-metallicity relation of \citet{Panter2008} (linearly extrapolated below [Fe/H]$=-1$). For the Galactic globular clusters we used the recent catalogue of \citet{Hilker2020}. 

The sky distribution of all considered objects is shown in Fig. \ref{extragal}. If a star's celestial coordinates coincide with those of an external galaxy or globular cluster within five half-light radii, we add to the Milky Way foreground prior an additional population corresponding to the characteristics of that object (stellar density, distance, metallicity). For the sake of simplicity, the density profile priors for all objects are assumed to be three-dimensional Gaussians.

\subsubsection{Update of the bar angle in the priors}\label{galupdates}

Our knowledge about the large-scale parameters of the Milky Way is constantly improving. In light of the growing evidence for a bar angle around $27\pm 2$ deg (see discussion in \citealt{Bland-Hawthorn2016}, reinforced also by \citealt{Queiroz2021}), we have updated the angle of the Galactic bar in our prior to that value.

\subsubsection{Taking into account evolution of surface metallicity}\label{zini}

We adopt here the latest version of the PARSEC1.2S + COLIBRI S37 stellar evolutionary model tracks \citep{Bressan2012, Marigo2017, Pastorelli2019}. Using these tracks in conjunction with the new CMD web interface allows us to take into account changes in the surface metallicity of stars during stellar evolution. While the effect is typically very small, element diffusion does introduce some small but appreciable decrease in the surface metal content for solar-mass stars before and around the turn-off (Fig. \ref{metallicity_effect}; see e.g. \citealt{BertelliMotta2018, Souto2019} for observational evidence). The effect is much stronger for low-metallicity stars. The opposite (i.e. a strong increase in the surface metallicity) happens for a fraction of the more evolved stars, such as those in the thermally pulsing asymptotic giant branch and Wolf-Rayet phases; this latter effect, however, is much less relevant to our results given that these evolutionary phases are much shorter-lived (and hence rarer) than main sequence stars, especially in nearby volume-limited samples.

\subsection{{\tt StarHorse} setup} \label{setup}

For the EDR3 run we used a grid of PARSEC 1.2S stellar models \citep{Marigo2017} in the 2MASS, Pan-STARRS1, SkyMapper, {\it Gaia} EDR3, and WISE photometric systems available on the PARSEC web page\footnote{\url{http://stev.oapd.inaf.it/cgi-bin/cmd_3.4}; see also \url{http://stev.oapd.inaf.it/cmd_3.4/faq.html}}. The model grid was equally spaced by 0.1 dex in log age as well as in initial metallicity [M/H]. The code explores distances within $\{ 1/(\varpi^{\rm cal}+3\cdot \sigma_{\varpi}^{\rm cal}), 1/(\varpi^{\rm cal}-3\cdot \sigma_{\varpi}^{\rm cal})\}$.

\begin{figure}\centering
 	\includegraphics[width=0.49\textwidth]{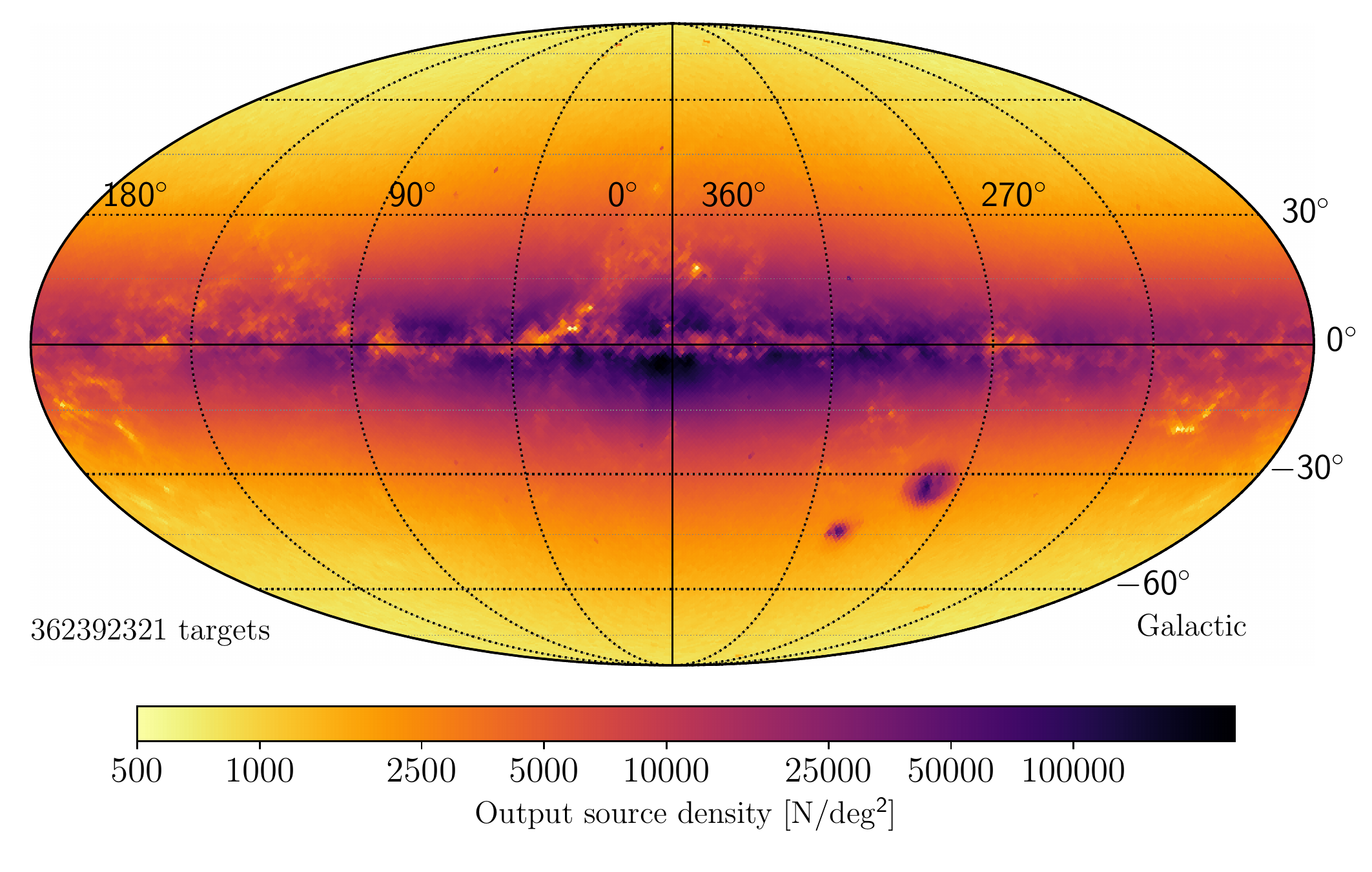}
 	\includegraphics[width=0.49\textwidth]{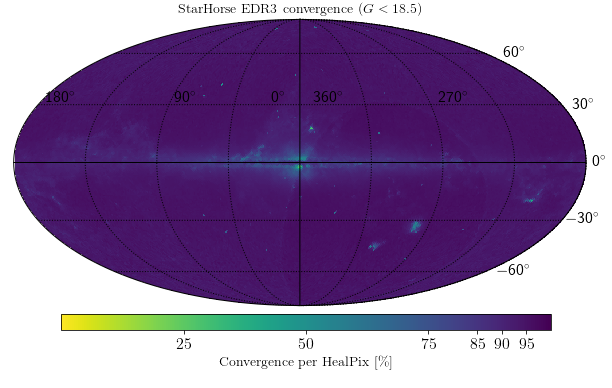}
 	\includegraphics[width=0.49\textwidth]{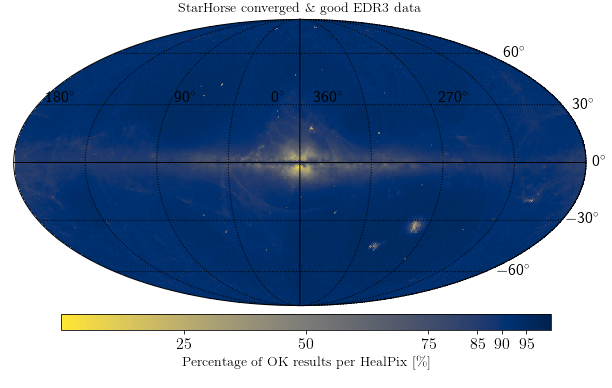}
 	\caption{Sky density map of all converged targets ($G<18.5$ mag; top panel). The middle and bottom panels show the relative fraction of converged stars and flag-cleaned results with respect to the input data.}
 	\label{skydensity}
 \end{figure}

For the present {\it Gaia} EDR3 run ($G<18.5$ mag, 400M stars), the code took on average 0.3 seconds per star to run (depending slightly on the position in the CMD and the number of photometric measurements available). In total, the computational cost for this {\tt StarHorse} run thus was $\sim 50,000$ CPU hours, reducing the CO$_2$ footprint of {\tt StarHorse} by a factor of $3$ with respect to the {\it Gaia} DR2 run presented in \citetalias{Anders2019}, while increasing the number of stars with reliable output parameters by more than a factor of 2.
The global statistics for our output results are summarised in Table \ref{summarytable} and discussed in detail in Sect. \ref{results}.

\begin{figure*}\centering
 	\includegraphics[width=0.33\textwidth, trim={0 2cm 0 1.5cm}]{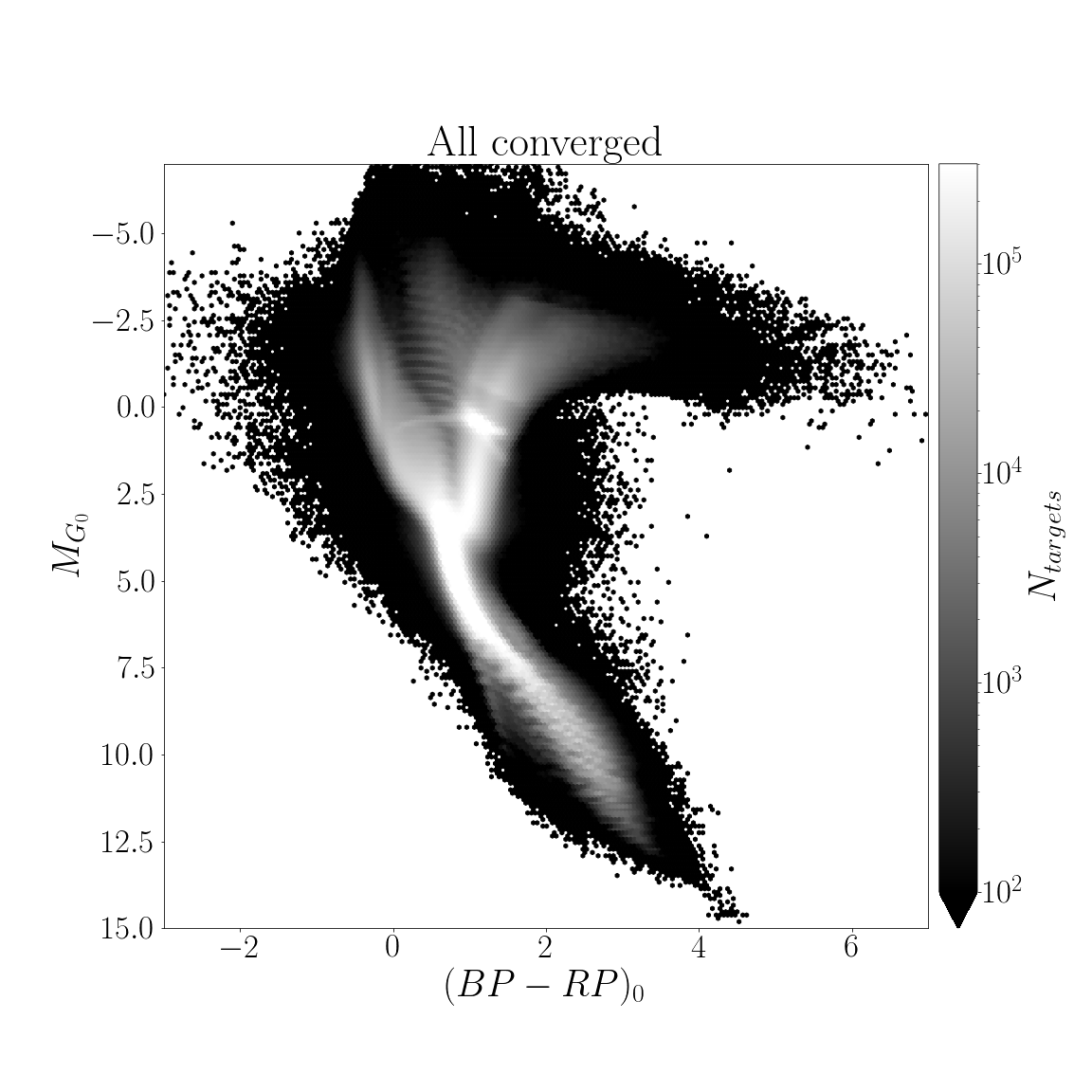}
 	\includegraphics[width=0.33\textwidth, trim={0 2cm 0 1.5cm}]{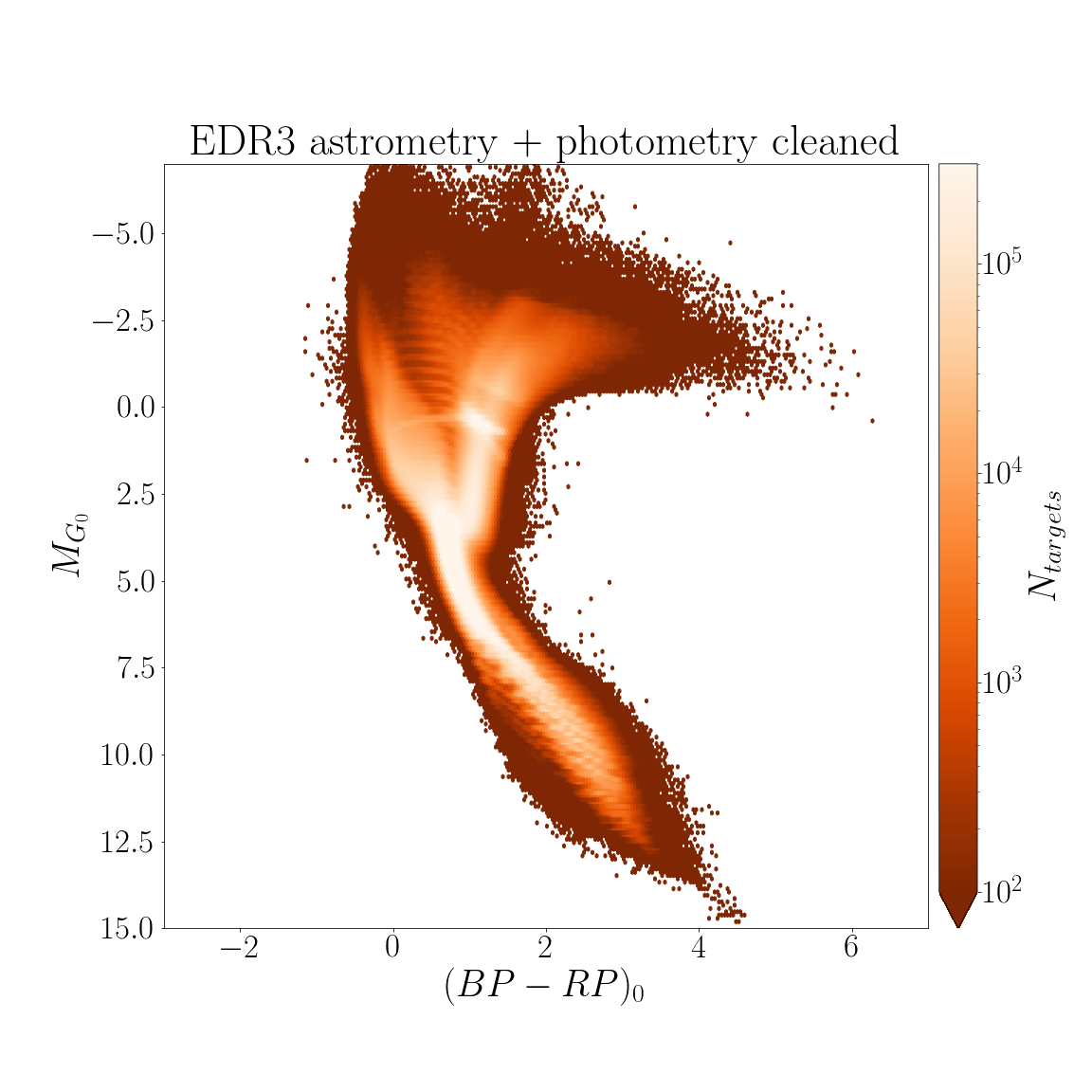}
 	\includegraphics[width=0.33\textwidth, trim={0 2cm 0 1.5cm}]{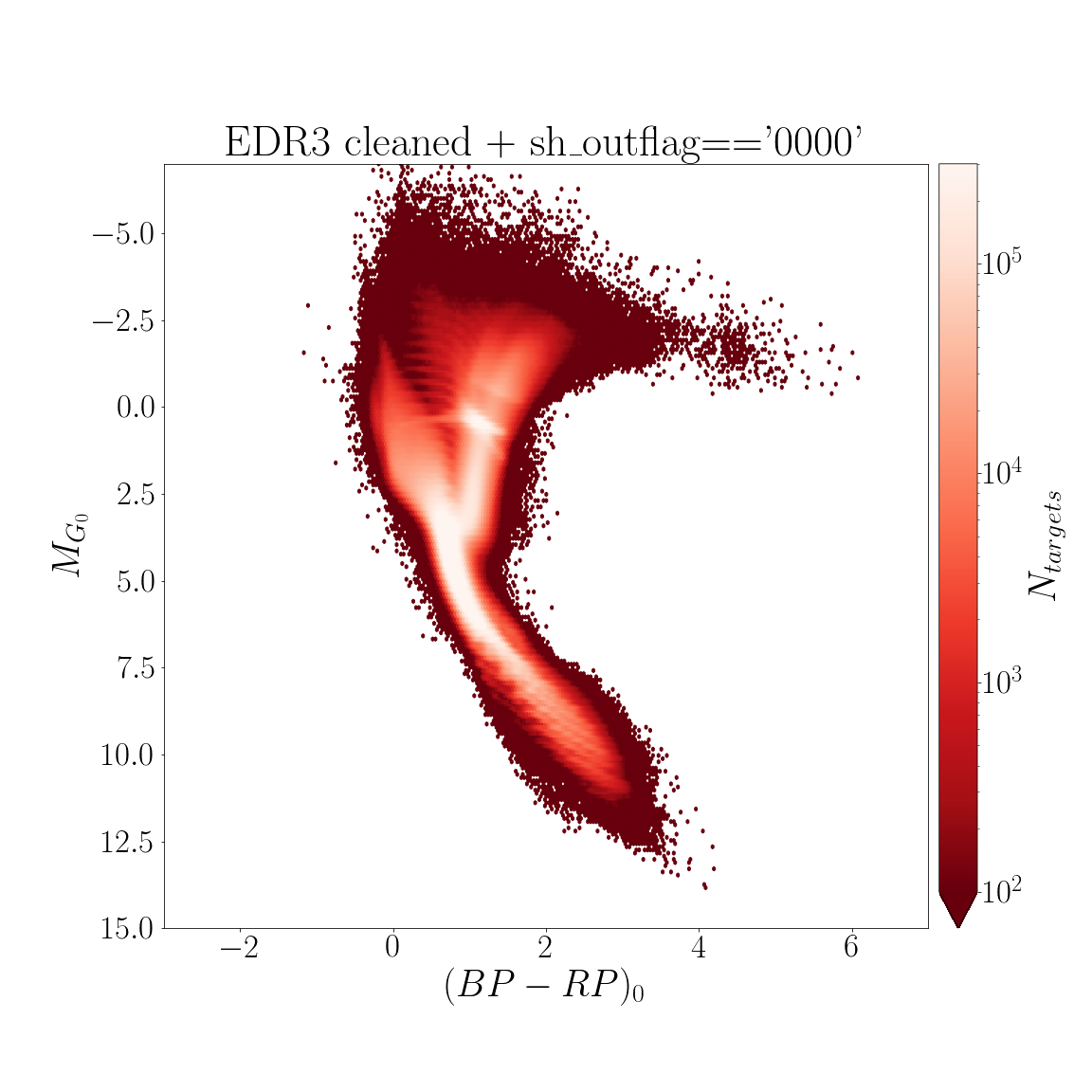}
 	\includegraphics[width=0.33\textwidth, trim={0 2cm 0 1.5cm}]{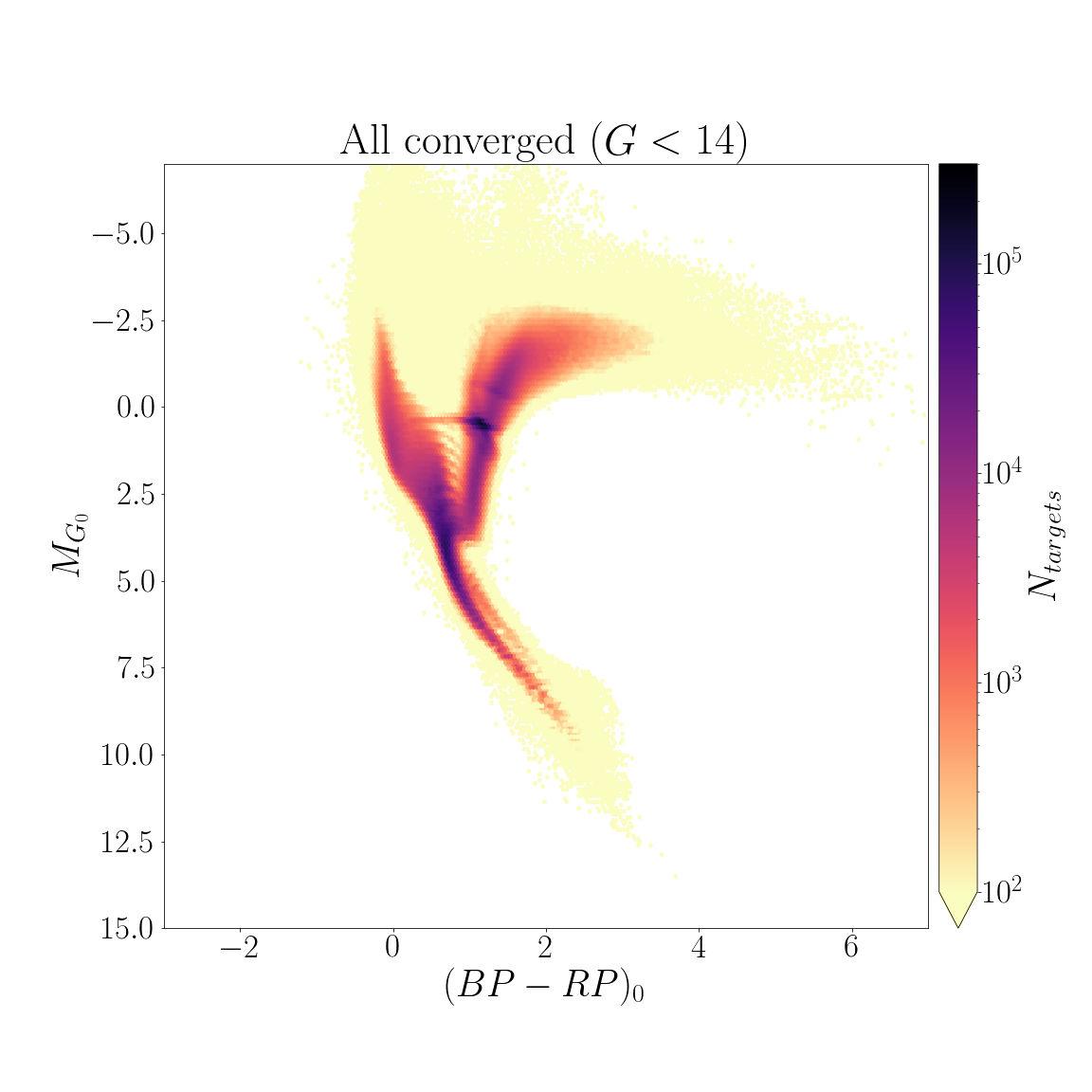}
 	\includegraphics[width=0.33\textwidth, trim={0 2cm 0 1.5cm}]{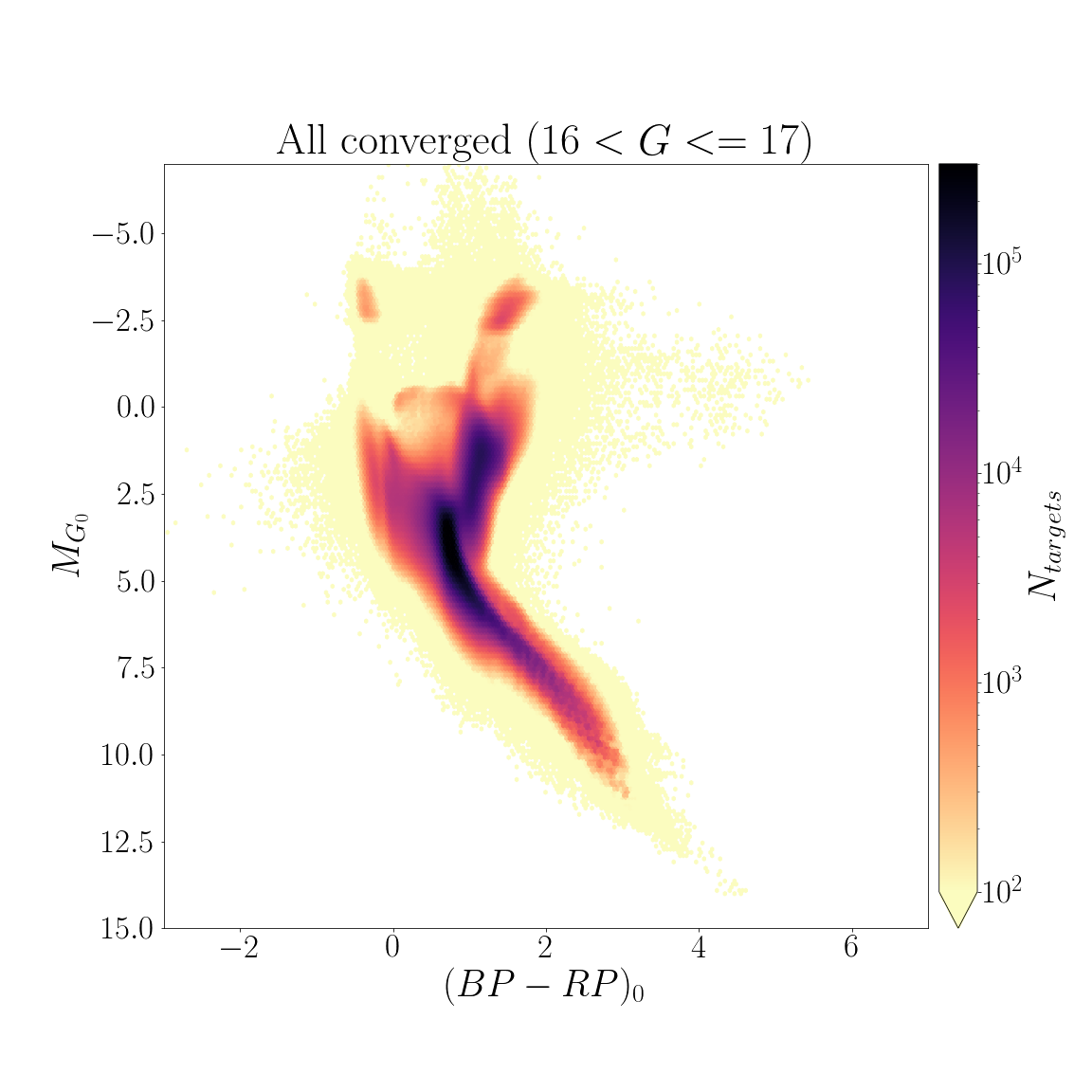}
 	\includegraphics[width=0.33\textwidth, trim={0 2cm 0 1.5cm}]{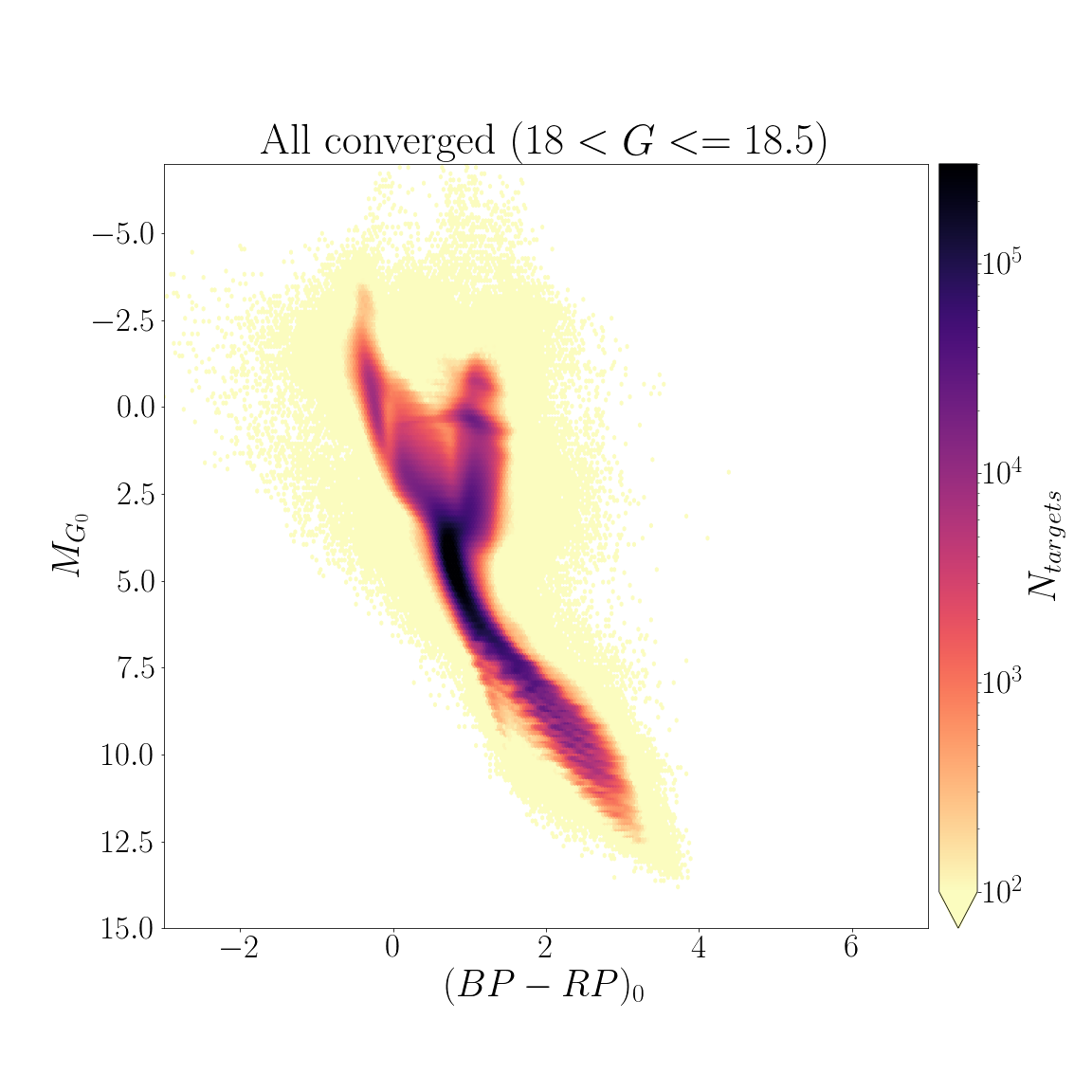}
 	\includegraphics[width=0.33\textwidth, trim={0 2cm 0 1.5cm}]{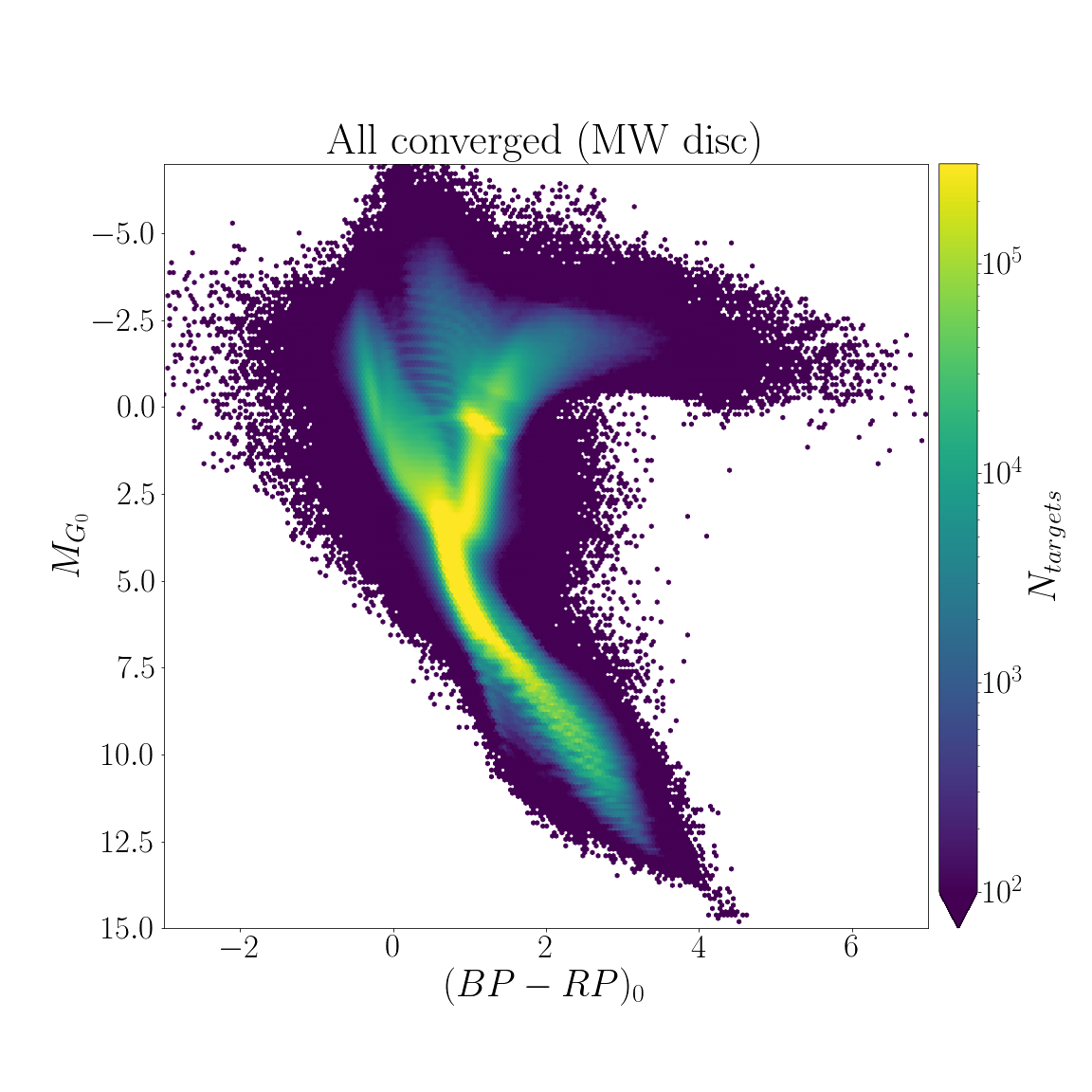}
 	\includegraphics[width=0.33\textwidth, trim={0 2cm 0 1.5cm}]{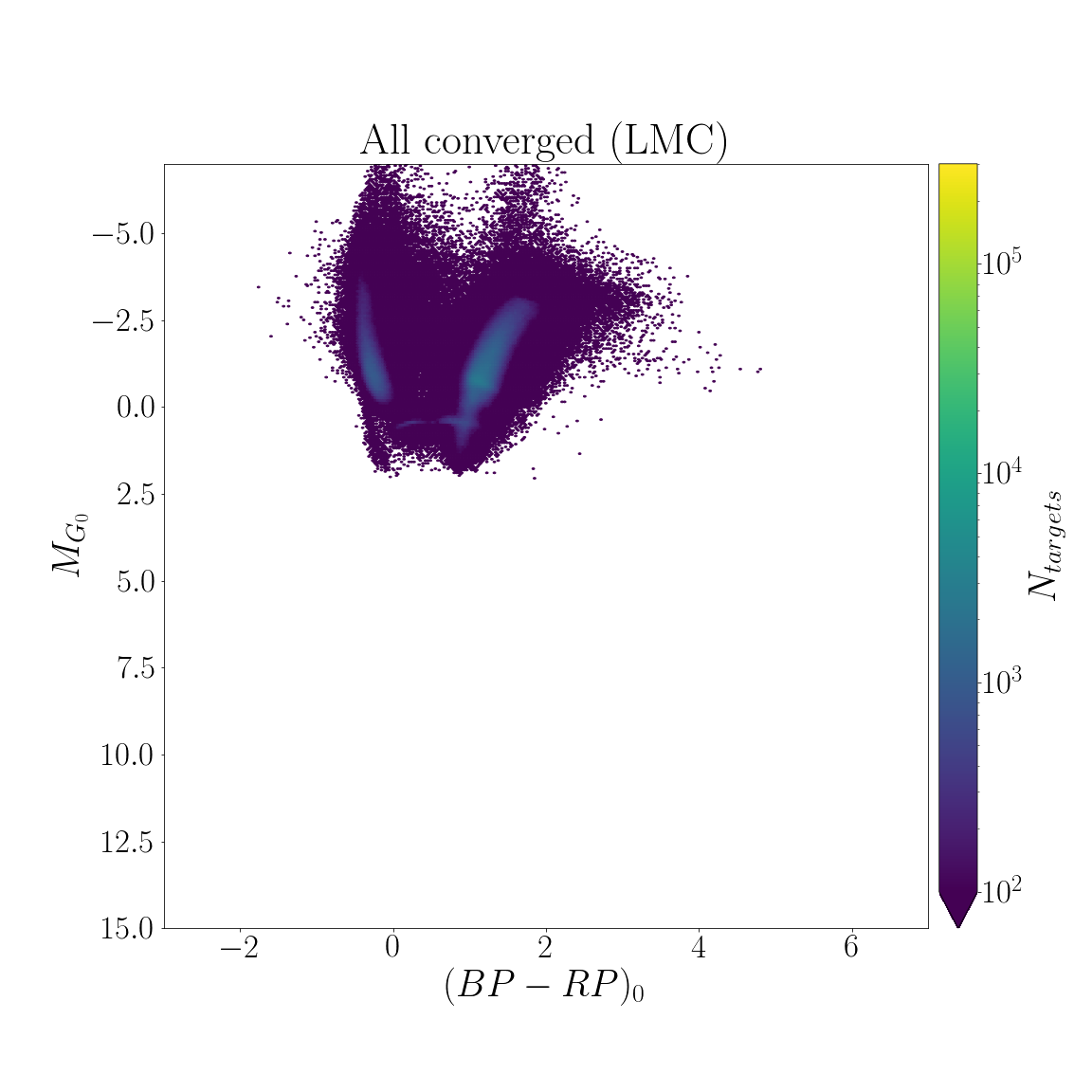}
 	\includegraphics[width=0.33\textwidth, trim={0 2cm 0 1.5cm}]{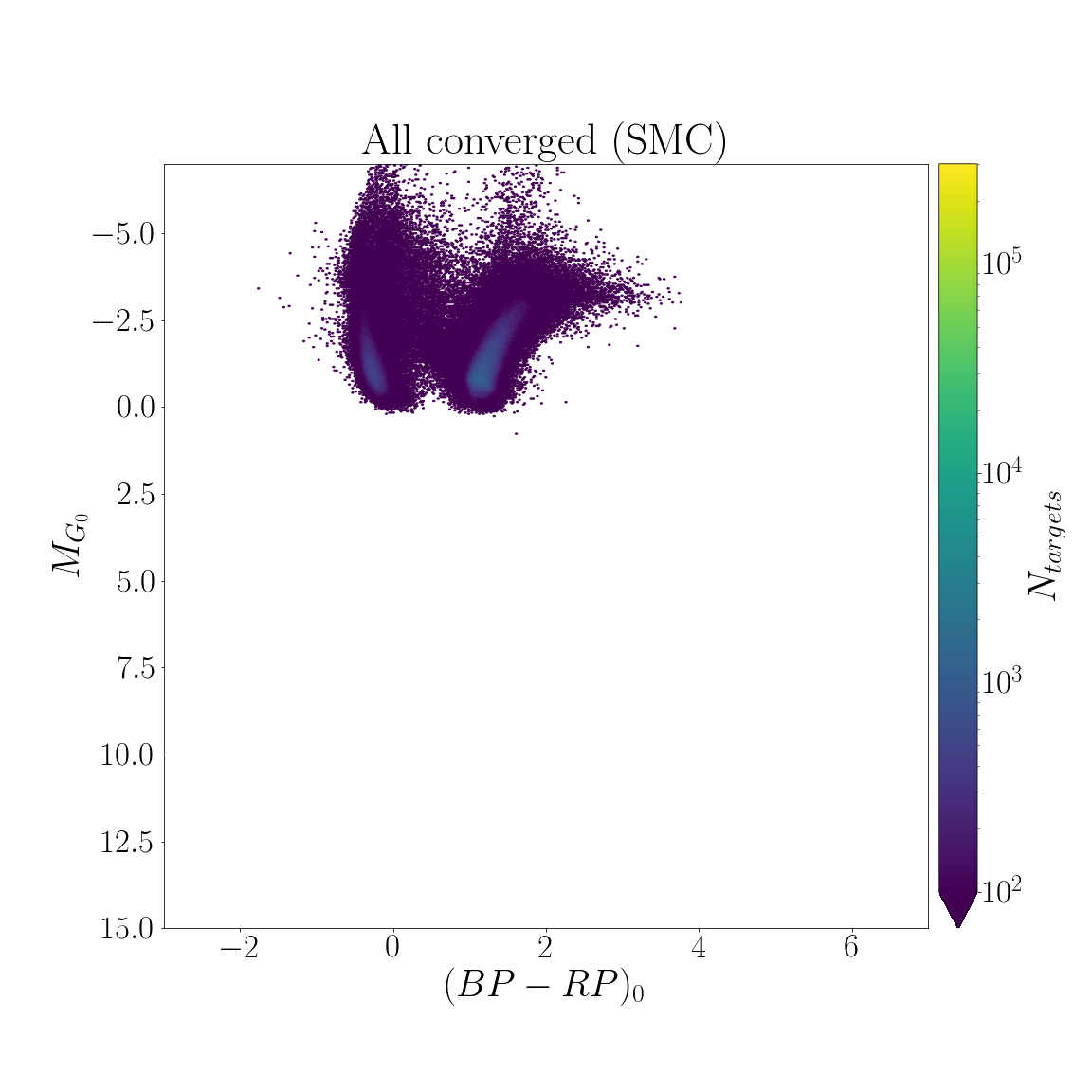}
 	\caption{{\tt StarHorse} posterior {\it Gaia} EDR3 CMDs. Top row, from left to right: All converged objects (362M), {\it Gaia} EDR3 cleaned sample (321M), EDR3- and flag-cleaned sample (282M). Middle row: CMDs for three broad magnitude bins, showing both the increasing mix of stellar populations (e.g. the giant-star populations of the Magellanic Cloud starting to appear around $M_G \sim -3$ in the $16<G<17$ panel) and the decreasing astrometric quality with increasing magnitude. Bottom row: Separate CMDs for the Milky Way disc (left; 339M stars), the LMC (middle; 1.09M stars), and the SMC (right; 94k stars). The abrupt absolute magnitude cut in the last two panels is caused by the $G<18.5$ magnitude cut.
    }
 	\label{cmds1}
 \end{figure*}

\begin{figure*}\centering
 	\includegraphics[width=0.33\textwidth]{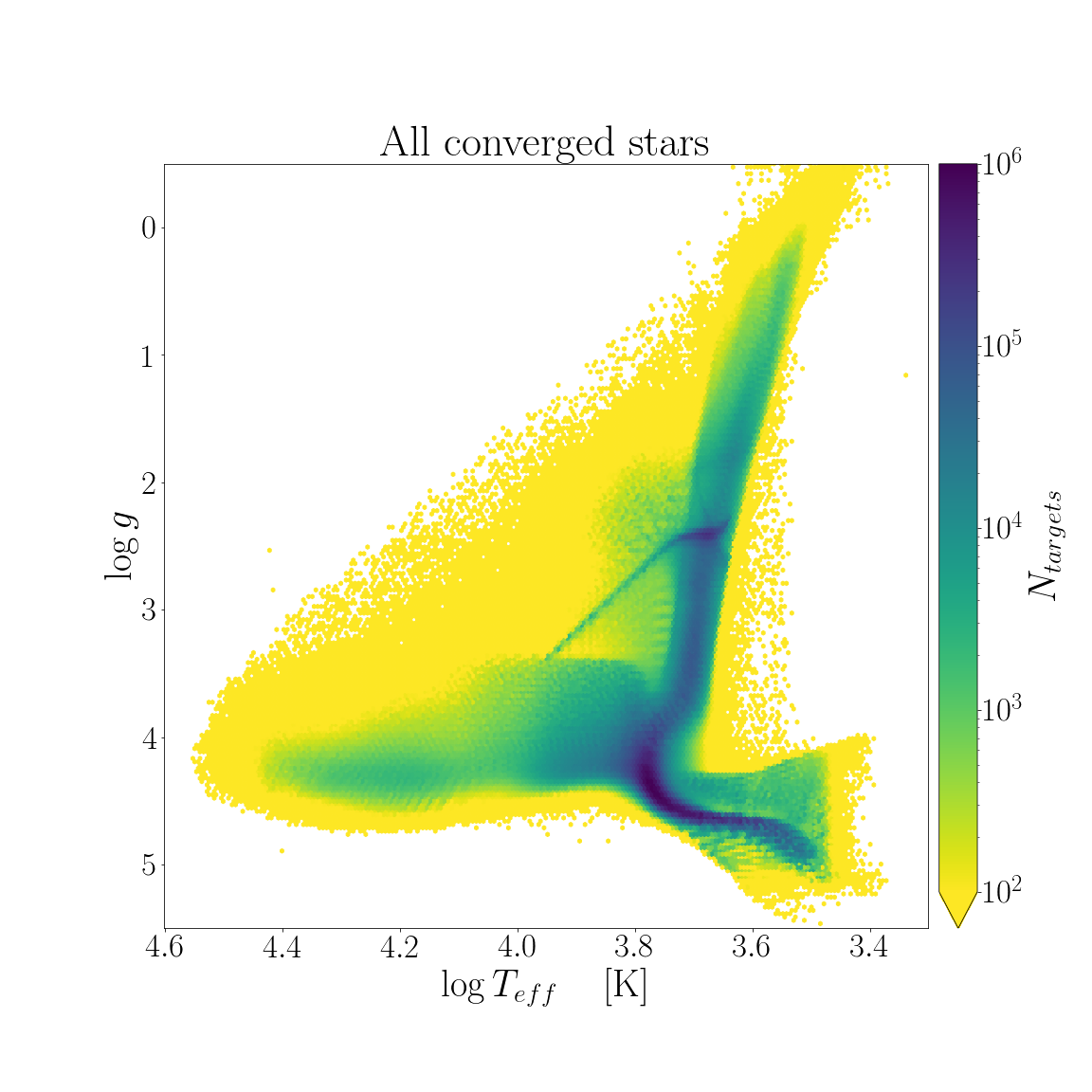}
 	\includegraphics[width=0.33\textwidth]{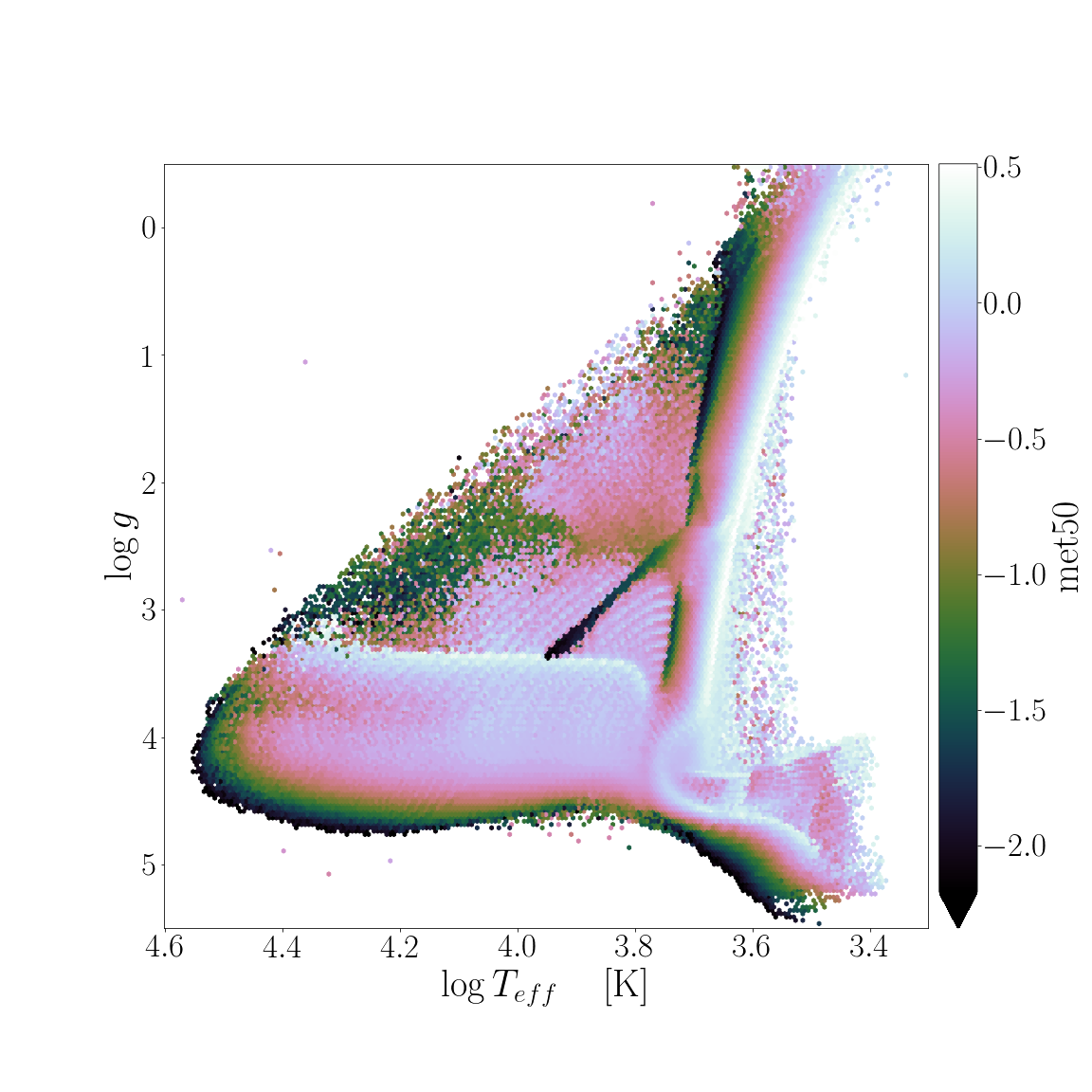}
 	\includegraphics[width=0.33\textwidth]{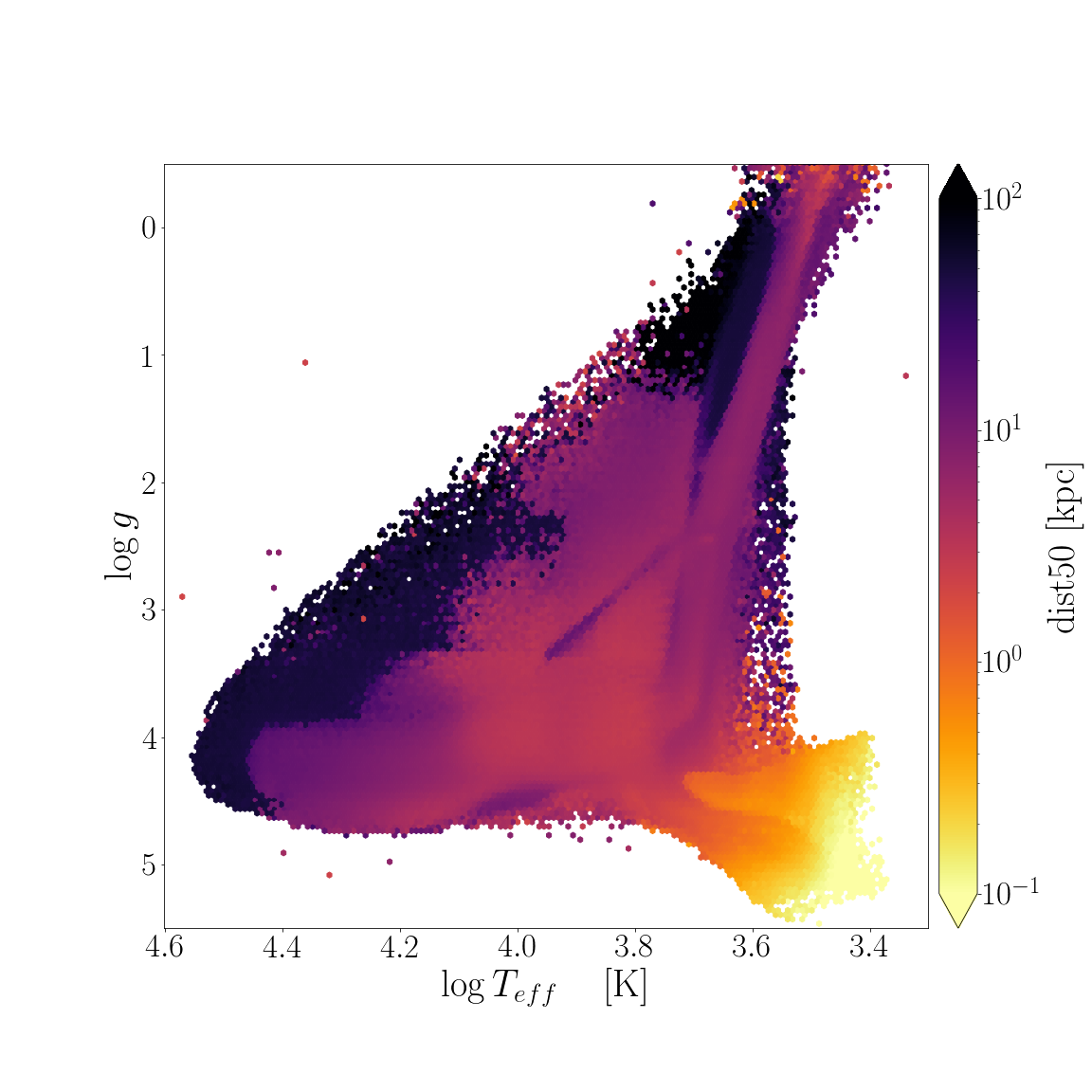}
    \caption{{\tt StarHorse}-derived {\it Kiel} diagrams (before applying any quality cuts). Left: Density plot. Middle: Colour-coded by median metallicity. Right: Colour-coded by median distance.
    }
 	\label{kieldiagrams}
\end{figure*}

\begin{figure*}\centering
 	\includegraphics[width=0.4\textwidth, trim={0 2cm 0 4cm}, clip=True]{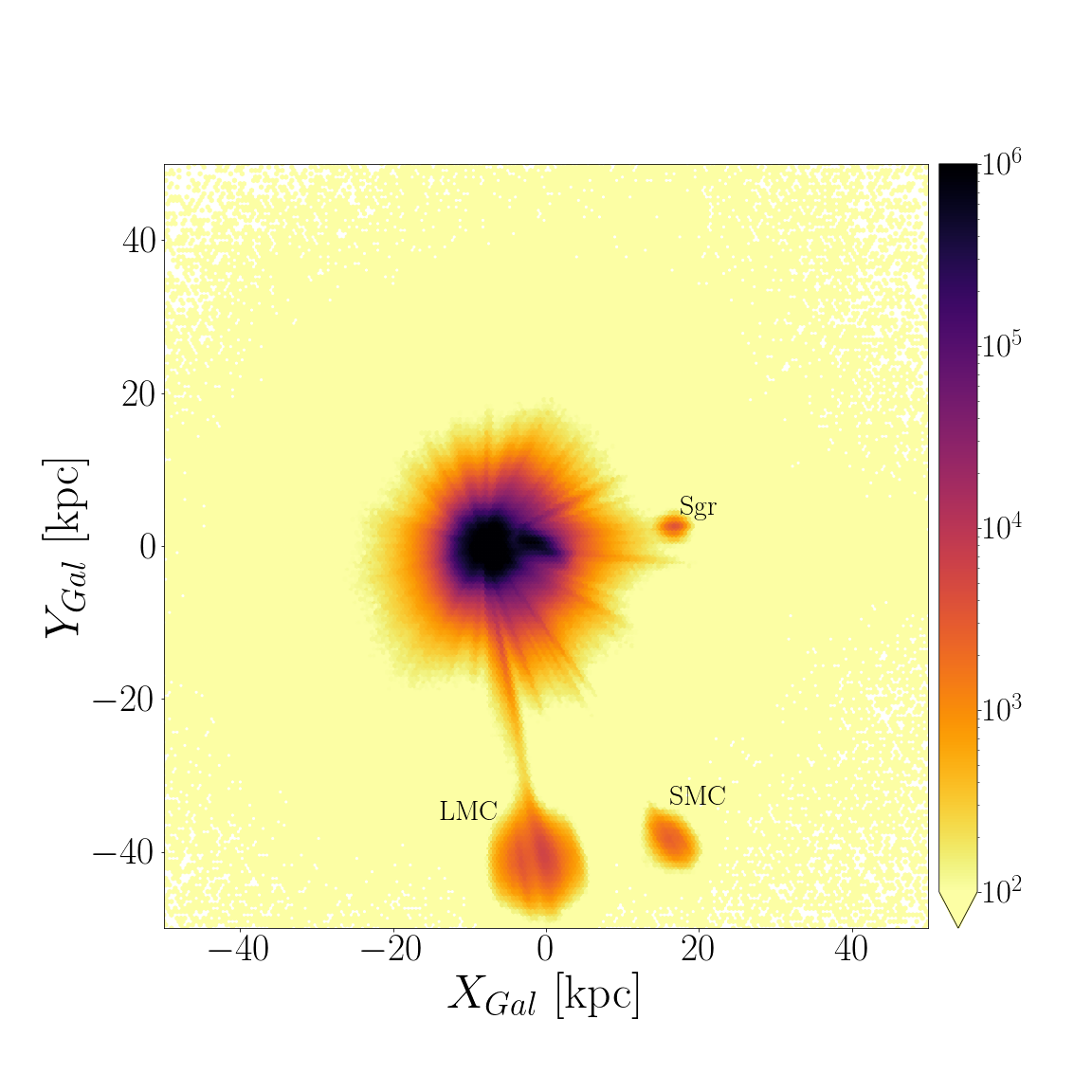}
 	\includegraphics[width=0.4\textwidth, trim={0 2cm 0 4cm}, clip=True]{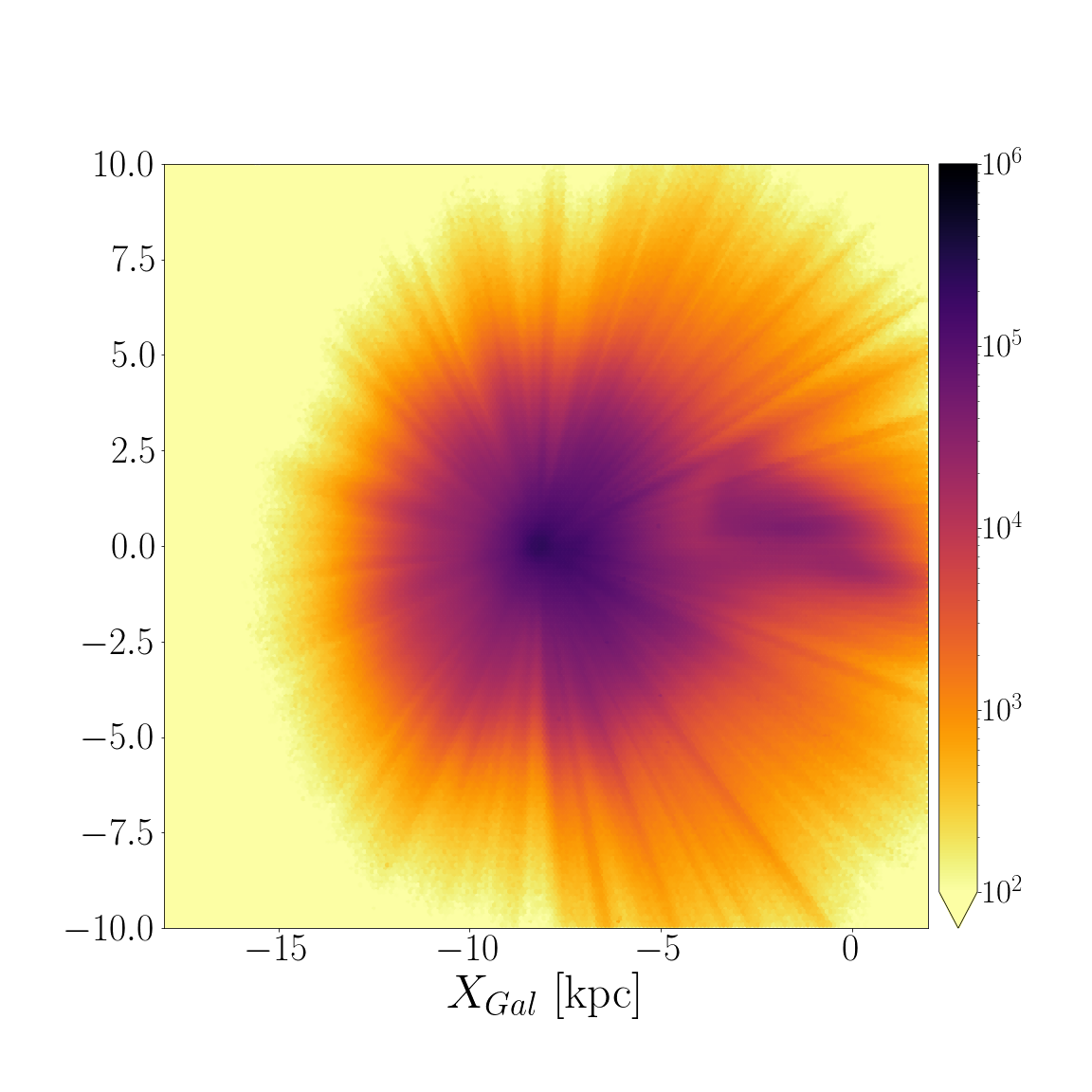}\\
 	\includegraphics[width=0.4\textwidth, trim={0 2cm 0 4cm}, clip=True]{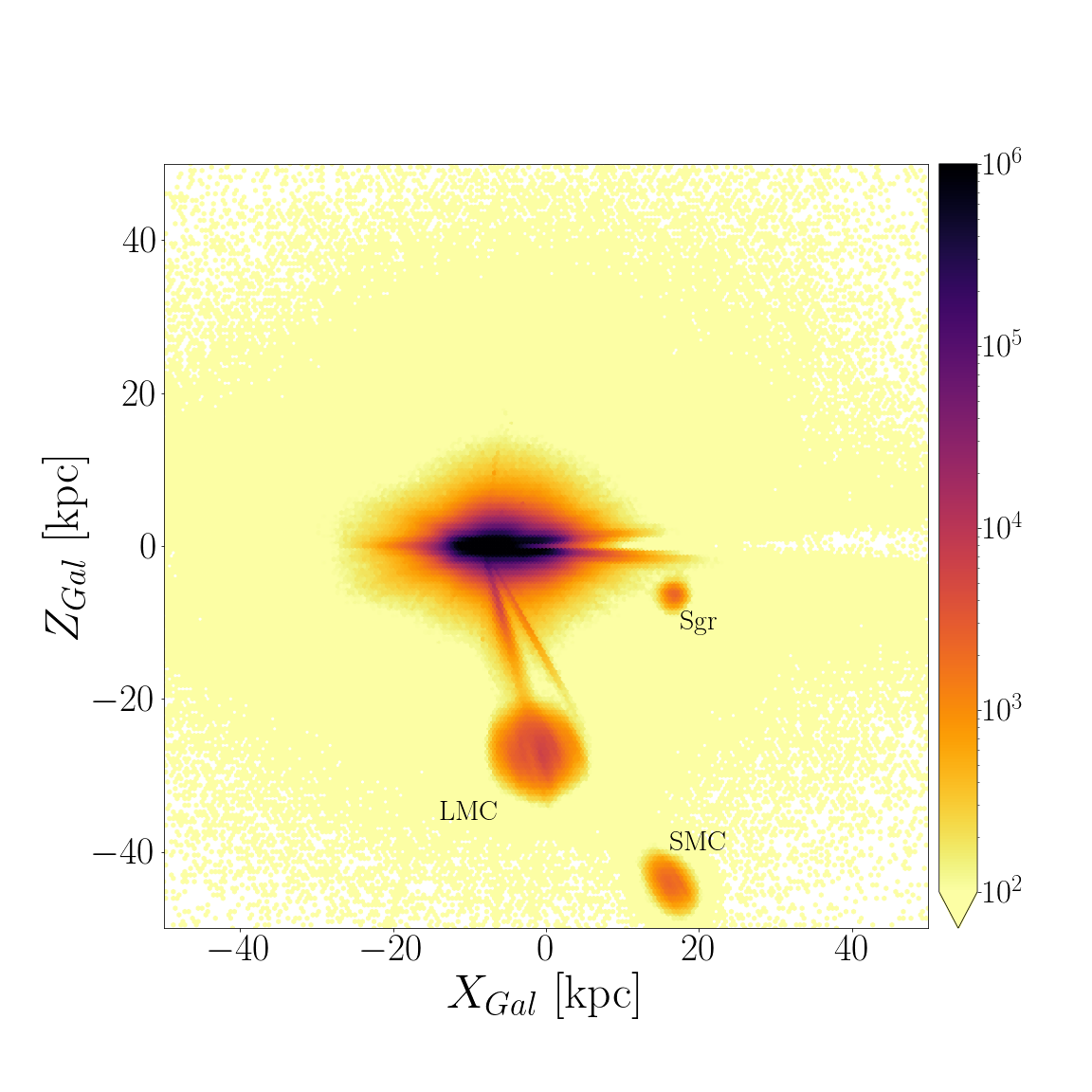}
 	\includegraphics[width=0.4\textwidth, trim={0 2cm 0 4cm}, clip=True]{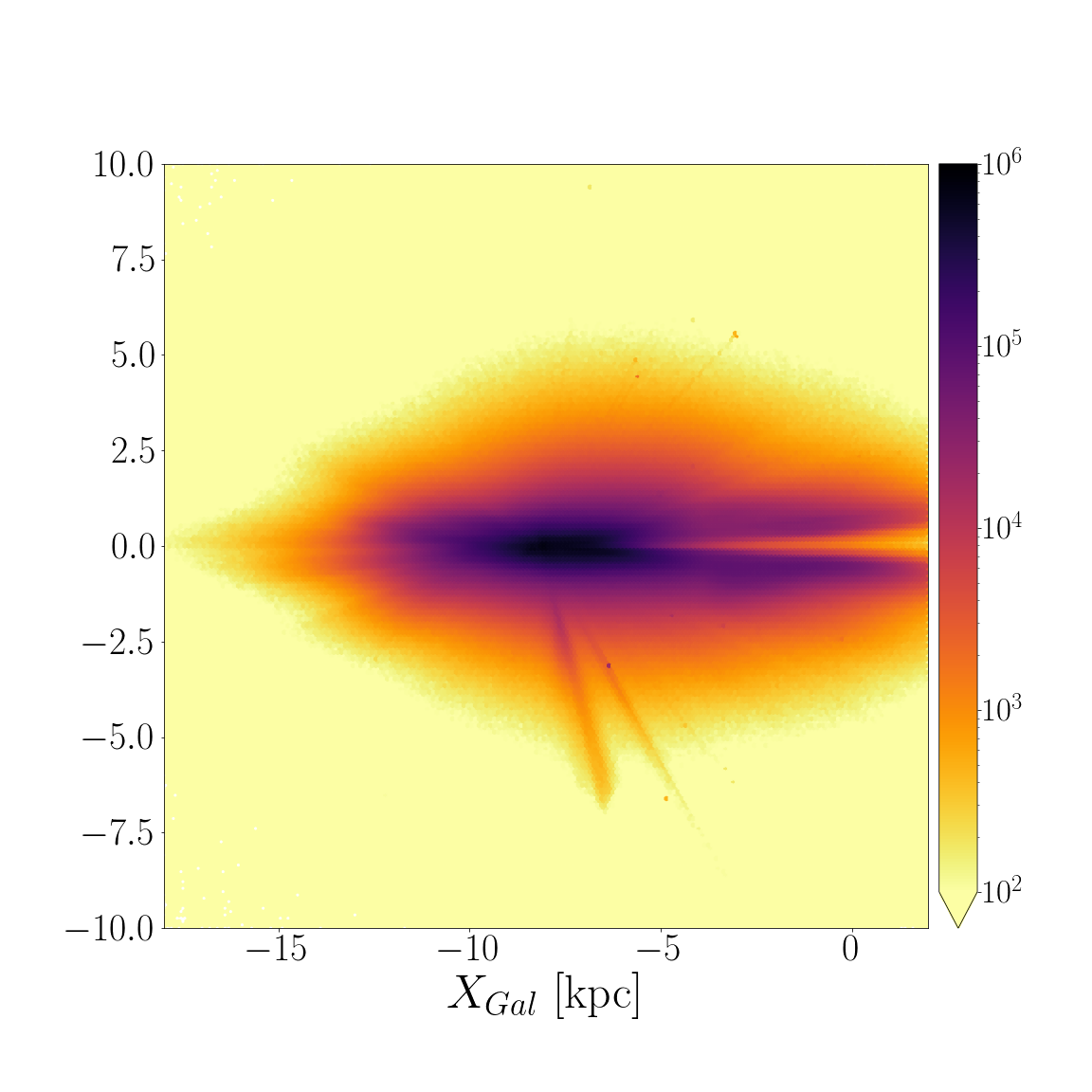}\\
 	\includegraphics[width=0.4\textwidth, trim={0 2cm 0 4cm}, clip=True]{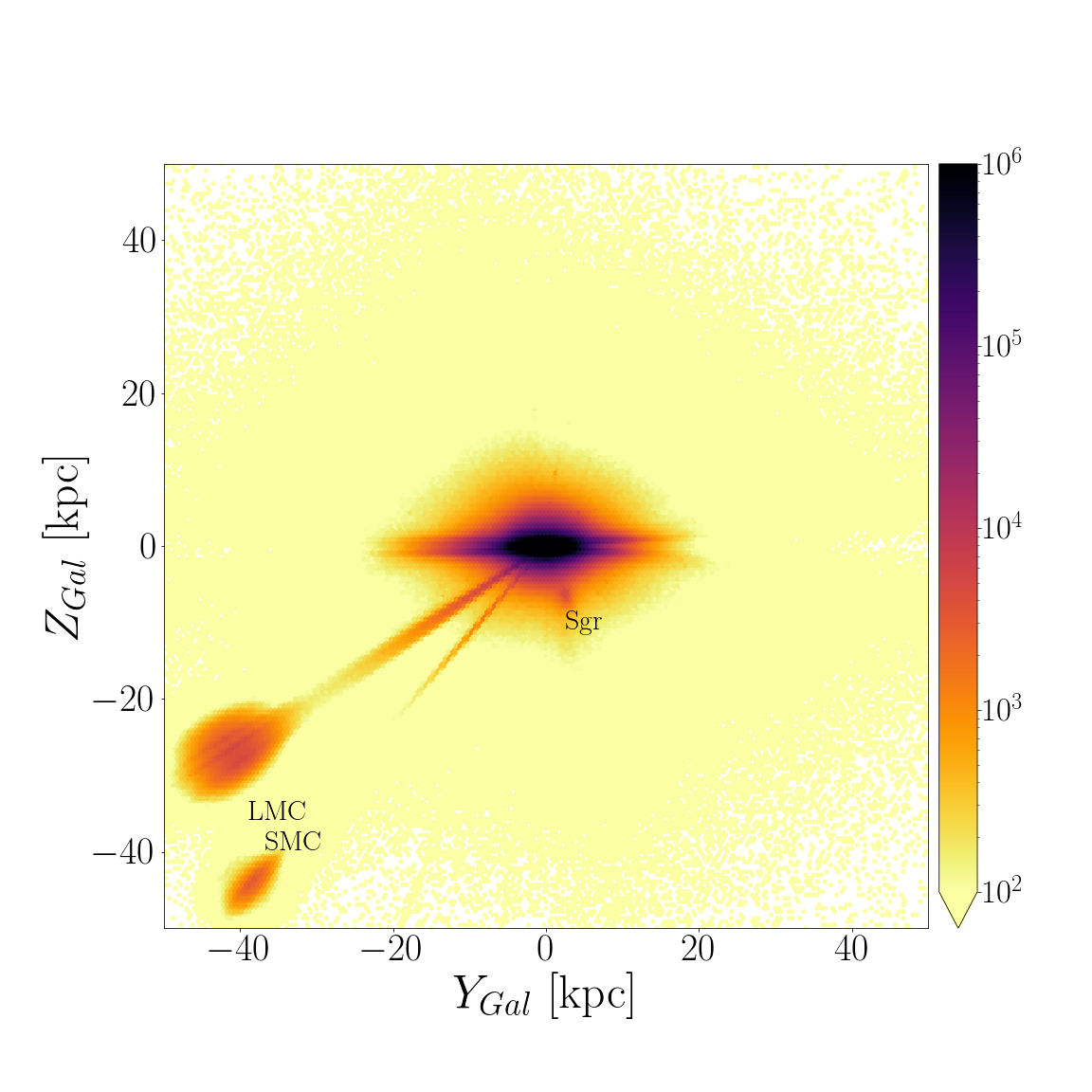}
 	\includegraphics[width=0.4\textwidth, trim={0 2cm 0 4cm}, clip=True]{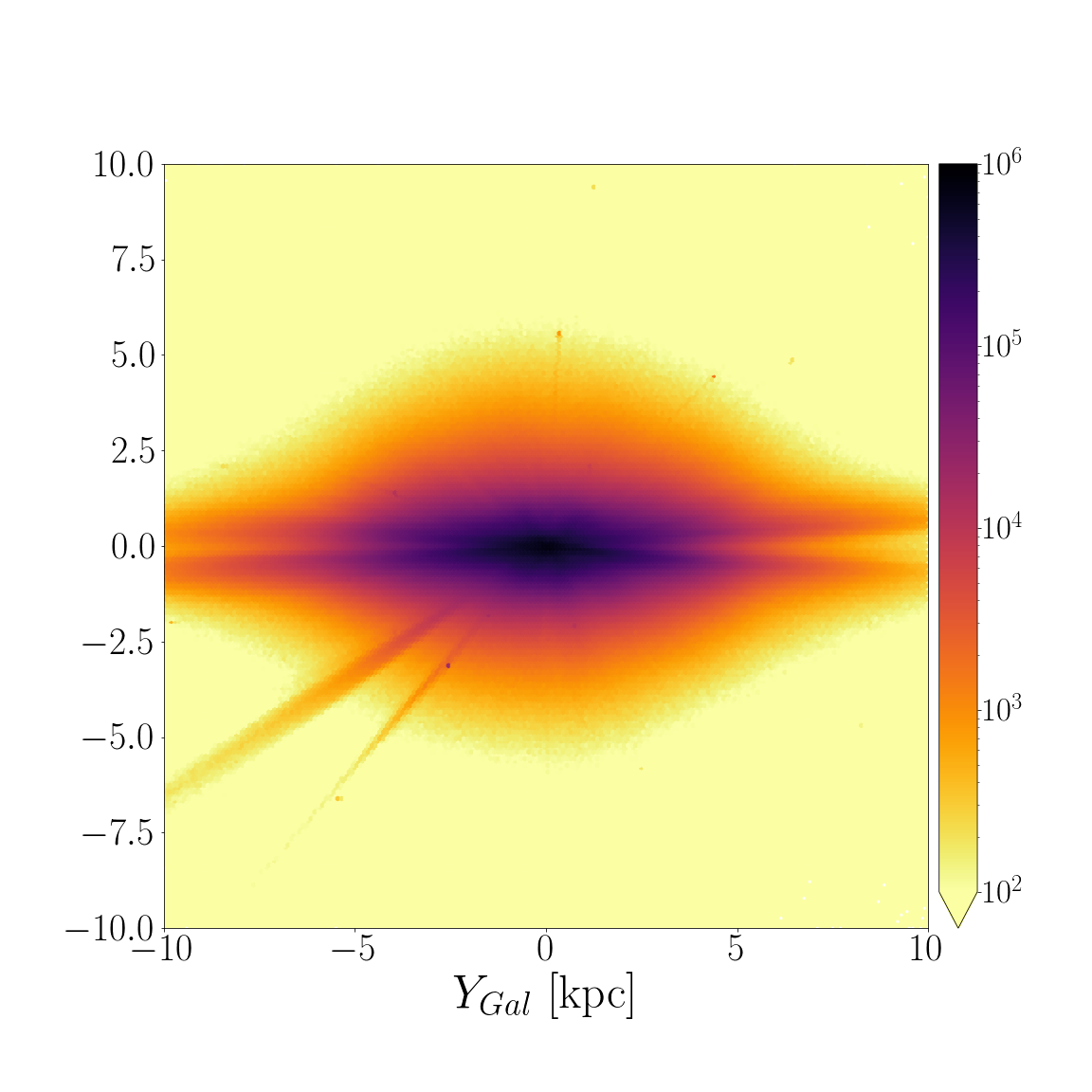}
 	\caption{{\tt StarHorse} density maps (from top to bottom: $XY$, $XZ$, and $YZ$) in galactocentric coordinates. The left column shows a 100 kpc wide cube centred on the Galactic centre, while the right column zooms into a 20 kpc wide cube centred on the Sun.}
 	\label{rzmaps}
\end{figure*}

\begin{table*}
\centering
\caption{Global statistics of some of the currently available astrometric and astro-photometric results based on {\it Gaia} DR2 and EDR3 data, in comparison to this work.}
\begin{tabular}{l|c|c|r|c|c|c}
Reference & Data used & mag limit & \# objects & $\sigma_d / d$\tablefootmark{G=17} & $\sigma_{A_V}$\tablefootmark{G=17}  & $\sigma_{T_{\rm eff}}$\tablefootmark{G=17}  \\
\hline
\citet{BailerJones2018}  &  DR2 parallaxes & $G\lesssim21$ & 1330M & 24\% & -- & -- \\
\hline
\multirow{2}{*}{\citet{Andrae2018}} & \multirow{2}{*}{DR2 photo-astrometry} &\multirow{2}{*}{$G\leq17$} &161M & -- & -- &324 K \\
                                    &                                       &                       & 88M & -- & 0.46 mag & -- \\
\hline
\citet{Anders2019}  &  DR2 + 2MASS+AllWISE  & \multirow{2}{*}{$G<18$} & 266M  & 40 \% & 0.25 mag & 350 K \\
\qquad flag-cleaned sample & +Pan-STARRS1 & & 137M & 18 \% & 0.23 mag & 230 K   \\
\hline
\citet{Green2019}  &  DR2 + 2MASS+Pan-STARRS1 & $z_{PS1}<20.9$ & 799M & 20 \% & 0.15 mag & -- \\
\hline
\citet{Bai2019, Bai2020}  &  DR2 photo-astrometry & $G\lesssim17$ & 133M & -- & 0.16 mag & 350 K \\
\hline
\multirow{2}{*}{\citet{BailerJones2021}}  &  EDR3 parallaxes & \multirow{2}{*}{$G\lesssim21$} & 1470M & 20 \% & -- & -- \\
                         &  EDR3 photo-astrometry & & 1310M & 16 \% & -- & -- \\
\hline
This work  &  EDR3 photo-astrometry & \multirow{5}{*}{$G<18.5$} & 402,431,354  &  &  &  \\
\qquad {\color{black} {\tt StarHorse} converged} &  +2MASS &  & 362,392,321  & \multirow{4}{*}{15 \%}  & \multirow{4}{*}{0.15 mag} & \multirow{4}{*}{183 K} \\
\qquad \& {\tt fidelity} $>0.5$ & +AllWISE & & 329,646,544 & &  &    \\
\qquad \& $|C^{\ast}|/\sigma_{C^{\ast}} < 5$ & +Pan-STARRS1 & & 321,131,855 & &  &    \\
\qquad \& {\tt sh\_outflag}$==$"0000" & +SkyMapper & & 281,501,963 & &  &   \\
\hline
\end{tabular}
\tablefoot{
\tablefoottext{G=17}{For comparability, we report here the median precision for stars at magnitude $G\approx17$.}
}
\label{summarytable}
\end{table*}

\subsection{Input and output flags}\label{flags}

Along with the output of our code (median statistics of the marginal posterior in distance, extinction, and stellar parameters), we provide a set of flags to help the user decide which subset of the data to use for their particular science case. These flags correspond to the columns defined in the next few subsections.

\subsubsection{{\it Gaia} EDR3 quality criteria used in this work}\label{gaiaflag}

In the previous {\tt StarHorse} {\it Gaia} DR2 run, we defined a set of input flags (summarised in the column {\tt SH\_GAIAFLAG}) based on the DR2 recommendations by the {\it Gaia} Collaboration (e.g. \citealt{Lindegren2018}). It contained three digits corresponding to astrometric fidelity (in particular the renormalised unit-weight error, {\tt ruwe}; \citealt{Lindegren2018a}), the photometric fidelity (indicated by the {\tt phot\_bp\_rp\_colour\_excess}), as well as the DR2-native {\tt variability\_flag}.

In this work we make use of the quality criteria established by \citet{Rybizki2021} and \citet{Riello2021} who have addressed these questions in detail and provide recipes to select high-quality EDR3 measurements. We thus follow their recommendations and use the following cuts:

{\it Astrometric fidelity:} We cross-matched our catalogue with the astrometric {\tt fidelity} flag defined by \citet{Rybizki2021}, based on a neural-network classifier for EDR3 objects. The classifier uses the twelve EDR3 astrometric columns identified by \citet{Smart2021} as containing most information about the fidelity of the EDR3 parallaxes and proper motions (and their uncertainties). It was trained on a set of bona fide trustworthy and bona fide bad EDR3 results. Bad astrometric results can be culled by requiring, for example, {\tt fidelity}$>0.5$.

{\it Colour excess factor:} The corrected version of the EDR3 {\tt phot\_bp\_rp\_colour\_excess} column, $C_{\ast}$ or {\tt bp\_rp\_excess\_corr} \citep[][see also Appendix B of \citealt{GaiaCollaboration2021}]{Riello2021}, indicates whether the BP/RP photometry of a {\it Gaia} source may be affected by background flux from neighbouring objects. When cleaning the {\tt StarHorse} results for potentially affected BP/RP photometry, we recommend using a cut of $|C_{\ast}| / \sigma_{C^{\ast}} < 5$, where $\sigma_{C^{\ast}}$ is a simple function of the $G$ magnitude, computed according to Eq. 18 in \citet{Riello2021}.

\subsubsection{\tt sh\_photoflag}

As in \citetalias{Anders2019}, we define the human-readable {\tt sh\_photoflag} that contains the information about the combination of photometric input data used for each object ({\it Gaia} EDR3, PS1, SkyMapper, 2MASS, AllWISE). For example, if only {\it Gaia} EDR3 $G, G_{RP}$ and 2MASS $HK_s$ magnitudes were available, the flag reads {\tt GRPHKs}. 
PS1 and SkyMapper photometry are separated by a slash ({\tt /}) in the {\tt sh\_photoflag}: for example, the flag {\tt Gg/riW1W2} means that the object in question has good {\it Gaia} $G$, PS1 $g$, SkyMapper $ri$, and AllWISE $W1W2$ measurements, while {\tt G/g} means that the object has only {\it Gaia} $G$ and SkyMapper $g$.

We note that with respect to \citetalias{Anders2019} we improved the quality filters especially for the input AllWISE and 2MASS data, as well as for the {\it Gaia} BP/RP photometry (see Sect. \ref{data}).

\subsubsection{\tt sh\_outflag}\label{outflag}

In \citetalias{Anders2019}, we defined a {\tt StarHorse} output flag, consisting of five digits that informed about the fidelity of the {\tt StarHorse} output parameters. The first digit served as the main quality indicator and filtered out stars with inconsistent median output parameters. Although the main caveats of the \citetalias{Anders2019} results have been rendered obsolete by EDR3, we still define an output flag for convenience. It contains the following four digits:

The first digit flags low number of consistent models. For some targets, the number of stellar models in our model grid found to be 3$\sigma$-consistent with the data is low, indicating either very precise results or (more likely) some tension in the input data. We consider a results unproblematic if the number of models is greater than 30, and apply a (strong) warning flag if this number is between 10 and 30 (below 10): IF ${\tt nummodels}>30$ THEN 0 ELIF ${\tt nummodels}>10$ THEN 1 ELSE 2.

The second digit flags negative extinction. Significantly negative extinctions should be treated with care: IF ${\tt AV95} > 0$ THEN 0 ELSE 1.

The third digit warns about very large uncertainties. Large uncertainties are not problematic per se, but the corresponding median values are not usually very informative, which is why we provide this flag to be able to filter out very uncertain results quickly. The definition is as follows: IF $0.5*{\tt (dist84-dist16)/dist50}>1$ OR $0.5*({\tt AV84-AV16})>1$ OR $0.5*({\tt teff84-teff16})>1000$ OR $0.5*({\tt logg84-logg16})>1$ OR $0.5*({\tt met84-met16})>1$ OR $0.5*{\tt (mass84-mass16)/mass50}>1$ THEN 1 ELSE 0.

The fourth digit flags very small uncertainties. Very small posterior uncertainties are most likely underestimated and probably indicate poor convergence. These results should also be used with care. The definition is as follows: IF $0.5*{\tt (dist84-dist16)/dist50}<0.001$ OR $0.5*({\tt av84-av16})<0.01$ OR $0.5*({\tt teff84-teff16})<20.$ OR $0.5*({\tt logg84-logg16})<0.01$ OR $0.5*({\tt met84-met16})<0.01$ OR $0.5*{\tt (mass84-mass16)/mass50}<0.01$ THEN 1 ELSE 0.

Unproblematic results from the point of view of {\tt StarHorse} can thus be filtered by requiring {\tt sh\_outflag}$==$"0000".

\begin{figure}\centering
 	\includegraphics[width=0.5\textwidth]{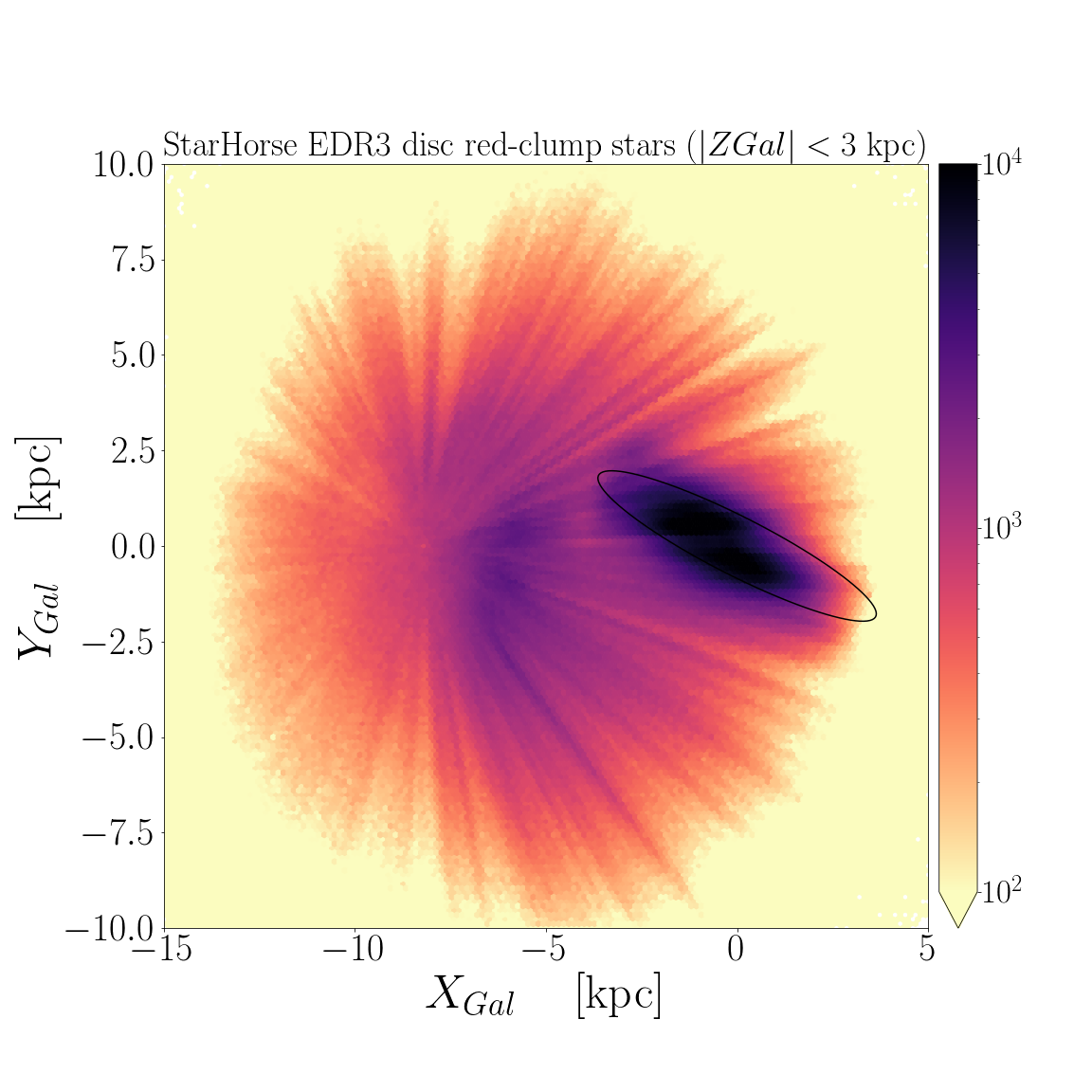}
 	\caption{XY density map, selecting all (13.8M) red-clump stars less than 3 kpc away from the Galactic midplane. The ellipse shows the orientation (27 deg with respect to the Sun-Galactic centre line) and approximate extent (semi-major axes $a=4.07$ kpc and $b=0.76$ kpc) of the Galactic bar assumed in the prior.}
 	\label{bar_rc_stars}
 \end{figure}

\begin{figure}\centering
 	\includegraphics[width=0.5\textwidth]{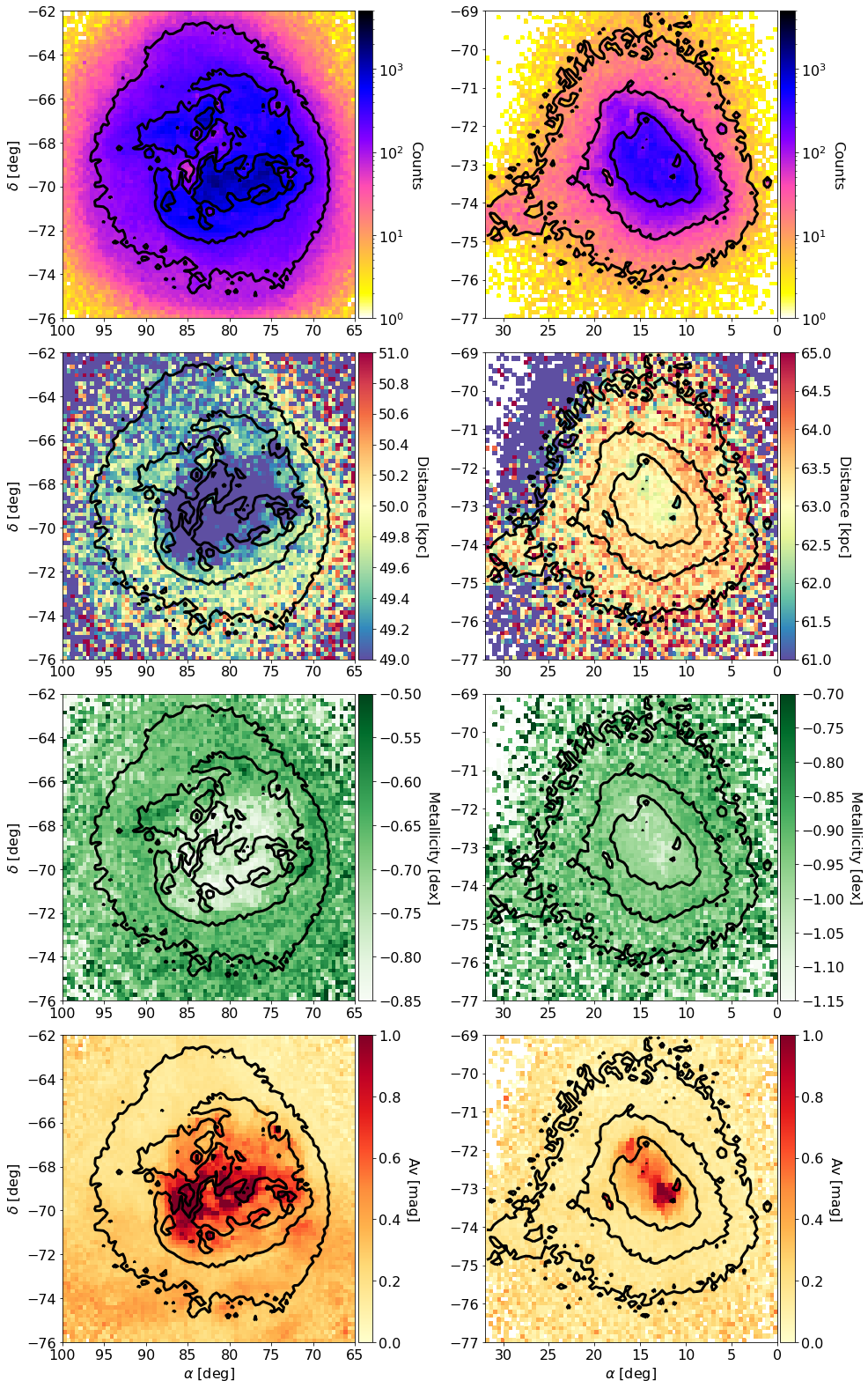}
 	\caption{Median sky density, distance, metallicity, and extinction maps (from top to bottom) of the Magellanic Clouds as seen by {\tt StarHorse} (in equatorial coordinates and only including objects with {\tt dist50} $> 25$ kpc). The left panels are centred on the LMC, the right panels on the SMC. The contour lines in each of the panels are derived from the sky density plots in the top panels. For the LMC, the contours are drawn at stellar densities of $[100, 300, 700]$ per pixel (from outside inwards), with $905,205$ sources within the outermost contour. For the SMC, the contour lines correspond to levels $[10, 50, 200]$, with $195,634$ sources contained inside the outermost contour.
    }
 	\label{magclouds1}
 \end{figure}

\begin{figure}\centering
 	\includegraphics[width=0.5\textwidth, trim={0 2cm 0 4cm}, clip=True]{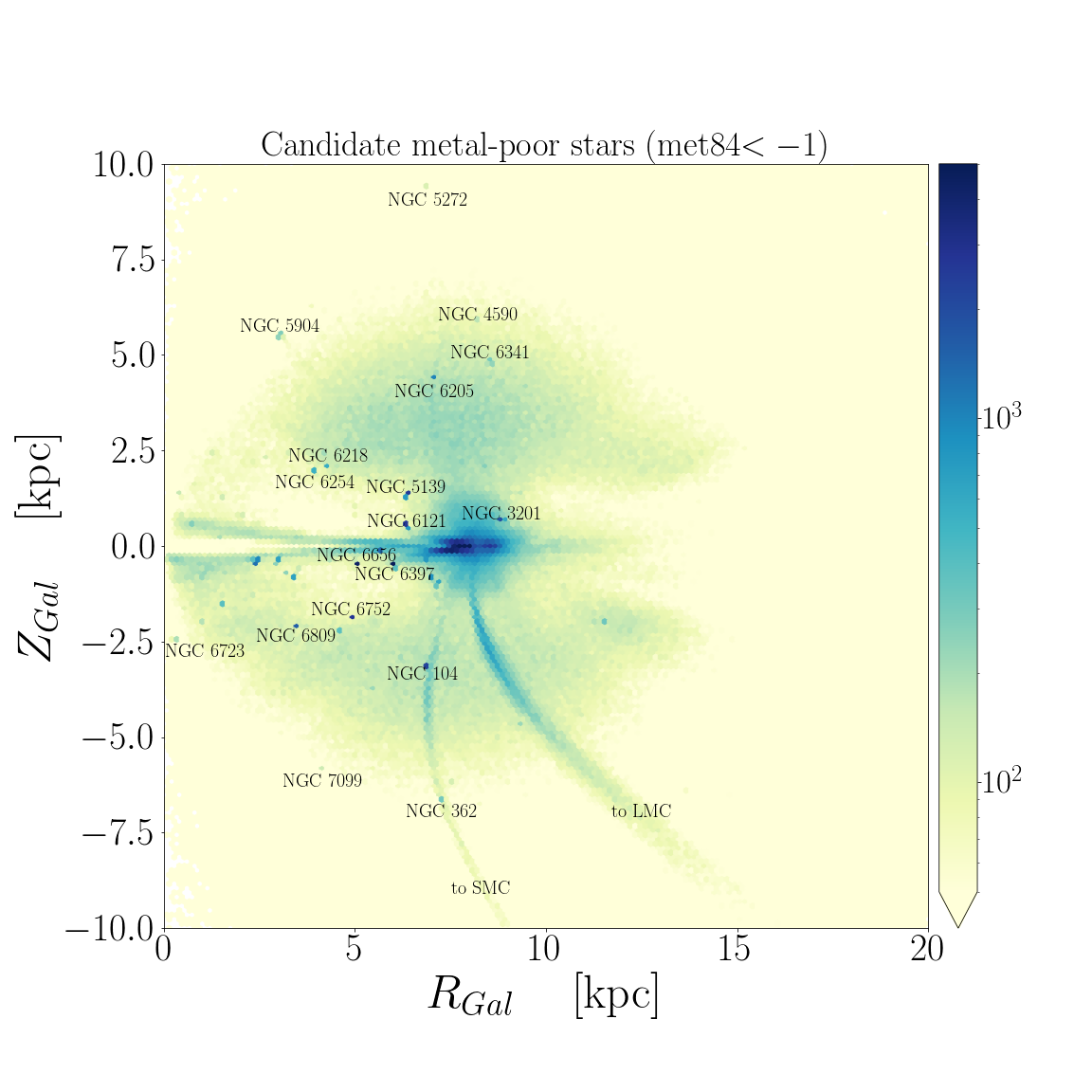}
 	\caption{Density map for bona fide candidate metal-poor stars ({\tt met84} $<-1$; 1.58M stars) in galactocentric coordinates. Some prominent overdensities corresponding to Galactic globular clusters and the direction towards the Magellanic System are annotated.}
 	\label{maps_mp}
\end{figure}

\begin{figure}\centering
 	\includegraphics[width=0.5\textwidth]{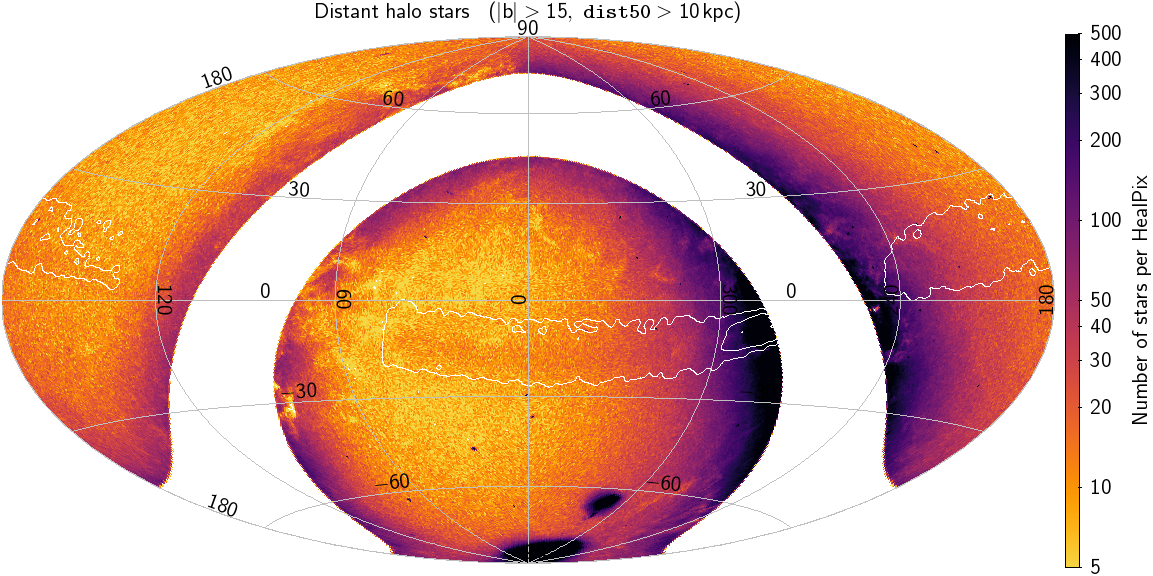}
 	\includegraphics[width=0.5\textwidth]{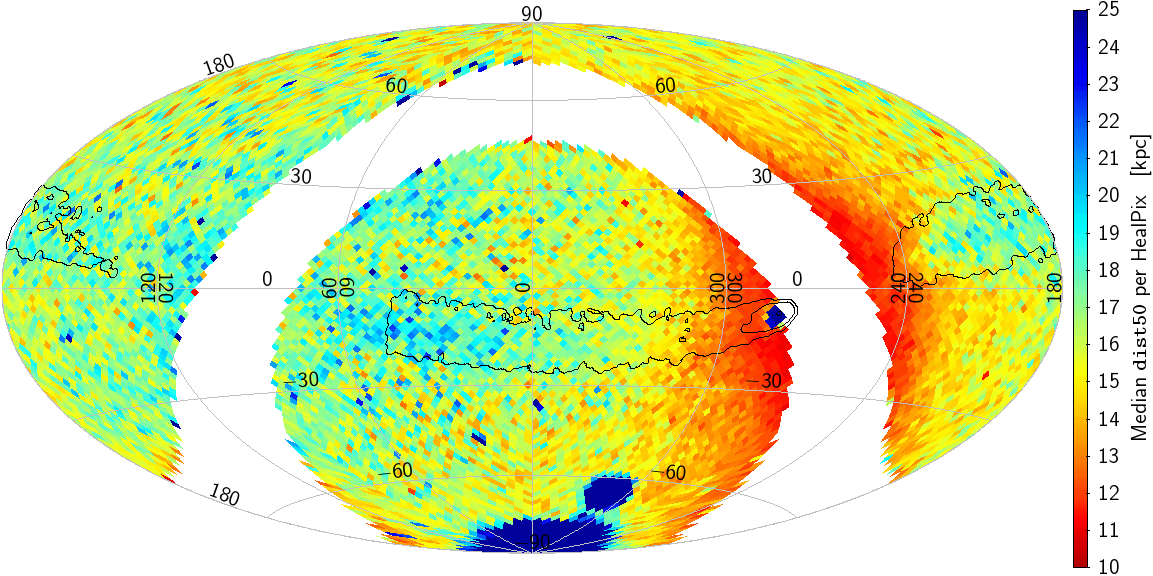}
 	\includegraphics[width=0.45\textwidth]{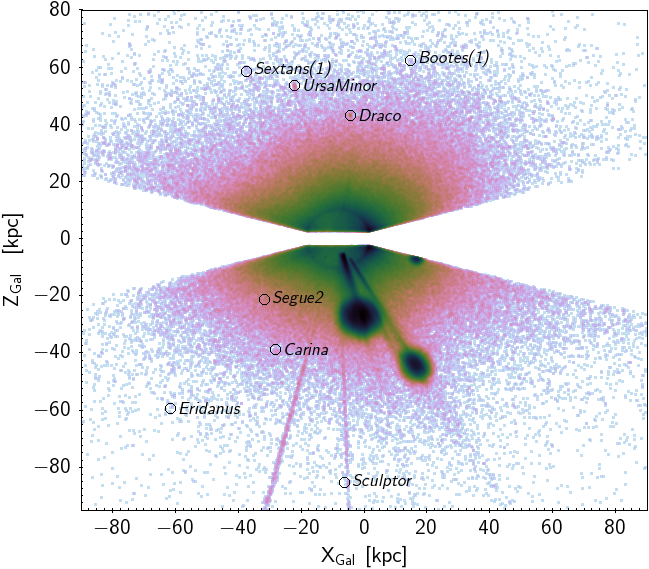}
 	\caption{Distribution of distant halo stars, selected by excluding the Galactic plane and a cut in median distance ($|b|>15$ deg, {\tt dist50} $>10$ kpc; 2.55M stars). The top panel shows the sky distribution in ecliptic coordinates, highlighting the presence of the Sagittarius stream close to the ecliptic plane. The middle panel shows the same projection, colour-coded by the median distance per HealPix. In both panels the contour overlay shows the location of the Sagittarius stream candidates from \citet{Antoja2020}. The bottom panel shows a Cartesian projection ($X_{\rm Gal}$ vs. $Z_{\rm Gal}$), highlighting some of the less prominent Local Group objects included in the priors.}
 	\label{halosample}
\end{figure}

\begin{figure}\centering
 	\includegraphics[width=0.49\textwidth]{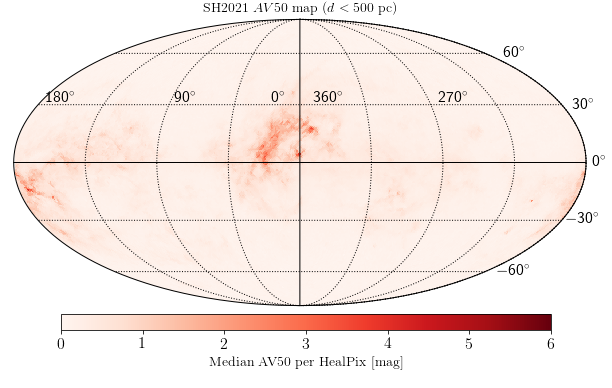}
 	\includegraphics[width=0.49\textwidth]{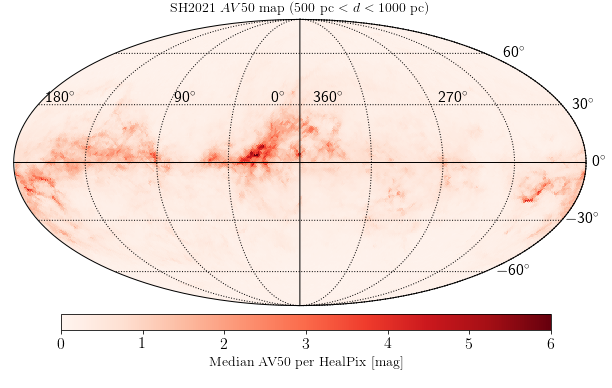}
 	\includegraphics[width=0.49\textwidth]{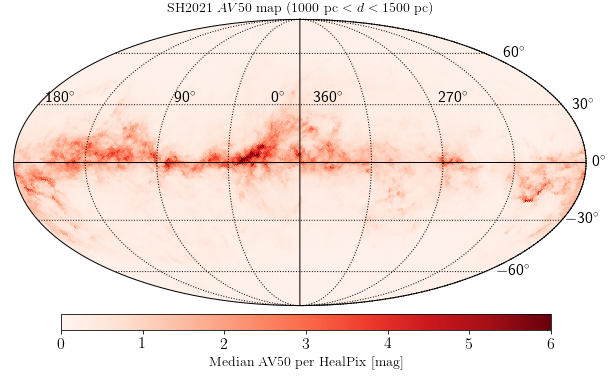}
 	\includegraphics[width=0.49\textwidth]{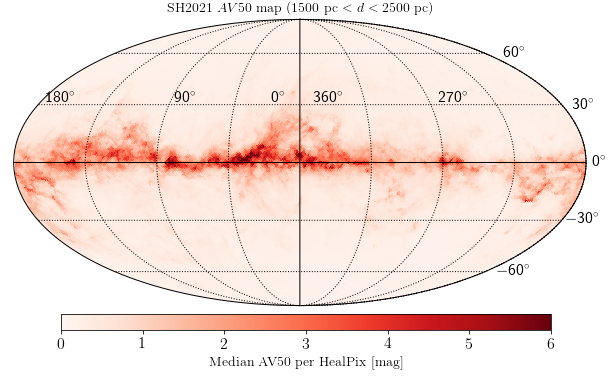}
 	\caption{All-sky median {\tt StarHorse} extinction map for four wide distance bins up to 2.5 kpc, as indicated in each of the subplots.
    }
 	\label{avmaps}
\end{figure}

\section{{\tt StarHorse} {\it Gaia} EDR3 results}\label{results}

\subsection{Summary}\label{summary}

Table \ref{summarytable} summarises the results of the {\tt StarHorse} run for {\it Gaia} EDR3 as well as previous results available from the recent literature. We observe that our new {\tt StarHorse} results compare favourably in terms of both sample size and parameter precision. For example, the results have notably improved in precision (typically shrinking the formal uncertainties by a factor of 2) with respect to \citetalias{Anders2019} (see Sect. \ref{comp} for a more detailed comparison). 

Figure \ref{corner} shows the distribution of the {\tt StarHorse} median posterior output values $T_{\rm eff}, \log g,$ [M/H], $M_{\ast}, d, $ and $A_V$ and their corresponding uncertainties, demonstrating the complexity of the dataset as well as the typical precision (discussed in more detail in Sect. \ref{precision}). Even the median output parameters are highly correlated, either intrinsically (enforced by the stellar models, e.g. $T_{\rm eff}$ vs. $\log g$), due to selection effects (e.g. $d$ vs. $M_{\ast}$), or because of degeneracies related to our method ($\sigma_{T_{\rm eff}}$ vs. $\sigma_{A_V}$).
The gridding effect in the metallicity panels of Fig. \ref{corner} is due to the finite resolution of the model grid.

Fig. \ref{skydensity} shows the sky distribution of the input sample (400M stars with $G<18.5$), as well as sky maps of the percentage of converged sample and the cleaned sample. The figure shows that the code convergence is lowest in the densest areas of the sky (the innermost bulge and Galactic plane as well as the centre of the Large Magellanic Cloud), and that cleaning the {\it Gaia} data enhances this effect. For example, the X shape in the inner bulge visible in the bottom panel of Fig. \ref{skydensity} is mainly produced by the quality cut in the {\it Gaia} EDR3 colour excess factor (compare to Fig. 21 of \citealt{Riello2021}).

In the following subsections, we present some immediate results that can be obtained from our catalogue, focussing on CMDs (Sect. \ref{cmds}), {\it Kiel} diagrams (Sect. \ref{kiel}), stellar density maps (Sect. \ref{maps}), and extinction maps (Sect. \ref{extmaps}).

\subsection{Extinction-corrected colour-magnitude diagrams}\label{cmds}

Since the stellar models used in our Bayesian inference have not changed much with respect to \citetalias{Anders2019}, the {\tt StarHorse} extinction-corrected CMDs are also similar. The top row of Fig. \ref{cmds1} shows the CMD of the total sample and two interesting subsamples (the {\it Gaia}-cleaned sample and the fully flag-cleaned sample). When comparing these panels to Fig. 5 in \citetalias{Anders2019}, we note that some of the previously noted unphysical features have disappeared (most notably, the 'nose' between the main sequence and the lower red-giant branch). On the other hand, new structure in the top parts of the full CMDs emerges from the explicit inclusion of the Magellanic Clouds in the priors. For illustration, the bottom row of Fig. \ref{cmds1} shows the populations of the Milky Way disc, the Large Magellanic Cloud (LMC), and the Small Magellanic Cloud (SMC).

The second row of Fig. \ref{cmds1} shows the CMDs for three bins in apparent magnitude. The overall appearance of the magnitude-binned CMDs in Fig. \ref{cmds1} resembles those of Fig. 4 in \citetalias{Anders2019}, with a few notable differences. For example, in addition to the sharp features of single-star evolution in the $G<14$ panel, we now also appreciate the unresolved binary sequence right above the low-mass main sequence. We also see the impact of the LMC and SMC populations on the CMD, already in the magnitude bin $16<G<17$. The rightmost middle panel, corresponding to $18<G<18.5$, shows already significant broadening in the CMD features. As in \citetalias{Anders2019}, this is a result of the growing uncertainty in the input parameters, especially the parallax. 
We also recall that the absolute magnitudes and de-reddened colours displayed in Fig. \ref{cmds1} are not a direct output of {\tt StarHorse}, but were computed from the observed magnitudes and the {\tt StarHorse} median distance, extinction, and effective temperatures\footnote{\bf \url{https://github.com/fjaellet/gaia_edr3_photutils}}.

\subsection{{\it Kiel} diagrams}\label{kiel}

Figure \ref{kieldiagrams} shows {\it Kiel} diagrams ($T_{\rm eff}$ vs. $\log g$) for the full {\it Gaia} EDR3 {\tt StarHorse} sample. The density plot (left panel) shows that most of the sample is classified as FGK stars, as expected. Also clearly visible (both in the left and the middle panel) are the stripe-like overdensities corresponding to the metallicity resolution of the stellar model grid already noted in Sect. \ref{summary}.

We note a much more defined horizontal branch with respect to \citetalias{Anders2019}, which is at least in part due to the metallicity prior for globular clusters. We also note a more populated pre-main sequence (region above the lower main sequence), since we now applied a slightly less restrictive age cut ($\log t>7$, as compared to $\log t>7.5 $ in \citetalias{Anders2019}).

The middle panel of Fig. \ref{kieldiagrams} ({\it Kiel} diagram colour-coded by metallicity) shows that the posterior metallicity information is consistent with the stellar model grid through most of the parameter space. The only few outliers from the space spanned by the stellar models  are stars whose median output parameters lie in-between the main sequence and the giant branch (due to a significantly bimodal posterior). The number of those stars (for which the median {\tt StarHorse} are unreliable) has diminished enormously with respect to \citetalias{Anders2019}.

Finally, the right panel of Fig. \ref{kieldiagrams} shows the typical distance range sampled for different regions of the {\it Kiel} diagram (also visible in Fig. \ref{corner}), showing the expected behaviour of large typical distances (even $>100$ kpc) for the most luminous stars and very small distances for the coolest and least massive dwarf stars ($<100$ pc; see e.g. \citealt{Smart2021}).

\subsection{Stellar density maps}\label{maps}

\subsubsection{Overall density distribution}

One of the main motivations for the {\tt StarHorse} project is Galactic cartography, and some of the newly implemented changes in the code (see Sect. \ref{updates}) result in a visible improvement of the stellar density maps. To illustrate this, Fig. \ref{rzmaps} shows two-dimensional projections of the stellar density distribution for the full {\tt StarHorse} sample in Cartesian galactocentric coordinates. The left column of the plot focuses on larger structures: the Galactic volume probed by {\it Gaia} and the neighbouring dwarf galaxies, as indicated in each panel. These populations are now clearly visible as overdensities in the maps, although a considerable amount of stars still has median distances that fall in between the Magellanic Clouds and the Milky Way - a result of the multimodal posterior distance distributions \citepalias[see e.g.][]{Anders2019}.

The right panels of Fig. \ref{rzmaps} zoom into a 20 kpc wide cube centred on the Sun. When we compare these maps to the ones presented in Fig. 7 of \citetalias{Anders2019}, we notice: 1. the increase in total stellar number density (from 137 million to 360 million stars), and 2. the greater volume probed by the {\it Gaia} EDR3 $G<18.5$ sample.

Direct consequences of the maps shown in Fig. \ref{rzmaps} for Galactic cartography, however, are not obvious, since these maps are the result of a complex convolution of the true stellar density distribution, interstellar extinction, the applied magnitude limit, the selection function, and the priors. In the following subsections, we discuss the density maps of some specific stellar populations that are arguably easier to interpret.

\subsubsection{Red-clump stars}

Core helium burning red-clump stars \citep[for a review see][]{Girardi2016} are often used as standard candles for mapping Galactic populations. They are numerous, relatively bright, and span a wide range of ages and metallicities.

Figure \ref{bar_rc_stars} shows the distribution of disc red-clump stars in the {\tt StarHorse} {\it Gaia} EDR3 catalogue. The stars have been selected using the {\it Kiel} diagram as in Sect. 4.4 of \citetalias{Anders2019}: 4500 K$<T_{\rm eff}<5000$ K, $2.35<\log g<2.55$, $-0.6<$[M/H]$<+0.4$, $|Z| < 3$ kpc. Figure \ref{bar_rc_stars} can thus be directly compared to Fig. 8 in \citetalias{Anders2019}. 

In Fig. 8 of \citetalias{Anders2019} we observed a very clear overdensity of red-clump stars tracing the Galactic bar. This result was all the more convincing since the bar angle used in the prior was significantly different from the one observed in the posterior distribution. However, Fig. 8 of \citetalias{Anders2019} also displayed some minor artefacts, such as underdensities of red-clump stars both in front of and behind the near side of the bar, or an underdense ring-like structure that arose from the quality cuts necessary to clean the DR2 {\tt StarHorse} data.

The EDR3 version of that figure, shown in Fig. \ref{bar_rc_stars}, shows that the result of \citetalias{Anders2019} (the detection of the Galactic bar in stellar density) is clearly maintained. The number of red-clump stars has become greater (13.8M vs. 10.8M), the underdensity artefacts of the map are greatly reduced, and the probed Galactic area now extends to regions beyond the (near side of the) bar.

The apparent bar angle is similar to the one in Fig. 8 of \citetalias{Anders2019} and thus still appears to be a few degrees higher than the one assumed in the prior (27 deg; see Sect. \ref{galupdates}). The main overdensity of the bar also appears relatively short compared to recent estimates of $\gtrsim5$ kpc. A quantitative analysis of the bar's structural parameters is, however, beyond the scope of this paper, as this requires careful modelling (e.g. \citealt{Wegg2015, Portail2017}) and taking into account selection effects.

Another feature in Fig. \ref{bar_rc_stars} is an overdensity appearing around $R_{\rm Gal}\sim6$ kpc that might correspond to the Sagittarius spiral arm (see e.g. \citealt{Reid2019}). This feature is much less clear in the red-clump stars than in maps of young stellar populations (e.g. \citealt{Castro-Ginard2021, Zari2021, Poggio2021}), and the map in Fig. \ref{bar_rc_stars} shows the underlying density distribution convolved with dust extinction and other selection effects. The clear overdensity in the (logarithmic) red-clump star count map is, however, a strong feature that deserves further investigation, since the strength of the spiral density signature in an intermediate-age population has implications on the modelling of the Milky Way's spiral arms. 

Recently, \citet{Nogueras2021} have used the high angular-resolution infrared photometric survey GALACTICNUCLEUS \citep{Nogueras2019} to determine the distances, extinctions, and stellar populations of the inner spiral arms in a small region of the sky containing the Galactic centre. While their data are of clearly superior quality, we suggest that similar mapping studies could be carried out using {\it Gaia} and multi-wavelength photometry (and possibly our {\tt StarHorse} catalogue) for the portions of the disc less affected by interstellar extinction.

\subsubsection{Magellanic Clouds}\label{magclouds}

The Magellanic Clouds as our immediate galactic neighbours represent a key laboratory to study gravitational interactions and their effects on the structure and kinematics of satellite galaxies. In this section, we analyse our results for the region of the Magellanic Clouds and compare them to the \citet{Luri2021} results. 

In Fig.~\ref{magclouds1} we show from top to bottom the sky density map, 2D distance distribution, metallicity and extinction maps for the sources around the Large Magellanic Cloud (LMC, left) and Small Magellanic Cloud (SMC, right), respectively, in equatorial coordinates.

For the LMC (left column of Fig.~\ref{magclouds1}), the sky density distribution highlights the main components of the galaxy. The innermost contour encloses the elongated bar, while the second contour highlights the spiral arm. We notice a small region with low star density between the bar and the spiral arm, in agreement with the star counts shown in \citet[e.g.][]{Luri2021}, but much less smooth, because of the relatively low convergence rate of {\tt StarHorse} in that region (due to crowding issues in the input data; see Fig. \ref{skydensity}).

The distance map (second row of Fig.~\ref{magclouds1}) indicates a median heliocentric distance of $49.4$ kpc (for comparison, the distance used in the prior is $d_{\rm prior}=50.58$ kpc; \citealt{McConnachie2012}), for the sources inside the outermost contour level, in agreement with previous estimations \citep[e.g.][]{Pietrzynski2019}. It also shows the expected distance gradient from the fact that the LMC is inclined about $34^{\circ}$, being the closer side the one towards larger declinations \citep[][and references therein]{Luri2021}. 

The LMC metallicity map (third row left panel of Fig.~\ref{magclouds1}) highlights a problematic result: In the inner parts of the LMC, we see a positive metallicity gradient from the bar region towards the outer disc, opposite to the trend observed with red-giant branch (RGB) stars from Magellanic Cloud Photometric Survey (MCPS) and OGLE-III \citep{Choudhury2016}, RR Lyrae stars from OGLE-IV \citep{Skowron2016}, or RGB stars from {\it Gaia} DR2 \citep{Grady2021}. The median metallicity in the bar region (inside the innermost contour level) is of $-0.77$ dex, while at the outer disc (between the innermost and outermost levels) is of $-0.68$ dex. This suggests that the little metallicity information contained the broad-band colours we use in this work is affected by significant systematics, at least for the very dense and complex regions of the Magellanic Clouds. The declining influence of the LMC prior biases the resulting median metallicities and inverts the expected trend (this can possibly be remedied when using the full posterior; see Appendix \ref{gmm}). 

Analogously, the right column of Fig.~\ref{magclouds1} shows the corresponding plots for the SMC sample. The sky distribution (top-right panel of Fig.~\ref{magclouds1}) highlights the irregular structure of the SMC and the beginning of the bridge towards the direction of growing right ascension (and decreasing declination). The distance map (second row right panel of Fig.~\ref{magclouds1}) provides a median distance to the SMC of $63.2$ kpc (prior: $d_{\rm prior}=63.97$ kpc), for the sources inside the outermost level, in agreement with previous estimations \citep[e.g.][]{Cioni2000}. The outer ring with closer distances may be partly an artefact due to the vanishing of the prior contribution towards the outer regions. No clear distance gradient is visible in the SMC. 

Two small blobs with slightly smaller distance are visible in the central parts of the SMC, which are also correlated with the metallicity. Again, a small positive metallicity gradient from the inner towards the outer parts of the galaxy is visible, opposite to the expected behaviour observed with the Red Giant Branch sources from MCPS and OGLE-III \citep{Choudhury2018} or {\it Gaia} DR2 \citep{Grady2021}. As in the LMC, the metallicity and extinction appear to be correlated, being the extinction higher towards the central more crowded region of the galaxy (see bottom-right panel of Fig.~\ref{magclouds1}).

\subsubsection{Candidate metal-poor stars}

The study of metal-poor stars provides a unique window into the formation and accretion history of our Galaxy, since the bulk of those stars were formed at high redshift and conserve abundance patterns unique to their site of formation \citep{Beers2005}.

Although the broad-/intermediate-band photometry used in this work is only marginally sensitive to metallicity (in fact, only when including optical $griz$ photometry can we expect to detect some metallicity information; see Sect. \ref{uncertainties}), low metallicities may manifest themselves in the broad-band colours (especially in the ultraviolet; e.g. \citealt{Norris1999}). We therefore venture to look at candidate metal-poor stars as determined by {\tt StarHorse}, by defining a candidate metal-poor sample as ${\tt met84}<-1$, corresponding to a 1$\sigma$ confidence-level cut. This selection yields 1.58 million objects (without applying any further quality cuts).

Figure \ref{maps_mp} shows the distribution of the metal-poor candidates in galactocentric cylindrical coordinates ($R_{\rm Gal}$ vs. $Z_{\rm Gal}$). We clearly see the imprint of the globular-cluster priors in this figure: All noticeable point-like overdensities correspond to prominent globular clusters, as annotated in the plot. We also note the overdensities in the direction of the Magellanic Clouds, corresponding to stars with bimodal distance probability density function (PDFs), resulting in a median distance in-between the inner halo and the Magellanic Clouds (see Sect. \ref{magclouds}). A similar, less obvious structure, is also visible in the direction of the core of the Sgr dSph galaxy (located towards $(l,b)\sim(5,-14)$, resulting in an elongated overdensity around $(R_{\rm Gal},Z_{\rm Gal})\sim(0$-$3, -1)$.

Apart from these expected features, we also note a very prominent overdensity of local dwarf stars, many of them also following a disc-like density profile, and a diffuse overdensity in the nearby Galactic halo.
The disc-like overdensity is likely mostly due to sample contamination, although even very metal-poor stars have been found on disc-like orbits recently in the Milky Way \citep{Sestito2020} as well as in simulations \citep{Sestito2021}. The diffuse overdensity at larger heliocentric distances is produced by more distant giant stars of the inner halo, expected from the combination of our selection function ($G<18.5$) and our halo prior. Its members can be regarded as potential targets for future/ongoing spectroscopic surveys. 
Another possible overdensity is seen in the central parts of the Galaxy, where indeed many of the Milky Way's oldest stars are expected to reside (e.g. \citealt{Tumlinson2010, Koch2016, Starkenburg2017, Horta2021, Queiroz2021}).

Although methods explicitly tailored to detect metal-poor star candidates from combined broad- and narrow-band colours can be expected to perform much better (e.g. \citealt{Beers1985, Youakim2017, daCosta2019, Thomas2019, Arentsen2019, Chiti2021, Huang2021}), our approach yields a large number of metal-poor star candidates for possible follow-up observations with multi-object spectroscopic surveys such as 4MOST \citep{deJong2019, Chiappini2019, Helmi2019}.

\subsubsection{Outer halo and Local Group}

Figure \ref{halosample} focuses on the density distribution of distant stars in the Galactic halo (defined by $|b|>15$ deg, ${\tt dist50}>10$ kpc). The two top panels (showing Aitoff projections of the sky in ecliptic coordinates) highlight the long tidal tails of the Sagittarius dSph galaxy, also called the Sgr stream (e.g. \citealt{Law2016}). This feature, although not included in our priors, appears clearly both in the density map (top panel of Fig. \ref{halosample}) and the median distance map (middle panel), superseding the extent of the previous membership maps of the Sgr stream, for example the one produced by \citet{Antoja2020} based on {\it Gaia} DR2 proper motions.

The lower panel of Fig. \ref{halosample} shows that the $G<18.5$ sample encompasses also a significant amount of individual stars in dwarf galaxies of the Local Group other than the Magellanic Clouds and the Sgr dSph. For many of them (e.g. Draco, Bootes I, Carina, Ursa Minor), the more informative extragalactic priors of the new {\tt StarHorse} results can help to improve membership probabilities. For others (e.g. Sculptor, Fornax), the prominent pencil-beams between the halo and the expected location of the respective dwarf galaxy hint a problematic prior (e.g. imprecise central coordinates or too low galaxy masses in the Local Group tables used) that results in typically bimodal distance posterior PDFs. 

\subsection{Extinction maps}\label{extmaps}

Figure \ref{avmaps} shows the median {\tt StarHorse}-derived line-of-sight extinction per HealPix cell in four consecutive distance bins out to 2.5 kpc (from top to bottom), illustrating the gradual increase in interstellar extinction as a function of distance and sky position. As expected, these maps are similar to the large-scale integrated dust extinction maps of, for example, \citet{Green2019}. Since we have used the three-dimensional extinction maps of \citet{Green2019} and \citet{Drimmel2003} (albeit convolved with quite broad Gaussians) in our prior, this is not too surprising. 

In principle, our extinction results can be used to infer precise distances to individual dust clouds (e.g. \citealt{Wolf1923, Zucker2020}) and to infer the three-dimensional distribution of dust \citep[e.g.][]{Lallement2019, Leike2020}. The top panel of Fig. \ref{avmaps} shows the presence of high-latitude dust within the 500 pc sphere around the Sun, confirming that the so-called North Polar Spur (the dust filament reaching up to $b\sim45$ deg at $l\sim0$ deg) is a local structure and not related to the Fermi bubbles produced by the Galactic centre (see \citealt{Das2020} for a comprehensive discussion).

\begin{figure*}\centering
 	\includegraphics[width=0.32\textwidth, page=16]{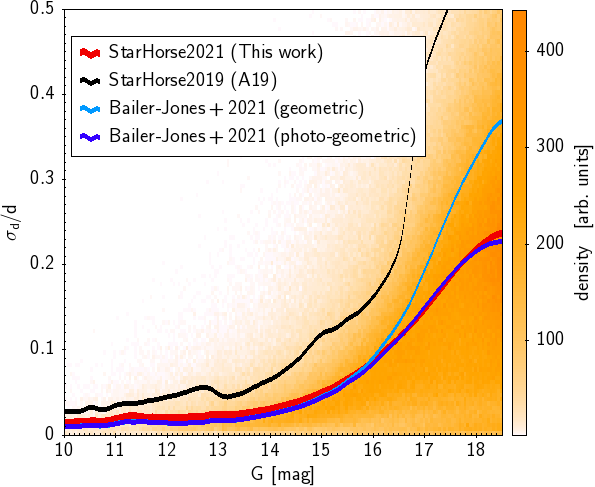}
 	\includegraphics[width=0.32\textwidth, page=16]{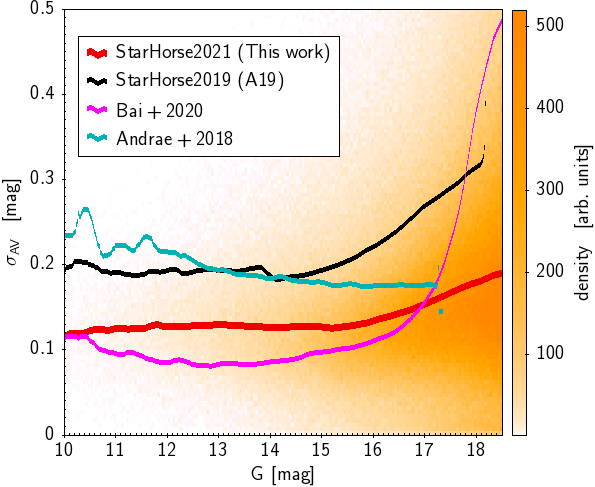}
 	\includegraphics[width=0.32\textwidth, page=16]{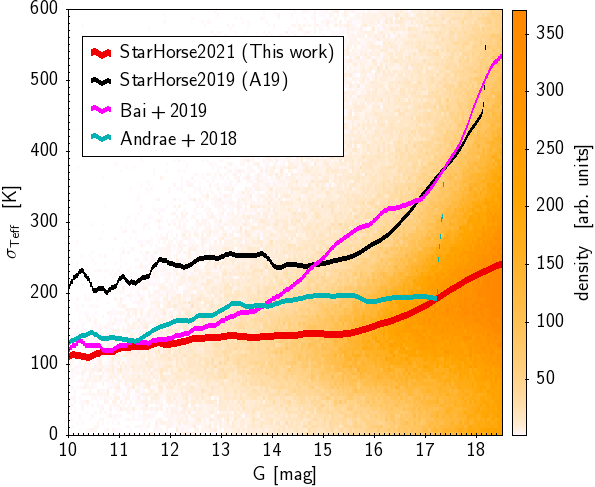}
 	\includegraphics[width=0.32\textwidth, page=16]{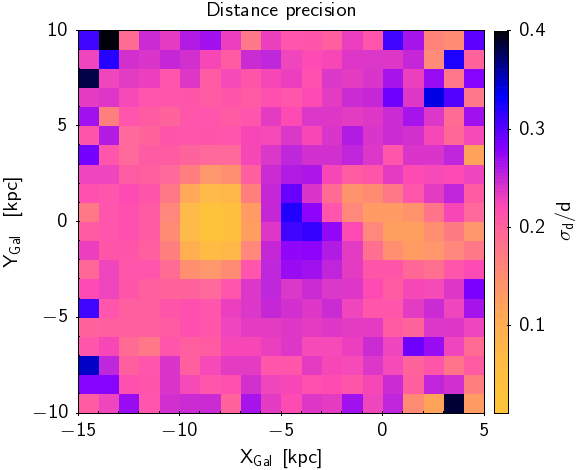}
 	\includegraphics[width=0.32\textwidth, page=16]{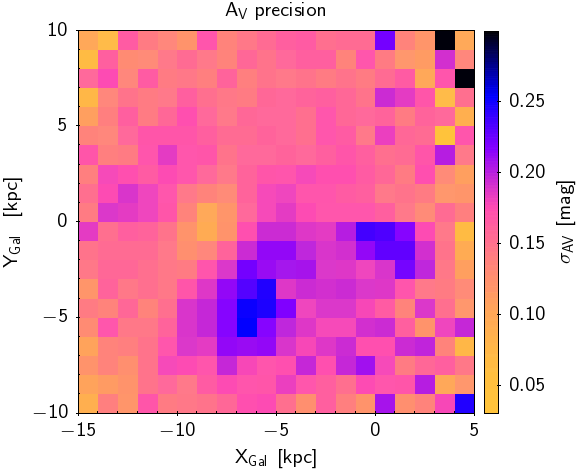}
 	\includegraphics[width=0.32\textwidth, page=16]{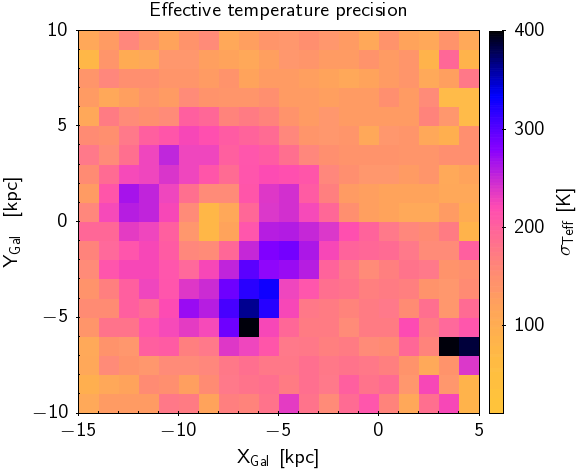}
 	\caption{{\tt StarHorse} formal output uncertainties. Top row: Uncertainties in distance (relative distance uncertainty; left), extinction (middle), and effective temperature (right) as a function of $G$ magnitude. In each top panel we show two-dimensional histograms of a random sample of 1 million {\it Gaia} EDR3 stars in orange, along with the running median smoothed by an Epanechnikov kernel (width $=0.2$; thick red line). For comparison we also show the corresponding values obtained from the (unfiltered) {\it Gaia} DR2 run \citepalias{Anders2019} in black, as well as the results from \citet{BailerJones2021} for distances, from \citet{Andrae2018} and \citet{Bai2020} for extinctions, and from \citet{Andrae2018} and \citet{Bai2019} for effective temperatures. Bottom row: Median formal output uncertainties as a function of Galactic position for the same random sample.
    }
 	\label{uncerts_vs_G}
\end{figure*}

\section{Precision and accuracy}\label{uncertainties}

\subsection{Internal precision}\label{precision}

Along with the median statistics of each output parameter, {\tt StarHorse} also delivers the corresponding confidence intervals (defined as the 16th and 84th percentile of the marginal posterior). The overall distribution of the output uncertainties (defined as, for example, $\sigma_{\rm Teff}=0.5\cdot({\tt teff84} - {\tt teff16})$, etc.) and the correlations between the output uncertainties are shown in the top-right corner plot of Fig. \ref{corner}. This plot shows the complete sample of converged stars and demonstrates that the output uncertainties are typically highly correlated (we note the logarithmic scaling of the plot axes). The highest correlations are seen, as expected, between effective temperature and extinction, and between distance and surface gravity.

The precision of the results, however, depends first and foremost on the quality of the {\it Gaia} EDR3 parallaxes and the availability of multi-band photometry for each source. Both these criteria are, to first approximation, functions of the {\it Gaia} $G$ magnitude. In Fig. \ref{uncerts_vs_G} we therefore show the formal uncertainties as a function of $G$ magnitude for a random sample of 1 million stars. The orange two-dimensional histogram in the background shows the uncertainty distribution of all objects, while the red line shows the smoothed median trend. We can appreciate that the distance uncertainties for stars with $G<14$ are typically around $2\%$, growing to about 8\% around $G\approx16$, and reaching 20\% at $G\approx18$. The improvement in precision with respect to our DR2 run \citepalias[][black line in Fig. \ref{uncerts_vs_G}]{Anders2019} is mainly due to the improvement in both precision and accuracy brought by the {\it Gaia} EDR3 parallaxes.

The bottom row of Fig. \ref{uncerts_vs_G} show the median formal uncertainties as a function of position in the Galaxy, again for a random set of 1 million stars. Many of the features in these uncertainty maps can already be appreciated (although at a different absolute scale) in Fig. 13 of \citetalias{Anders2019}. Apart from the overall precision improvement (by a factor of $\sim2$) the major changes are: 1. a slight increase of the 'parallax sphere' (the region for which parallaxes are determined with a precision of $\lesssim20\%$), 2. the disappearance of the bulk of stars with very high distance uncertainties that had to be flagged because they were compatible with both dwarf- and giant-star solutions, and 3. a slightly lower impact of the missing PS1 photometry on the $T_{\rm eff}$ and $A_V$ precisions below a declination of $-20\deg$ ($Y_{\rm Gal}<0, X_{\rm Gal}\gtrsim-10$ kpc) thanks to the use of SkyMapper data (and, in fact, a higher precision in the region where both catalogues overlap).

The precision of the secondary output parameters ($\log g$, [M/H], and $M_{\ast}$), not shown in Table \ref{summarytable} and Fig. \ref{uncerts_vs_G}, behave similarly as a function of $G$, although the improvement in precision with respect to the DR2 results is slightly less pronounced (by a factor of 1.5). At magnitude $G\approx17$, the median uncertainties for the secondary output parameters amount to $\sigma_{\log g}^{G=17}=0.23$ dex, $\sigma_{\rm [M/H]}^{G=17}=0.10$ dex, and $\sigma_{M_{\ast}}/M_{\ast}^{G=17}=9.5\%$.

\subsection{Comparison to open clusters}\label{clusters}

\begin{figure*}\centering
 	\includegraphics[width=0.9\textwidth]{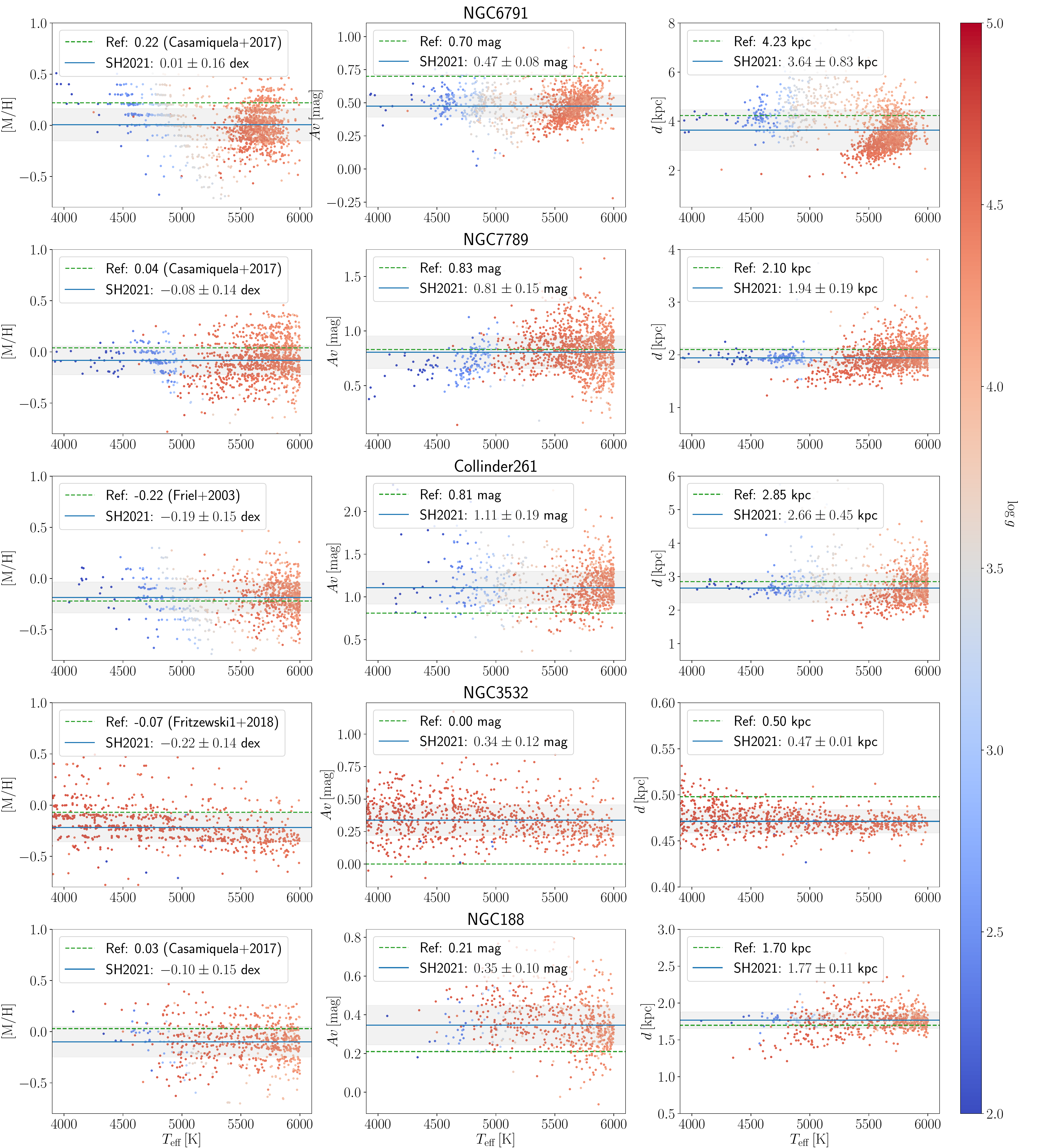}
 	\caption{Metallicity, extinction, and distance results for FGK star members of the five most populated Galactic OCs in the \citet{Cantat-Gaudin2020} catalogue, reflecting the typical precision of the {\tt StarHorse} results as well as some systematic trends with effective temperature and surface gravity. The blue lines refer to the cluster median and the blue-shaded area to the median absolute deviation, while reference values are plotted as dashed orange lines. We note that the reference cluster metallicities are in fact iron abundances [Fe/H], which are only approximately equal to the total metallicity [M/H] determined by {\tt StarHorse}.}
 	\label{clusters1}
\end{figure*}

Member stars of an OC are expected to have, to first order, the same age, metallicity, distance, and interstellar extinction. They thus constitute excellent samples fro evaluating the precision and accuracy of our astrophysical parameters.

Figure \ref{clusters1} shows comparisons of the {\tt StarHorse} distance, extinction, and metallicity scales for the five most populated and well-studied OCs (NGC 6791, NGC 7789, Collinder 261, NGC 3532, and NGC 188) in the {\it Gaia} DR2 OC catalogue of \citet{Cantat-Gaudin2020}. These OCs are those with the most identified members, mainly by virtue of being relatively massive and nearby (but not so nearby that they are extended in the sky and in proper-motion space, like the Hyades). 
Each panel of Fig. \ref{clusters1} shows a {\tt StarHorse} output parameter as a function of effective temperature in comparison to the literature values for the particular cluster. The five clusters are diverse enough in their physical characteristics to appreciate some first trends as a function of effective temperature, surface gravity, metallicity, and cluster age.

For example, for the old metal-rich cluster NGC 6791 the {\tt StarHorse}-derived metallicities are clearly underestimated (with respect to the spectroscopically derived cluster metallicity ([Fe/H]; \citealt{Casamiquela2017}) and show a quite large scatter (which is, however, both expected and reflected in the quoted uncertainties). Similarly, the code finds a slightly lower extinction and distance than derived by \citet{Cantat-Gaudin2020}.

\begin{figure*}\centering
 	\includegraphics[width=0.95\textwidth]{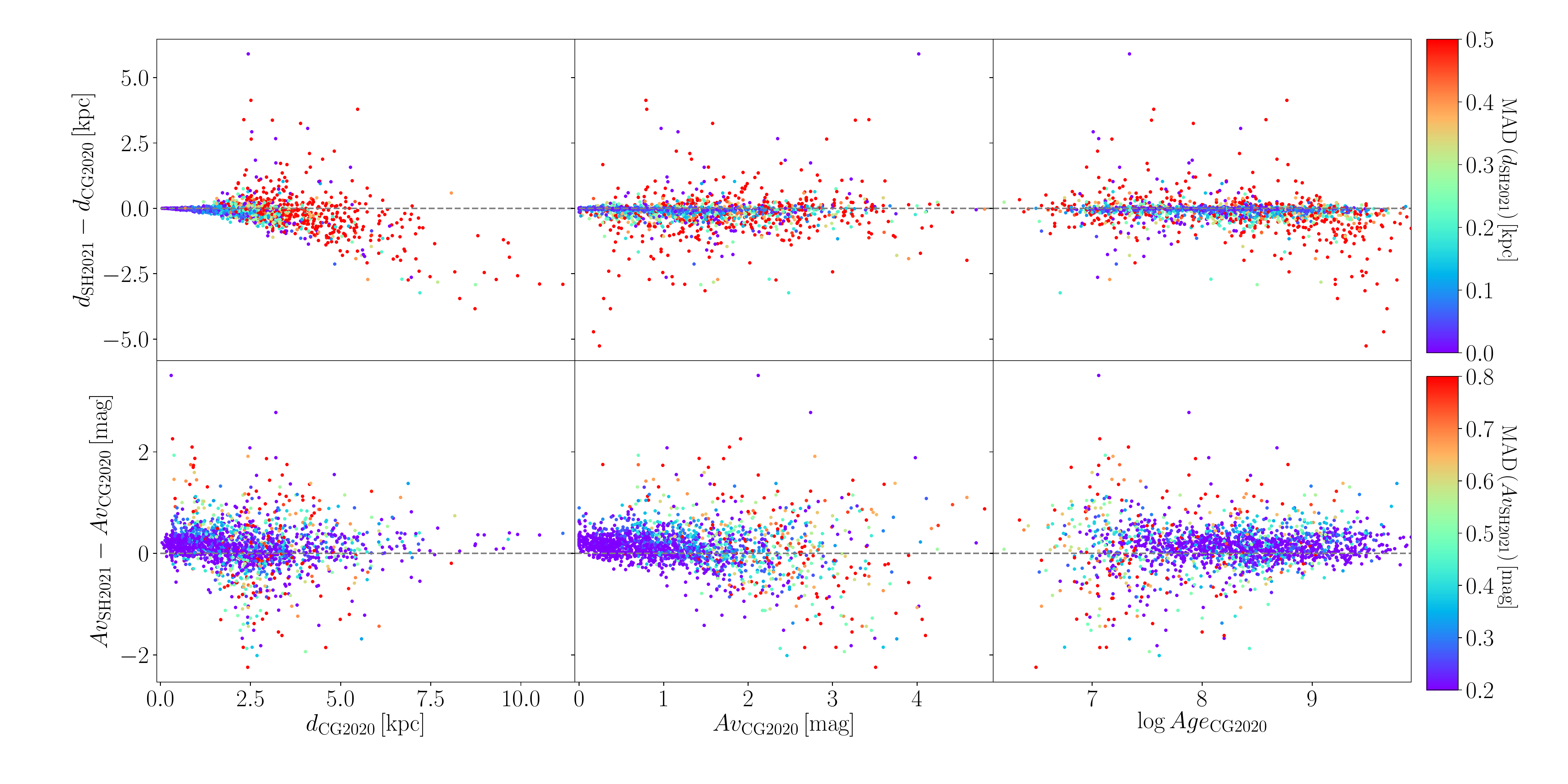}\\
 	\caption{Distance (top row) and extinction (bottom row) comparison with the OC parameter catalogue of \citet{Cantat-Gaudin2020}. Each point represents one star cluster. We show the systematic difference between {\tt StarHorse} (calculated as the median of all FGK member stars) and the reference value as a function of distance, $A_V$, and log age. The colour denotes the intrinsic dispersion (MAD) within each cluster.}
 	\label{clusters3}
\end{figure*}

\begin{figure*}\centering
 	\includegraphics[width=0.49\textwidth]{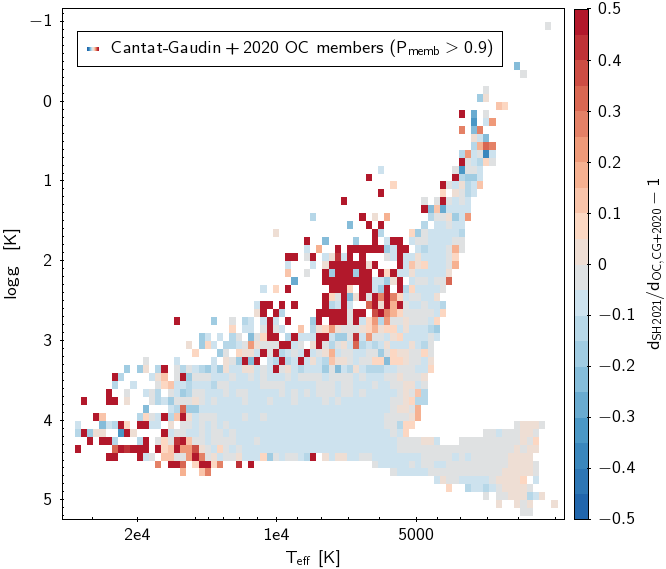}
 	\includegraphics[width=0.49\textwidth]{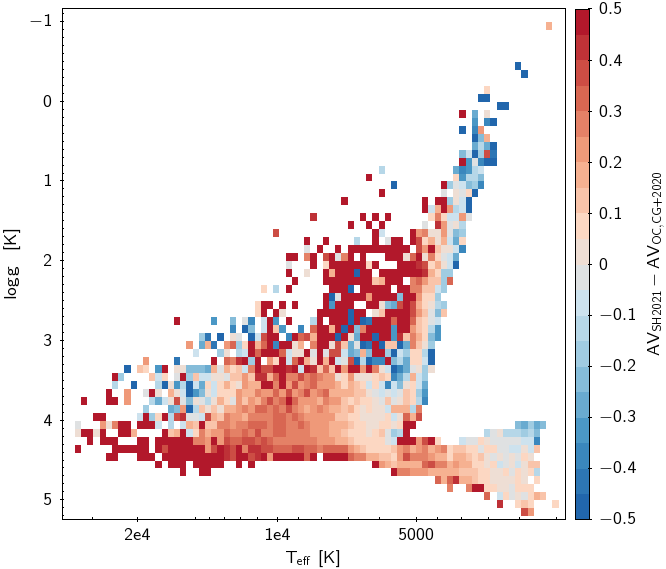}
 	\includegraphics[width=0.49\textwidth]{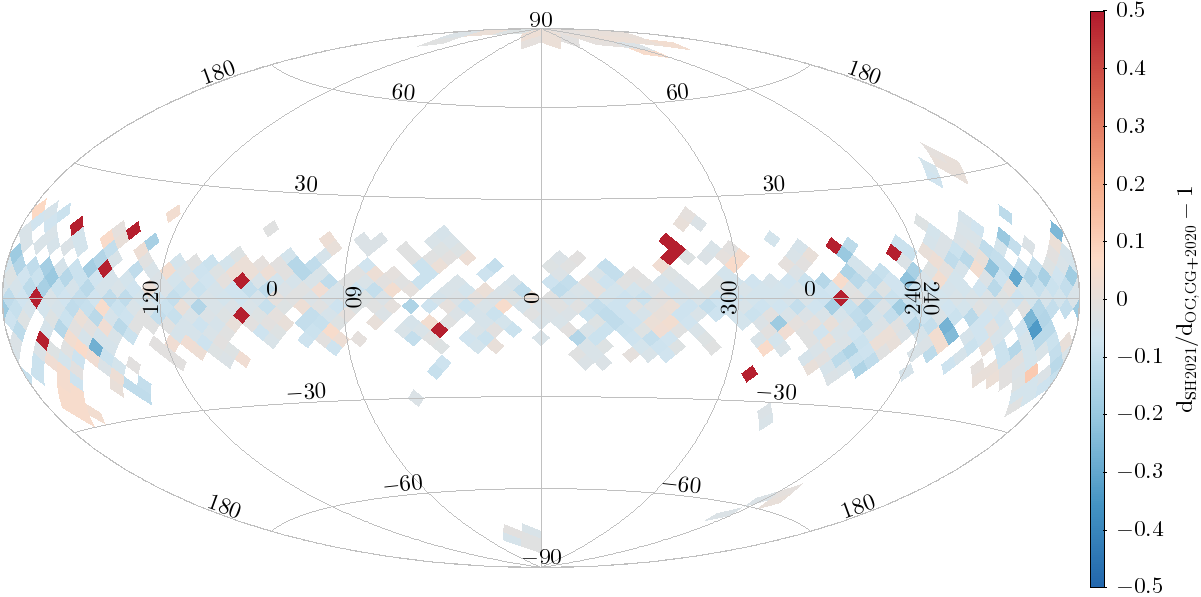}
 	\includegraphics[width=0.49\textwidth]{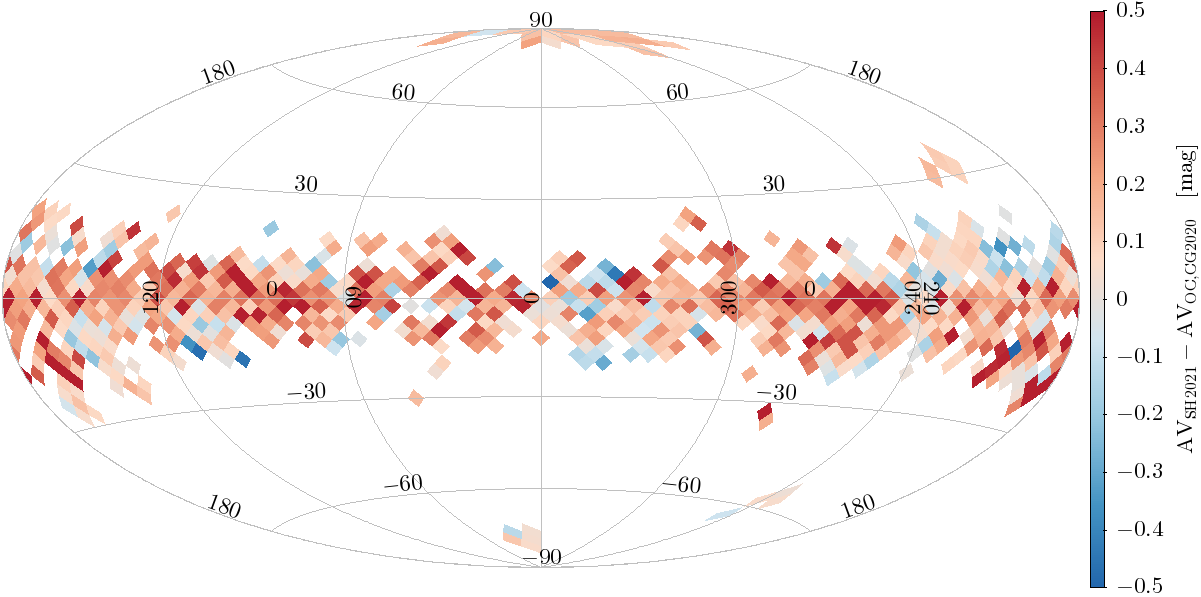}
 	\caption{{\tt StarHorse} results for OC members: comparison to the distance and extinction scale of \citet{Cantat-Gaudin2020}. Top panels: {\it Kiel} diagrams colour-coded by median relative distance difference (left) and absolute extinction difference (right) per pixel. Bottom panels: Sky distribution colour-coded by median differences.}
 	\label{clusters4}
\end{figure*}

A more quantitative comparison for the bulk of the known Galactic OC population (the 1867 OCs with astrophysical parameters from the \citealt{Cantat-Gaudin2020} catalogue) is shown in Figs. \ref{clusters3} and \ref{clusters4}.
\citet{Cantat-Gaudin2020} determined the distance, extinction, and age of each cluster homogeneously with an artificial neural network trained on a set of high-quality measurements (mostly relying on \citealt{Bossini2019}). 
In Fig. \ref{clusters3} we plot the {\tt StarHorse} median values per cluster (for FGK-type stars, 3800 K $<T_{\rm eff}<6000$ K) compared with the \citealt{Cantat-Gaudin2020} determinations of the distance and extinction. The colour represents the median absolute deviation (MAD) obtained for the cluster members in SH.
Figure \ref{clusters3} shows that the OCs cover a broad range of physical parameters: 90\% of the clusters are nearer than 4.4 kpc and have less than 2.5 magnitudes of extinction, and the age range (10-90th percentile) covers $\log \tau$ from 7.2 to 9.1. 

A complementary catalogue of astrophysical parameters for OCs was recently presented by \citet{Dias2021}. It contains parameters of 1743 OCs (in their vast majority also contained in \citealt{Cantat-Gaudin2020}) determined by isochrone fitting of {\it Gaia} DR2 photometry. We used this catalogue as an additional reference to test if the discrepancies between {\tt StarHorse} and the \citet{Cantat-Gaudin2020} catalogue can partly be attributed to systematics in the OC catalogues as well. We only show the comparison  with \citet{Cantat-Gaudin2020}; the comparison with the \citet{Dias2021} catalogue leads to the very similar general conclusions.

The top row of Fig. \ref{clusters3} shows that the concordance with the OC distance scale is reasonable. The majority of both the clusters and the member stars present less than 20\% deviation. The deviating clusters are mostly distant objects with very few member stars, partly uncertain membership, and thus a large internal dispersion of {\tt StarHorse} parameters (red dots). We see a trend of negative differences with respect to the OC catalogue: on average, our EDR3 distances are shorter by $-3.5\%$. An opposite trend of similar magnitude, however, is seen in the comparison with the \citet{Dias2021} catalogue: our distances are larger than theirs by $+3.8\%$ on average. No significant trends of distance difference with neither extinction nor age are found.

For extinction, on the other hand, some systematics similar to those seen in \citetalias{Anders2019} can be appreciated: in particular, a slight systematic overestimation for nearby, low-extinction objects. This may in part be due to the fact that {\tt StarHorse} treats every object as a single star and tries to adjust its parameters to a PARSEC isochrone. For similar-mass unresolved binaries on the main sequence this typically leads to an overestimated effective temperature, an underestimated $\log g$ (moving the object towards the sub-giant or lower red-giant branch), and an overestimated extinction to compensate for the extra brightness (compared to a single star). We also refer to Appendix \ref{caveats}.

In Fig. \ref{clusters4} we further investigate possible systematic biases depending on sky position and spectral type.
We find a rather uniform sky distribution of relative distance differences that is consistent with Fig. \ref{clusters3}. The distance systematics are typically very small and lightly negative ($\lesssim5\%$; see also the last row of Fig. \ref{clusters3}). A sky pattern is hardly discernable, but may be related to the parallax bias present in {\it Gaia} DR2 (and thus also in the \citealt{Cantat-Gaudin2020} and \citealt{Dias2021} catalogues), which has been largely accounted for in EDR3 (using the corrections proposed by \citealt{Lindegren2021}).

Both the parallax improvement with respect to {\it Gaia} DR2 and the inclusion of a dust map in the new priors allow a slightly smoother distribution of extinction differences than in \citetalias{Anders2019}. However, we see that extinction is generally 0.1-0.2 mag higher than the one estimated by \citet{Cantat-Gaudin2020}. Our extinction estimates are, on the other hand, slightly lower than the ones in the catalogue of \citealt{Dias2021}, so that the absolute scale is far from well defined. 

Furthermore, the top-right panel of Fig. \ref{clusters4} shows significant systematic trends of extinction with position the {\it Kiel} diagram (being most severe in sparsely populated areas). For example, it seems that {\tt StarHorse} tends to slightly underestimate extinctions for metal-rich (redder) giant stars, while it overestimates extinctions for metal-poor giants. For dwarf stars, extinction biases are generally low, except for (probable) binary stars close to the turn-off phase (see Appendix \ref{binaries}).

{\tt StarHorse} also tends to severely overestimate the extinction of the stars hotter than $7000$ K \citep{Pantaleoni2021}. Due to the initial-mass-function prior, stellar models with $T_{\rm eff} \gtrsim10^4$ K are highly suppressed in the posterior - which leads to significantly biased results for massive stars (see Appendix \ref{obstars} for details).

\subsection{Comparison to asteroseismology}\label{seismo}

\begin{figure*}\centering
 	\includegraphics[width=0.33\textwidth]{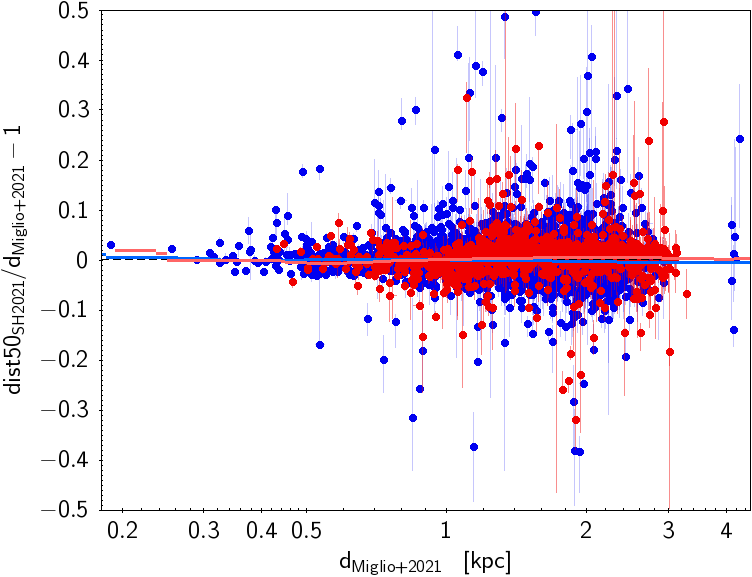}
 	\includegraphics[width=0.33\textwidth]{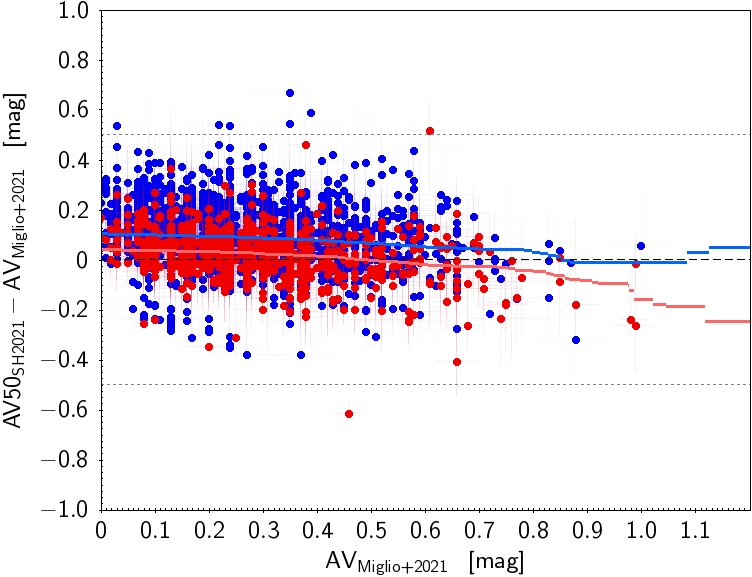}
 	\includegraphics[width=0.33\textwidth]{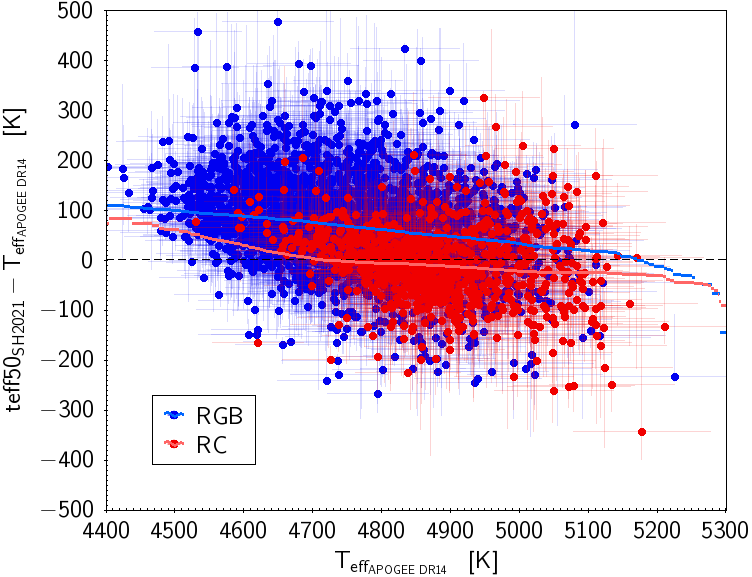}
 	\includegraphics[width=0.33\textwidth]{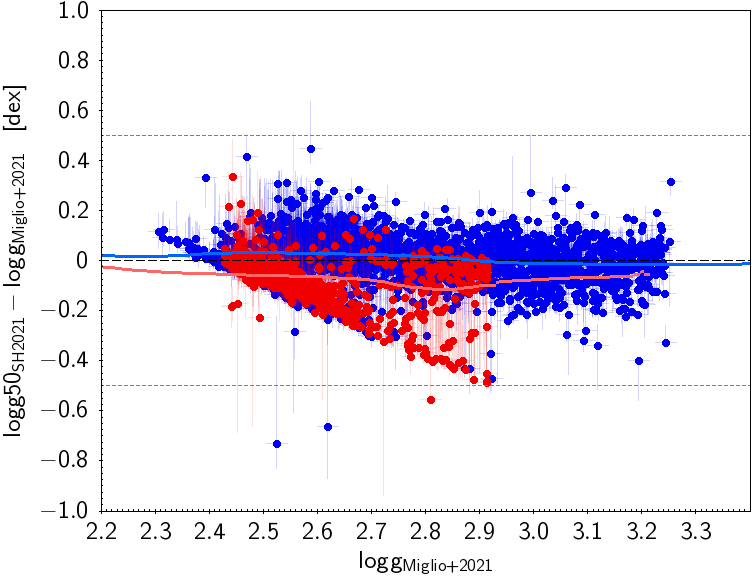}
 	\includegraphics[width=0.33\textwidth]{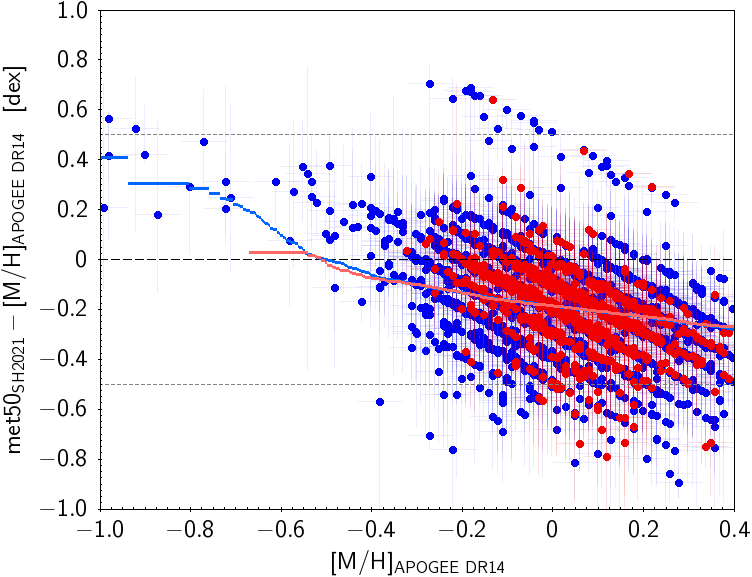}
 	\includegraphics[width=0.33\textwidth]{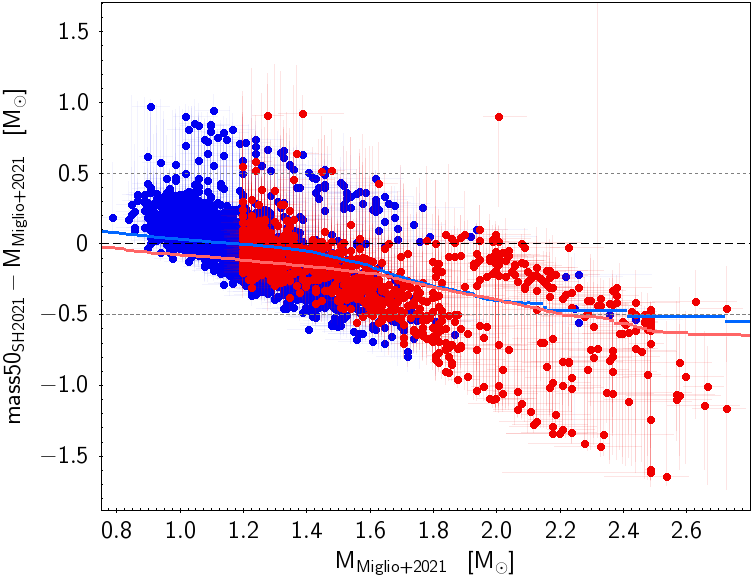}
 	\caption{Comparison of {\tt StarHorse} EDR3 distances, extinctions, and effective temperatures (top row), as well as surface gravities, metallicities, and masses (bottom row) with the high-precision asteroseismic+spectroscopic red-giant catalogue of \citet{Miglio2021}. In each panel we show the parameter difference as a function of the parameter itself, where the blue dots refer to RGB stars, while the red dots refer to core He-burning red-clump stars.}
 	\label{miglio}
\end{figure*}

Asteroseismology provides a unique way to peer into the interior of stars (for a recent review see \citealt{Aerts2021}) and thus also to test the accuracy of our derived stellar parameters. In particular, asteroseismology of solar-type and red-giant stars \citep{Chaplin2013} can provide very precise stellar surface gravities and masses \citep{Ulrich1986, Kjeldsen1995}, evolutionary stages \citep{Bedding2011, Mosser2011}, and thereby also distances and extinctions \citep{Rodrigues2014}.

In Fig. \ref{miglio} we compare our photo-astrometric results with the most precise and accurate parameters obtained for red-giant field stars (outside the immediate solar vicinity) to date. \citet{Miglio2021} combined asteroseismic observations by {\it Kepler} \citep{Gilliland2010} with APOGEE DR14 spectroscopy \citep{Majewski2017, Abolfathi2018} and used the PARAM tool \citep{daSilva2006, Rodrigues2017} to determine precise stellar parameters, distances, and extinctions. The authors also tested the influence of different stellar modelling assumptions (atomic diffusion, initial He abundance, [$\alpha$/Fe]-enhancement, etc.)

The sample comprises 3,195 stars in the {\it Kepler} field, and thus the systematic trends seen in Fig. \ref{miglio} do not necessarily apply to the full sky, but the plots give a fair impression of the typical precision and accuracy that can be expected for giant stars. Similar to the comparison shown in \citetalias{Anders2019} for the {\it Kepler} field, we do not see any significant trend in terms of distances, indicating again the improvement of the parallax zero-point calibration achieved by {\it Gaia} EDR3 and the corrections proposed by \citet{Lindegren2021}. For extinctions, we detect a slight overestimation ($\sim 0.1$ mag) with respect to \citet{Miglio2021} for the RGB stars, while no significant offset is seen for the red-clump stars. As expected, a similar behaviour is seen in the effective temperature differences: the RGB $T_{\rm eff}$ scale of APOGEE DR14 is typically 100 K cooler than our inferred effective temperatures.
We suggest that these slightly different trends for RGB stars and red-clump stars can probably be generalised to the full sky. We caution, however, that the absolute $T_{\rm eff}$ scales of both spectroscopy and stellar models are uncertain to within similar levels \citep[e.g.][]{Rodrigues2017, Miglio2021}.

The second row of Fig. \ref{miglio} shows the comparison for the secondary (naturally less reliable) output parameters $\log g$, [M/H], and mass. In each of these panels we see negative offsets and trends for red-clump stars, and similar, but typically milder ones for the RGB stars.

\begin{figure*}\centering
 	\includegraphics[width=0.33\textwidth]{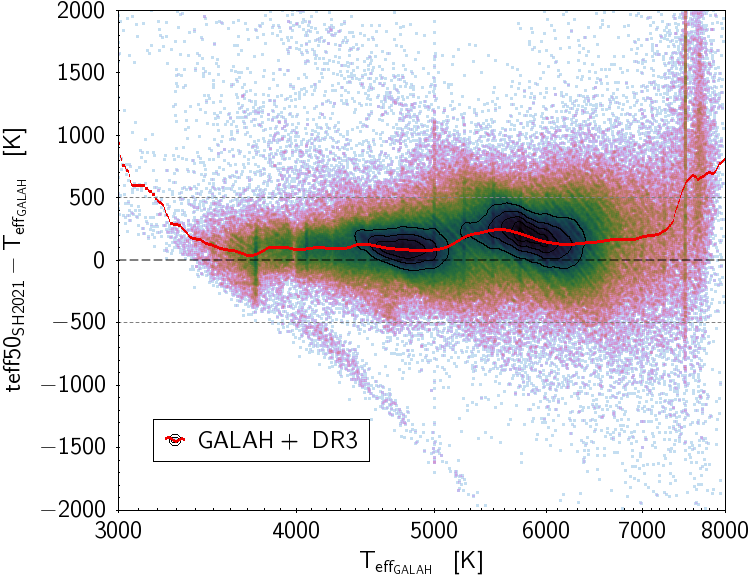}
 	\includegraphics[width=0.33\textwidth]{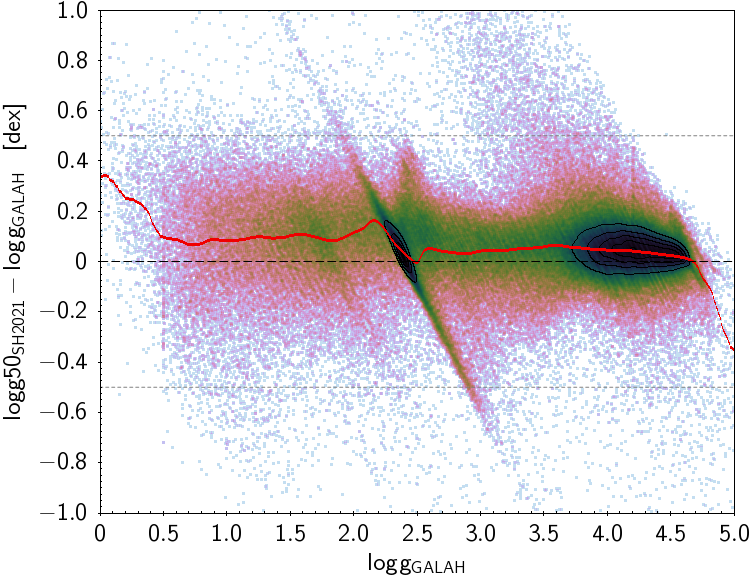}
 	\includegraphics[width=0.33\textwidth]{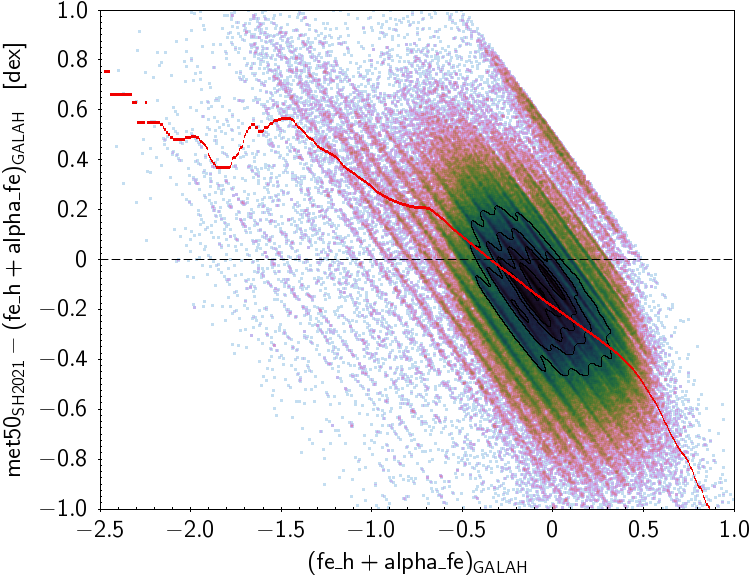}
 	\caption{Comparison of {\tt StarHorse} EDR3 effective temperatures (left), surface gravities (middle), and metallicities (right) with the spectroscopically derived labels from GALAH+ DR3 \citep{Buder2021}. In each panel the red line corresponds to the running median.}
 	\label{galahcomp}
\end{figure*}

\subsection{Comparison to GALAH DR3}\label{galah}

Large-area spectroscopic stellar surveys like RAVE \citep{Steinmetz2006, Steinmetz2020}, LAMOST \citep{Deng2012, Zhao2012}, or APOGEE \citep{Majewski2017} are ideal to detect possible stellar parameter trends for astro-photometrically derived results. In \citetalias{Anders2019} we showed a comparison with APOGEE DR14 (see also previous subsection); here we choose another example: the stellar parameters from the third data release \citep{Buder2021} of the GALAH survey \citep{DeSilva2015}.

Figure \ref{galahcomp} shows the parameter comparison to GALAH DR3 for effective temperature, surface gravity, and metallicity. In line with Fig. \ref{miglio}, we see some slight trends (typically an overestimation by 100-200 K) for effective temperatures in the range of FGK (both dwarfs and giants) stars, where most of the common stars are located and for which the GALAH pipeline works best \citep{Buder2018}. Perhaps with the exception of the large $\log g$ spread generated by the red clump (indicating some impurity of the {\tt StarHorse} red-clump sample), a similar pattern as for $T_{\rm eff}$ is observed for $\log g$, with the difference that both the median offset and the dispersion decrease with $\log g$, because for dwarf stars the {\tt StarHorse} surface gravity is typically well constrained by the {\it Gaia} EDR3 parallax.

The least constrained parameters, as expected for our technique (combining parallaxes and broad-band photometry), are certainly mass and metallicity. The right panel of Fig. \ref{galahcomp} shows clearly how our metallicity estimates are dominated by the (broad) Galactic metallicity priors. They clump around zero and exhibit a tail towards negative metallicity, but show little concordance trend (even for metal-poor stars) with metallicities determined from the high-resolution GALAH survey. We therefore remind the reader to use these estimates with caution.

\subsection{Caveats}\label{caveats0}

In \citetalias{Anders2019} we enumerated a list of caveats that applied for the {\tt StarHorse} {\it Gaia} DR2 results. While many of them have been addressed by the improvements presented in this work, some important caveats remain and should be taken into account when using the results presented here. We discuss these in some detail in Appendix \ref{caveats}.

\section{Comparison to previous results}\label{comp}

In this section we compare our results to some previous attempts to determine astrophysical parameters for massive amounts of {\it Gaia} stars. A comprehensive comparison to all such datasets is beyond the scope of this paper, so we choose some illustrative examples. In particular we compare to the {\it Gaia} EDR3 distances of \citet{BailerJones2021}, the astrophysical parameters of \citetalias{Anders2019}, and the effective temperatures of \citet{Bai2019} and the extinctions of  \citet{Bai2020}, both obtained by machine-learning algorithms.

\begin{figure}\centering
 	\includegraphics[width=0.49\textwidth]{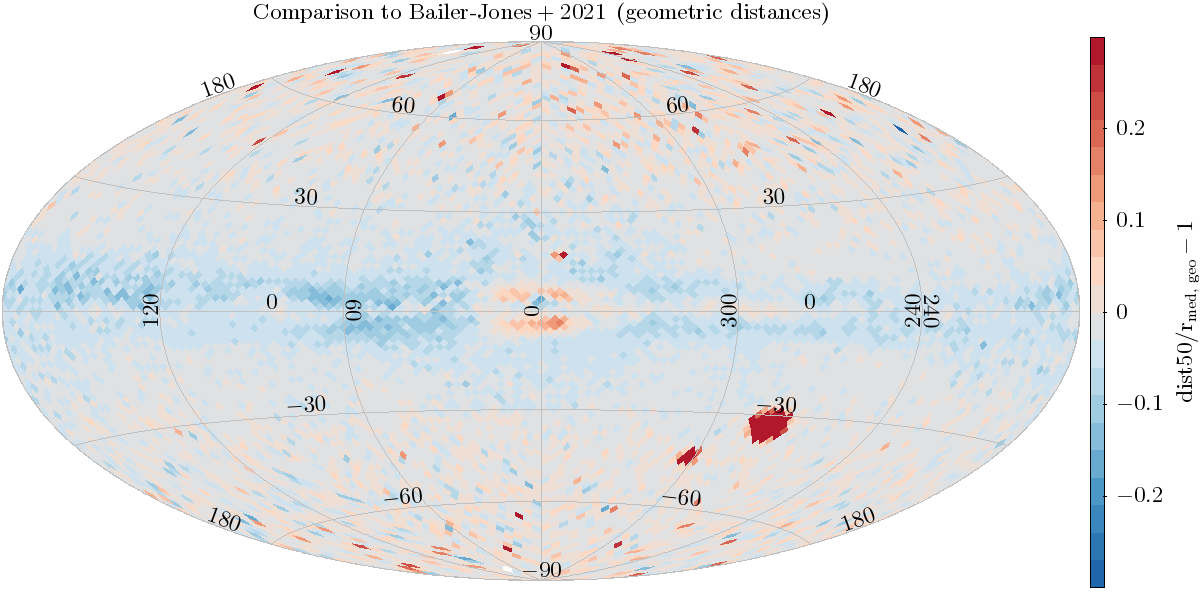}
 	\includegraphics[width=0.49\textwidth]{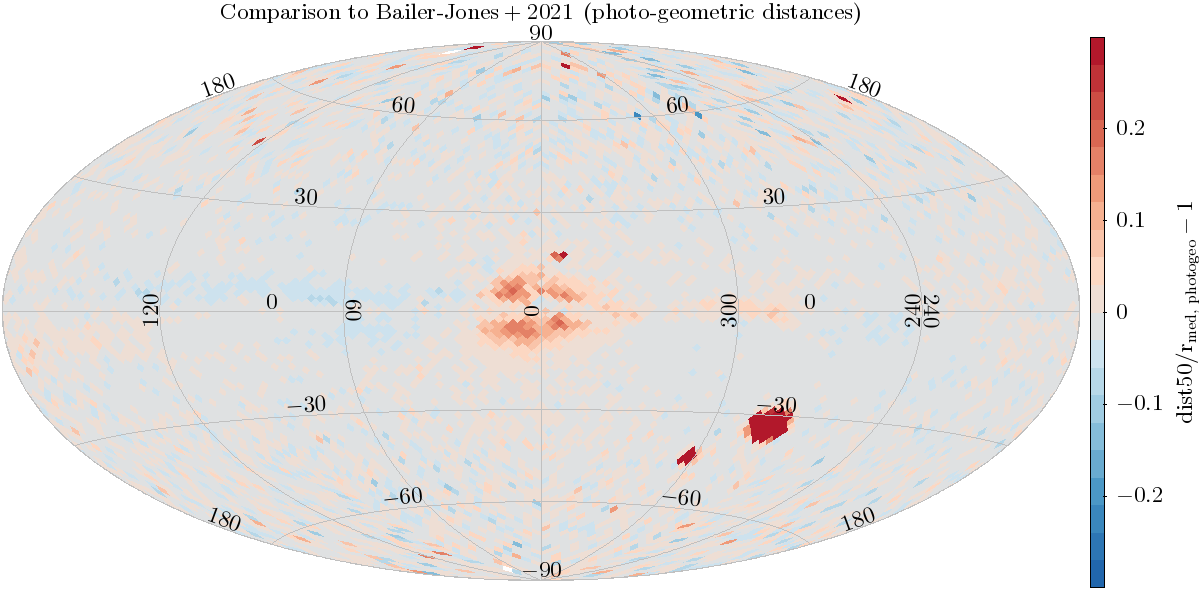}
 	\includegraphics[width=0.24\textwidth]{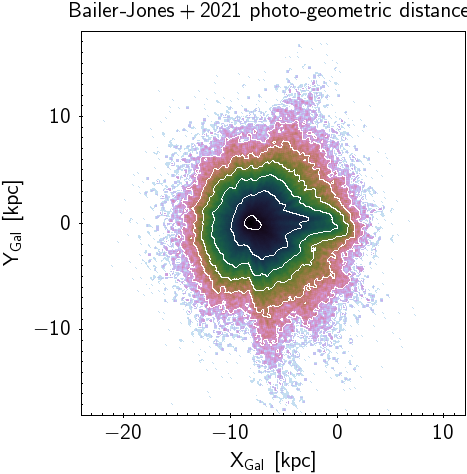}
 	\includegraphics[width=0.24\textwidth]{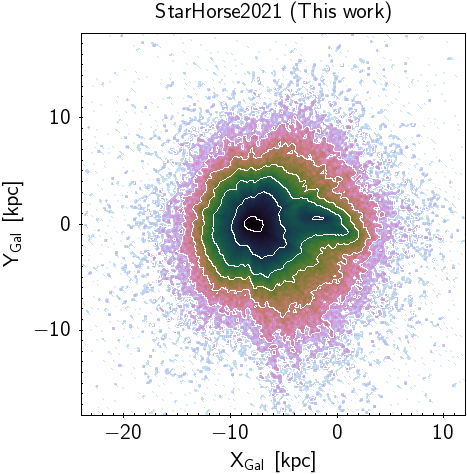}
 	\caption{Comparison of {\tt StarHorse} EDR3 distances with the EDR3 distances from \citet{BailerJones2021} using 1 million random stars. Top panel: Sky map showing the relative distance difference with respect to the geometric distances (computed using only the {\it Gaia} EDR3 parallaxes). Middle panel: Same for the photo-geometric distances (using also the EDR3 photometry in the distance inference). Bottom panels: Visual appearance of the Cartesian Galactic maps derived from the photo-geometric distances of \citet{BailerJones2021} (left) and {\tt StarHorse} (right) for the same random sample. In both bottom panels the contour lines are logarithmically spaced.}
 	\label{bj21}
\end{figure}

\begin{figure}\centering
 	\includegraphics[width=0.49\textwidth]{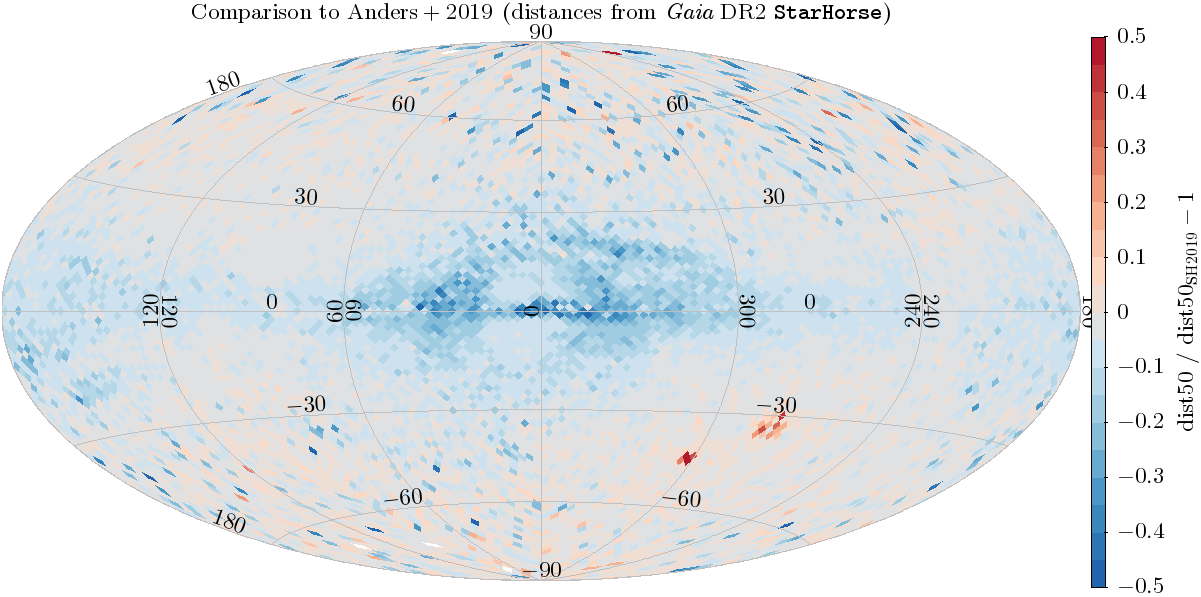}
 	\includegraphics[width=0.49\textwidth]{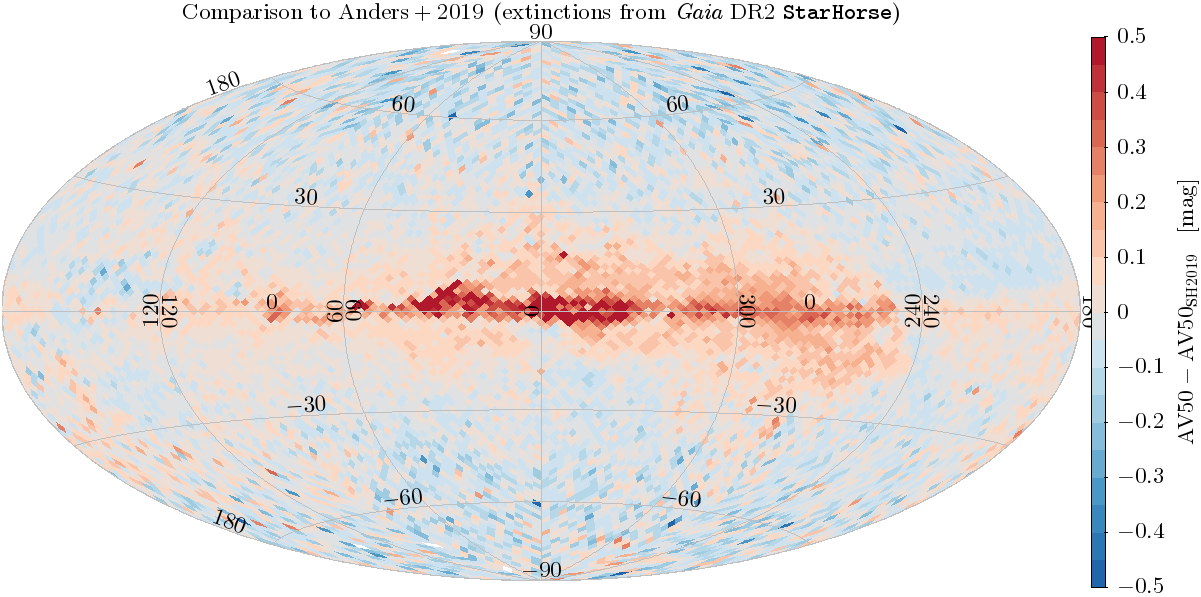}
 	\includegraphics[width=0.49\textwidth]{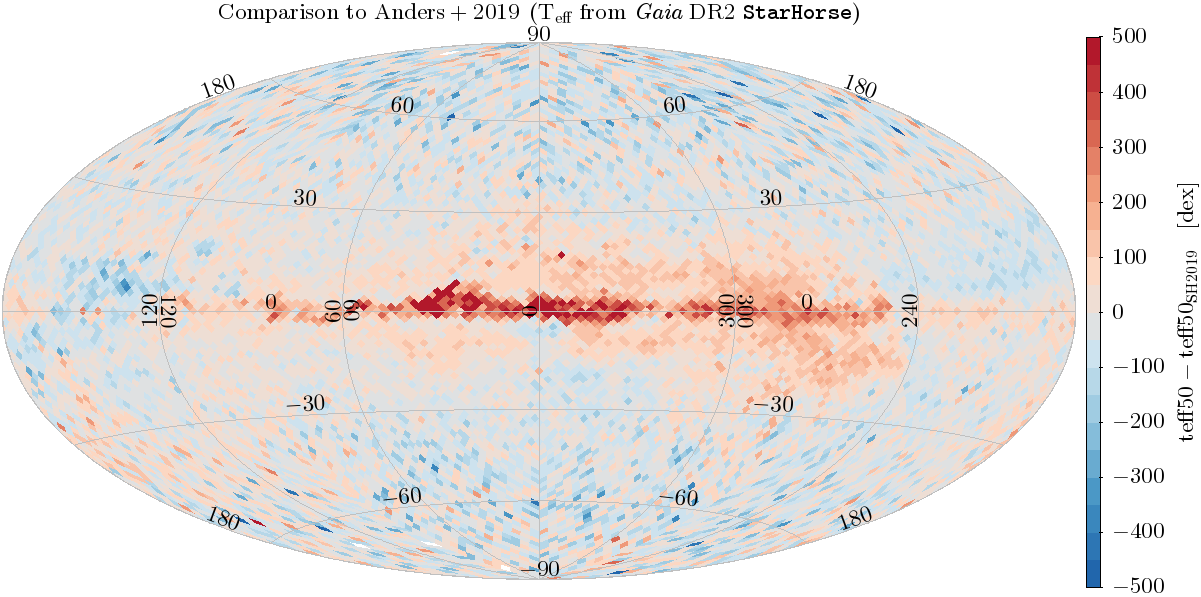}
 	\includegraphics[width=0.49\textwidth]{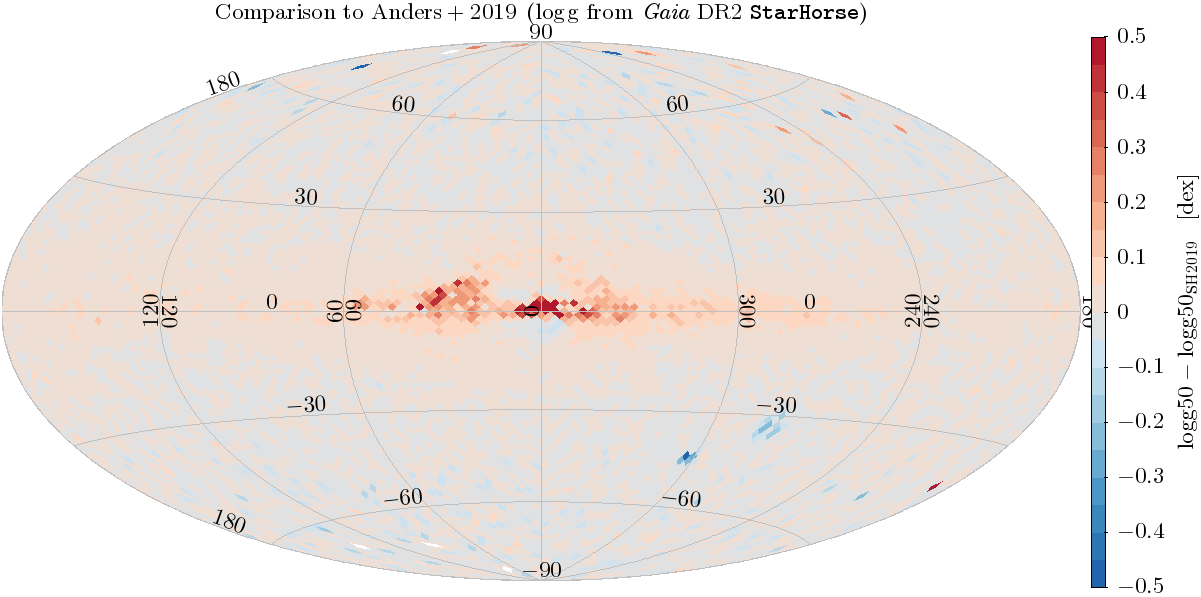}
 	\includegraphics[width=0.49\textwidth]{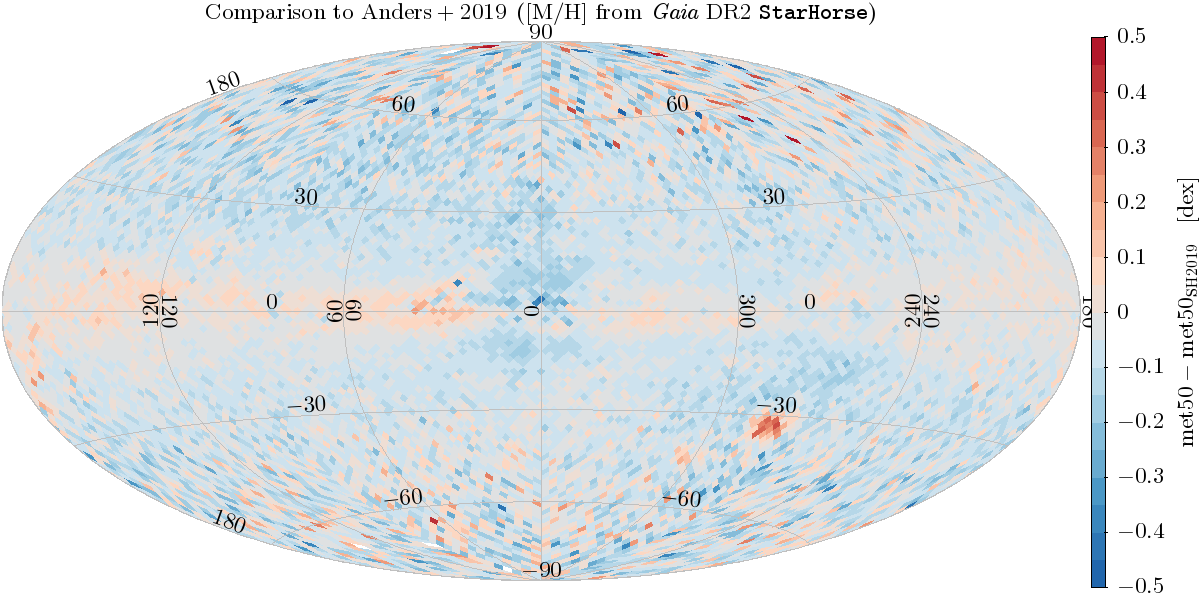}
 	\caption{Comparison of the {\tt StarHorse} EDR3 results with the {\tt StarHorse} DR2 results from \citet{Anders2019}, for a random sample of 1 million stars. From top to bottom: Sky distribution of median distance, extinction,  effective temperatures, surface gravity, and metallicity differences.}
 	\label{sh19}
\end{figure}

\subsection{Comparison to the \citet{BailerJones2021} EDR3 distances}

Shortly after {\it Gaia} EDR3, \citet{BailerJones2021} published two sets of distance estimates for 1.47 billion objects based on EDR3 data. The first set, dubbed geometric distances, used solely {\it Gaia} parallaxes and a sophisticated prior for the stellar density distribution in the Milky Way \citep{Rybizki2020}, analogous to the {\it Gaia} DR2 catalogue published by the same group \citep{BailerJones2018}. The second set, dubbed photo-geometric distances, also used the {\it Gaia} EDR3 photometry to refine the distance prior for each star, thus providing more precise (and arguably also more accurate) results.

Figure \ref{bj21} shows a comparison of our distance estimates with the two sets of distances obtained by \citet{BailerJones2021} for random sample of 1 million stars. We see a remarkable concordance over almost the entire sky, especially with the set of photo-geometric distances (excluding only the Magellanic Clouds and the centremost Galactic regions; see the second panel in Fig. \ref{bj21}). Slightly larger differences (in the sense that {\tt StarHorse} typically delivers smaller distances) are present in the comparison to the purely geometric distances, for the region around the Galactic plane. The high differences in the Magellanic Clouds are expected, since the priors of \citet{BailerJones2021} do not include any extragalactic stellar populations, while we explicitly included these in our prior.

This is reassuring, since the method of \citet{BailerJones2021} is quite different in both its prior assumptions (distance scale lengths are derived from a synthetic Milky Way model) and in the implementation of the posterior calculation (Markov chain Monte Carlo sampling). Also, the derived stellar density maps (bottom panels of Fig. \ref{bj21}) are very similar for \citet{BailerJones2021} and {\tt StarHorse}: both show almost the same density contours, perhaps with the exception that the Galactic bar appears slightly more prominent in our map, and that the {\tt StarHorse} halo priors seem to allow for slightly more distant objects to be present.
In addition, Fig. \ref{uncerts_vs_G} (left panel) shows that the internal uncertainties obtained by \citet{BailerJones2021} for their distance estimates are very similar to our {\tt StarHorse} results.

\subsection{Comparison to DR2-derived parameters}
\subsubsection{{\it Gaia} DR2 {\tt StarHorse} \citepalias{Anders2019}}

We have described the methodological differences with respect to our previous {\tt StarHorse} results derived from {\it Gaia} DR2 in Sect. \ref{updates}. The most important difference is, however, the clearly superior quality of the {\it Gaia} EDR3 catalogue (see \citealt{Fabricius2021} for numerous examples showing the increased precision and accuracy of EDR3 compared to DR2). The direct comparison to the \citetalias{Anders2019} catalogue (shown in Fig. \ref{sh19}) is therefore interesting, but of limited value as a true benchmark test; it is highlighting mostly the shortcomings of the previous catalogue.

As a first example of the improvements made since \citetalias{Anders2019}, Fig. \ref{bar_rc_stars} in this paper shows the spatial distribution of the RC stars selected as in Fig. 8 of \citetalias{Anders2019}. In the \citetalias{Anders2019} plot the astrometric quality flag to select 'good' astrometric sources was applied, while for EDR3 (our Fig. \ref{bar_rc_stars}) no quality cut was applied. The final EDR3 sample of red-clump stars contains 13,640,423 sources. Even without applying any quality cuts, improvements from DR2 to EDR3 are clearly evident in this figure. For example, the ring-like feature between 2-3 kpc reported in \citetalias{Anders2019} and \citet[][their Fig. 8]{Rybizki2021}, mostly provoked by bad astrometric solution, has almost disappeared. Also the unphysical paucity of red-clump stars in front of the Galactic bar visible in Fig. 8 of \citetalias{Anders2019} has vanished.

Furthermore, Fig. \ref{uncerts_vs_G} (uncertainties as a function of $G$) shows that the precision of the EDR3 results is significantly improved: typically the internal EDR3 {\tt StarHorse} uncertainties in all output parameters (red lines in Fig. \ref{uncerts_vs_G}) are smaller than the corresponding \citetalias{Anders2019} ones (black lines) by a factor of 2 at any given magnitude.

Figure \ref{sh19} shows a direct comparison with the \citetalias{Anders2019} results. In each panel we show a HealPix map of the median differences as a function of sky position. The top panel (distance comparison) shows the relative difference ({\tt dist50}/{\tt dist50}$_{A19}-1$), while the other panels show absolute differences (this work $-$ \citetalias{Anders2019}).

\begin{figure}\centering
 	\includegraphics[width=0.49\textwidth]{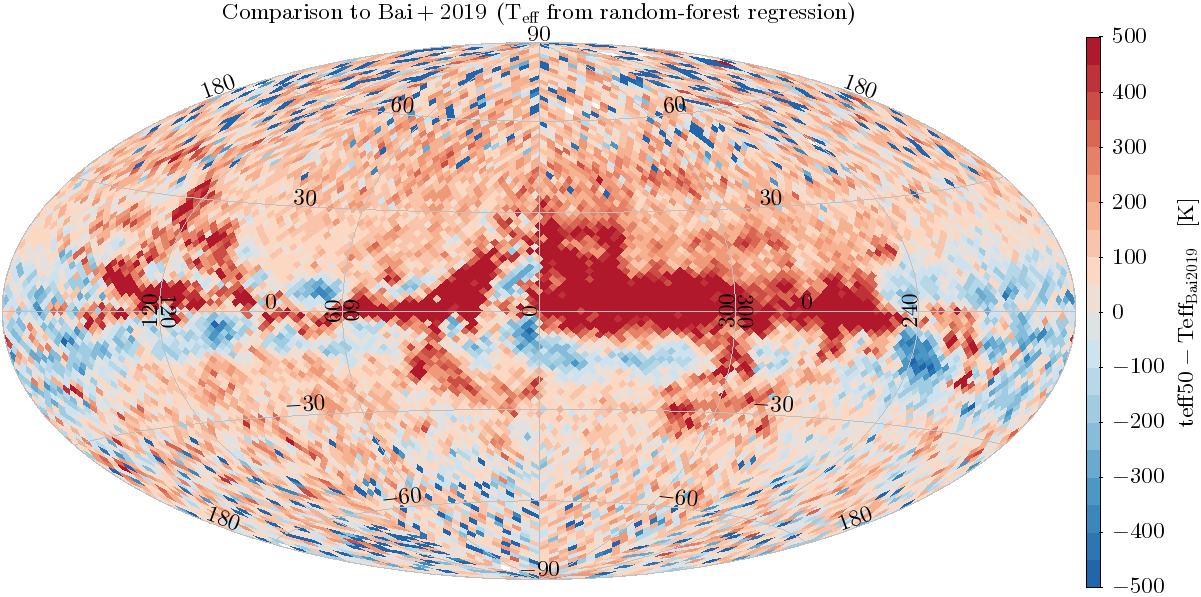}
 	\includegraphics[width=0.49\textwidth]{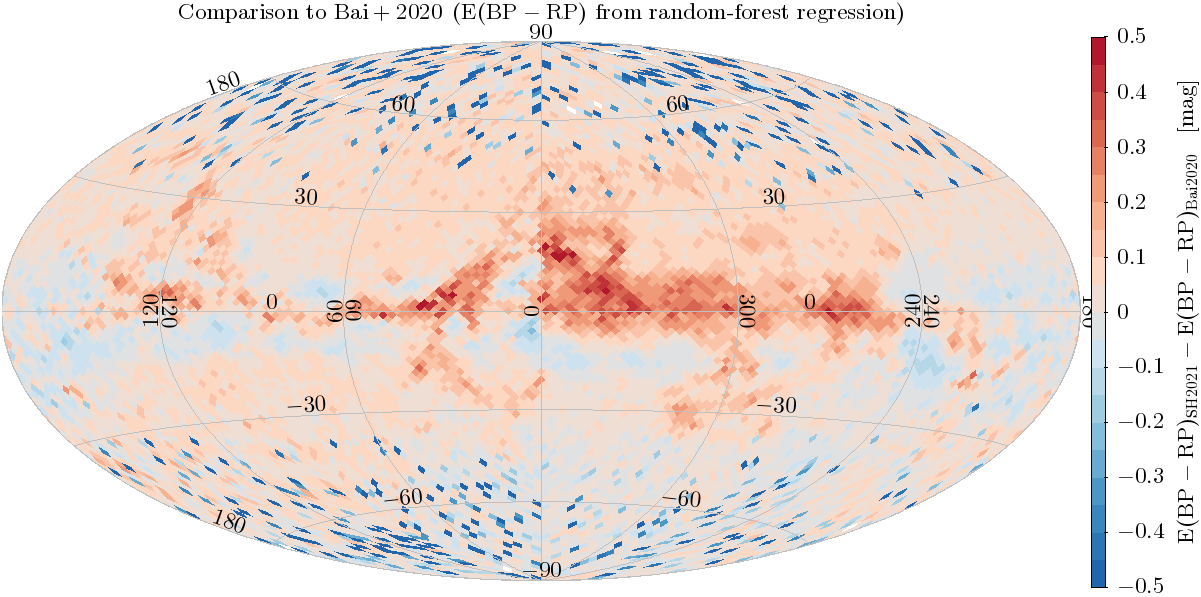}
 	\includegraphics[width=0.24\textwidth]{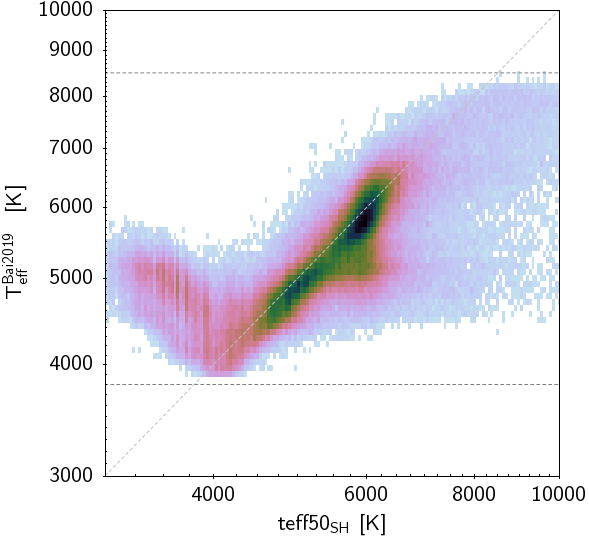}
 	\includegraphics[width=0.24\textwidth]{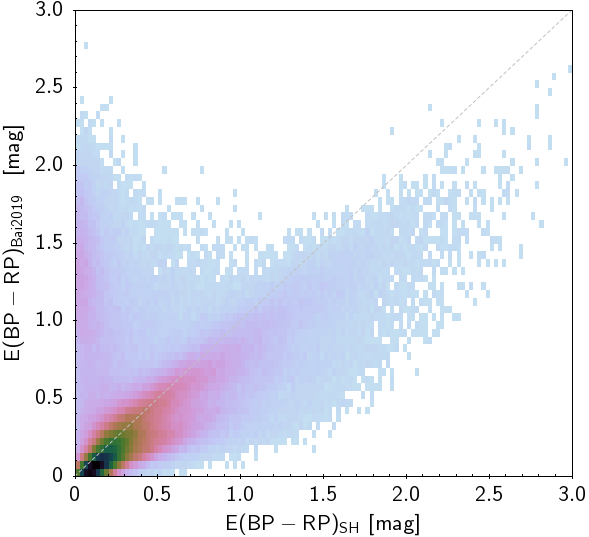}
 	\caption{Comparison of the {\tt StarHorse} EDR3 results with the effective temperatures from \citet[][top panel]{Bai2019}, and the reddenings from \citet[][middle panel]{Bai2020}. The colour scale is the same as in Fig. \ref{sh19}. The bottom panels show the one-to-one comparisons for both parameters.}
 	\label{bai}
\end{figure}

\subsubsection{\citet{Bai2019, Bai2020} effective temperatures and extinctions}

Shortly after the release of {\it Gaia} DR2, \citet{Bai2019} produced a catalogue of stellar effective temperatures for 133 million {\it Gaia} DR2 stars, using a random-forest regressor trained on spectroscopically measured temperatures from a variety of stellar surveys, achieving precise ($\sigma_{\rm Teff}=191$ K) results for stars in the test and control samples. In a subsequent paper \citep{Bai2020} the authors used a similar regression technique to determine precise $E(B-V)$ reddening values for the same stars (also using their previously derived effective temperatures).

In Fig. \ref{bai} we compare our new {\tt StarHorse} results to the values of Bai et al. For effective temperature (lower-left panel of Fig. \ref{bai}), we find relatively good mean concordance for FGK stars (that comprise the training set of \citealt{Bai2019}), while for hotter and cooler stars (according to {\tt StarHorse}) the machine-learning pipeline of \citet{Bai2019} seems to force the $T_{\rm eff}$ values into the range of the training set. We also find significant systematics with sky position for both effective temperature and reddening (top and middle panels of Fig. \ref{bai}) that seem to correlate partly with Galactic extinction and partly with the sky coverage of the training set used by \citet{Bai2019}.

Based on the previous comparisons, we suggest that systematics in the results of Bai et al. are likely causing the bulk of the differences seen in Fig. \ref{bai}. This comparison also reminds us that extreme caution is due when interpreting the results of a machine-learning algorithm outside the (multi-dimensional) range of training data.

\section{Conclusions}\label{conclusions}

We present a catalogue of 362 million stellar parameters, distances, and extinctions based on {\it Gaia} EDR3, Pan-STARRS1, SkyMapper, 2MASS, and AllWISE. The new data and computational updates in our code serve to substantially improve the accuracy and precision over previous photo-astrometric stellar-parameter estimates (typically by a factor of 2 compared to \citetalias{Anders2019}). 

The typical precisions, at magnitude $G=14$ (17), amount to 3\% (15\%) in distance, 0.13 mag (0.15 mag) in $V$-band extinction, and 140 K (180 K) in effective temperature. Our results are validated by comparisons with OCs, as well as with asteroseismic and spectroscopic measurements, indicating systematic errors smaller than the nominal uncertainties for the vast majority of objects. We also provide distance- and extinction-corrected CMDs, extinction maps, and extensive stellar density maps that reveal detailed substructures in the Milky Way and beyond.

The new density maps now probe a much greater volume, extending to regions beyond the Galactic bar and to Local Group galaxies, with a larger total number density. The Galactic bar remains a very prominent feature in the density maps, especially when focussing on red-clump stars. Other subtler features, such as spiral arms or the Sagittarius stream, also start to appear in the density maps.

Our {\it Gaia} EDR3 {\tt StarHorse} catalogue can be queried through the {\it Gaia} mirror archive \url{gaia.aip.de} hosted by the Leibniz-Institut f\"ur Astrophysik Potsdam (AIP). In addition, we also provide approximations to the full posterior PDFs for download in HDF5 format and instructions for bulk data download at \url{data.aip.de/starhorse21} (see Appendix \ref{gmm} for details).

In the near future, {\it Gaia} DR3 (planned for Q2 2022)\footnote{\url{https://www.cosmos.esa.int/web/gaia/dr3}} will provide new {\it Gaia} astrophysical parameters for $\sim500$M stars, in part determined using also the {\it Gaia} BP/RP and RVS spectra, allowing for a further increase in precision for many millions of stars that might possibly supersede some parts of this catalogue. However, we expect that our Bayesian multi-wavelength approach will continue to be relevant and useful for fainter sources without BP/RP spectra ($G\gtrsim 17$), including after the release of the {\it Gaia} DR3 stellar parameters.

\bibliographystyle{aa}
\bibliography{EDR3_SH}

\begin{acknowledgements}
We warmly thank Anthony Brown (Leiden) and the referee for comments on the manuscript.

During the analysis, we have made extensive use of the astronomical {\tt java} software TOPCAT and STILTS \citep{Taylor2005}, Aladin Lite \citep{Bonnarel2000, Boch2014}, as well as the {\tt python} packages {\tt numpy} and {\tt scipy} \citep{Oliphant2007}, {\tt astropy} \citep{AstropyCollaboration2013}, {\tt healpy} \citep{Gorski2005, Zonca2019}, {\tt dustmaps} \citep{Green2018}, {\tt pomegranate} \citep{Schreiber2018}, {\tt dask} \citep{DDT2016}, {\tt HoloViews} (\url{http://holoviews.org}), {\tt matplotlib} \citep{Hunter2007}, and {\tt corner} \citep{Foreman-Mackey2016}. This research has made use of the SVO Filter Profile Service (\url{http://svo2.cab.inta-csic.es/theory/fps/}; \citealt{Rodrigo2020}) supported from the Spanish MINECO through grant AYA2017-84089.
\\

This work has made use of data from the European Space Agency (ESA) mission {\it Gaia} (\url{http://www.cosmos.esa.int/gaia}), processed by the {\it Gaia} Data Processing and Analysis Consortium (DPAC, \url{http://www.cosmos.esa.int/web/gaia/dpac/consortium}). Funding for the DPAC has been provided by national institutions, in particular the institutions participating in the {\it Gaia} Multilateral Agreement. 
FA acknowledges funding from the European Union's Horizon 2020 research and innovation programme under the Marie Sk\l{}odowska-Curie grant agreement No. 800502 H2020-MSCA-IF-EF-2017 and from MICINN (Spain) through the Juan de la Cierva-Incorporación program under contract IJC2019-04862-I. This work was partially funded by the Spanish MICIN/AEI/10.13039/501100011033 and by "ERDF A way of making Europe" by the “European Union” through grant RTI2018-095076-B-C21, and the Institute of Cosmos Sciences University of Barcelona (ICCUB, Unidad de Excelencia ’Mar\'{\i}a de Maeztu’) through grant CEX2019-000918-M. TA acknowledges the grant  RYC2018-025968-I funded by MCIN/AEI/10.13039/501100011033 and by "ESF Investing in your future".
LC acknowledges support from "programme national de physique stellaire" (PNPS) and the "programme national cosmologie et galaxies" (PNCG) of CNRS/INSU. AM acknowledges support from the European Research Council Consolidator Grant funding scheme (project ASTEROCHRONOMETRY, G.A. n. 772293, \url{http://www.asterochronometry.eu}). PR acknowledges the support of the Agence Nationale de la Recherche (ANR project SEGAL ANR-19-CE31-0017) and the European Research Council (ERC grant agreement No. 834148).

\end{acknowledgements}

\newpage
\appendix

\section{Data model}\label{tables}

Table \ref{datamodel1} provides the data model for the provided {\tt StarHorse} output tables.

\begin{table*}
\centering
\caption{Data model of the {\it Gaia} EDR3 {\tt StarHorse} catalogue released via the {\it Gaia} mirror at {\tt gaia.aip.de}.}
\begin{tabular}{rlll}
ID & Column name & Unit & Description \\
\hline
0 & {\tt source\_id} &  & {\it Gaia} EDR3 unique source identifier \\ 
1 & {\tt dist05} & kpc & distance, 5th percentile \\ 
2 & {\tt dist16} & kpc & distance, 16th percentile \\ 
3 & {\tt dist50} & kpc & distance, 50th percentile \\ 
4 & {\tt dist84} & kpc & distance, 84th percentile \\ 
5 & {\tt dist95} & kpc & distance, 95th percentile \\ 
6 & {\tt av05} & mag & line-of-sight extinction at $\lambda=5420\, \AA$, $A_{\rm V}$, 5th percentile \\ 
7 & {\tt av16} & mag & line-of-sight extinction at $\lambda=5420\, \AA$, $A_{\rm V}$, 16th percentile \\ 
8 & {\tt av50} & mag & line-of-sight extinction at $\lambda=5420\, \AA$, $A_{\rm V}$, 50th percentile \\ 
9 & {\tt av84} & mag & line-of-sight extinction at $\lambda=5420\, \AA$, $A_{\rm V}$, 84th percentile \\ 
10 & {\tt av95} & mag & line-of-sight extinction at $\lambda=5420\, \AA$, $A_{\rm V}$, 95th percentile \\ 
11 & {\tt teff16} & K & effective temperature, 16th percentile \\ 
12 & {\tt teff50} & K & effective temperature, 50th percentile \\ 
13 & {\tt teff84} & K & effective temperature, 84th percentile \\ 
14 & {\tt logg16} & [cgs] & surface gravity, 16th percentile \\ 
15 & {\tt logg50} & [cgs] & surface gravity, 50th percentile \\ 
16 & {\tt logg84} & [cgs] & surface gravity, 84th percentile \\ 
17 & {\tt met16} &  & metallicity, 16th percentile \\ 
18 & {\tt met50} &  & metallicity, 50th percentile \\ 
19 & {\tt met84} &  & metallicity, 84th percentile \\ 
20 & {\tt mass16} & $M_{\odot}$ & stellar mass, 16th percentile \\ 
21 & {\tt mass50} & $M_{\odot}$ & stellar mass, 50th percentile \\ 
22 & {\tt mass84} & $M_{\odot}$ & stellar mass, 84th percentile \\ 
23 & {\tt ag50} & mag & line-of-sight extinction in the $G$ band, $A_{\rm G}$, 50th percentile, derived from {\tt av0} and {\tt teff50} \\ 
24 & {\tt abp50} & mag & line-of-sight extinction in the $G_{\rm BP}$ band, $A_{\rm BP}$, 50th percentile, derived from {\tt av50} and {\tt teff50} \\ 
25 & {\tt arp50} & mag & line-of-sight extinction in the $G_{\rm RP}$ band, $A_{\rm RP}$, 50th percentile, derived from {\tt av50} and {\tt teff50} \\ 
26 & {\tt bprp0} & mag & dereddened colour, derived from {\tt phot\_bp\_mean\_mag, phot\_rp\_mean\_mag, abp50, arp50} \\ 
27 & {\tt mg0} & mag & absolute magnitude, derived from {\tt phot\_g\_mean\_mag} (recalibrated), {\tt dist50}, and {\tt ag50} \\ 
28 & {\tt xgal} & kpc & Galactocentric Cartesian X coordinate, derived from {\tt dist50} and assuming $X_0 = -8.2$ kpc \\ 
29 & {\tt ygal} & kpc & Galactocentric Cartesian Y coordinate, derived from {\tt dist50} and assuming $X_0 = -8.2$ kpc \\ 
30 & {\tt zgal} & kpc & Galactocentric Cartesian Z coordinate, derived from {\tt dist50} and assuming $Z_0 = 0$  \\ 
31 & {\tt rgal} & kpc & Galactocentric planar distance, derived from {\tt XGal} and {\tt YGal} \\ 
32 & {\tt fidelity} &  & {\it Gaia} EDR3 astrometric fidelity flag \citep{Rybizki2021} \\ 
33 & {\tt bp\_rp\_excess\_corr} &  & {\it Gaia} EDR3 photometric BP/RP flux excess factor (corrected following \citealt{Riello2021}) \\ 
34 & {\tt sh\_photoflag} &  & {\tt StarHorse} photometry input flag \\ 
35 & {\tt sh\_outflag} &  & {\tt StarHorse} output quality flag \\ 
\end{tabular}
\label{datamodel1}
\end{table*}

\section{Approximation of the full posterior}\label{gmm}

\begin{figure*}\centering
 	\includegraphics[width=0.49\textwidth]{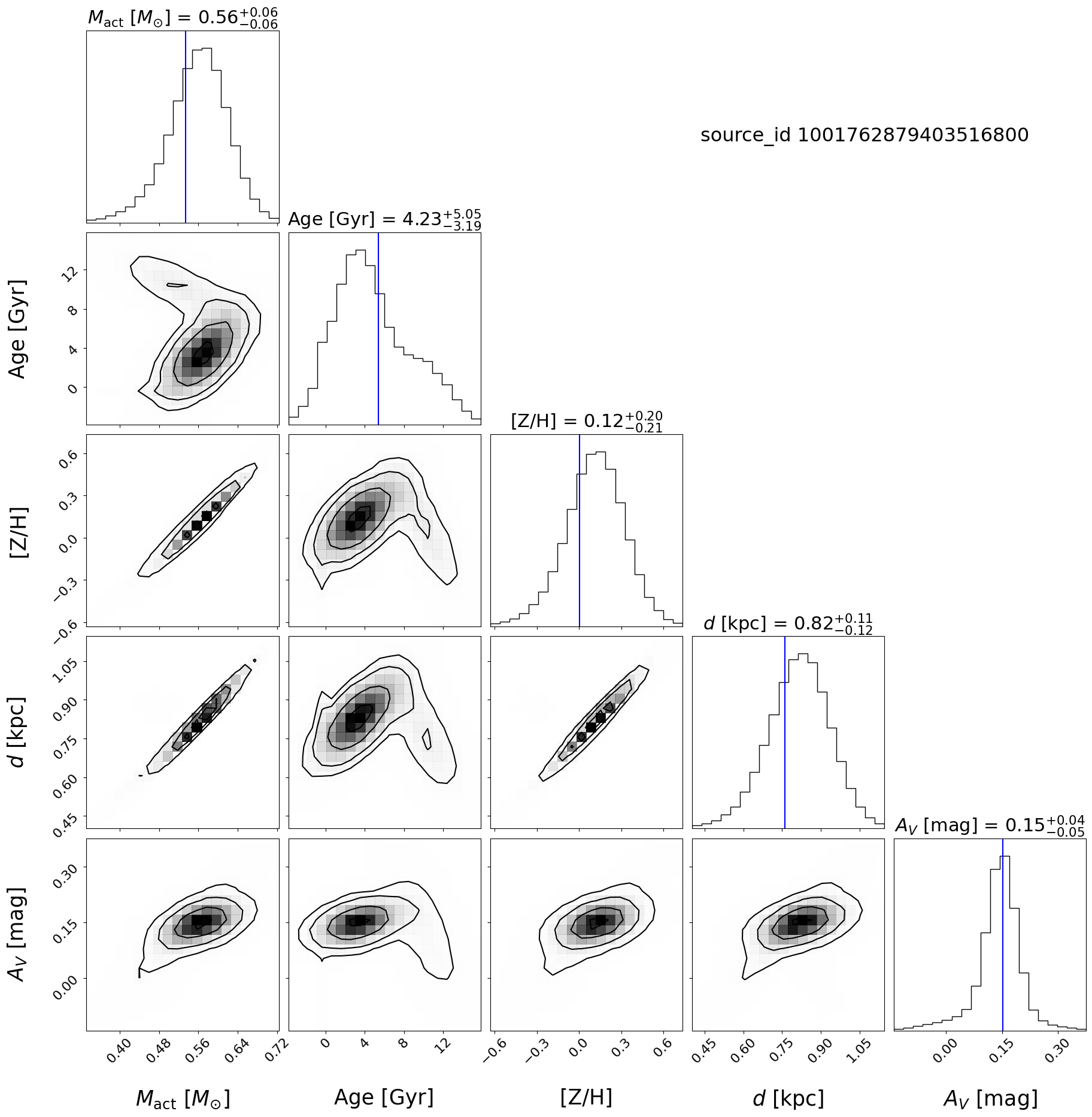}
 	\includegraphics[width=0.49\textwidth]{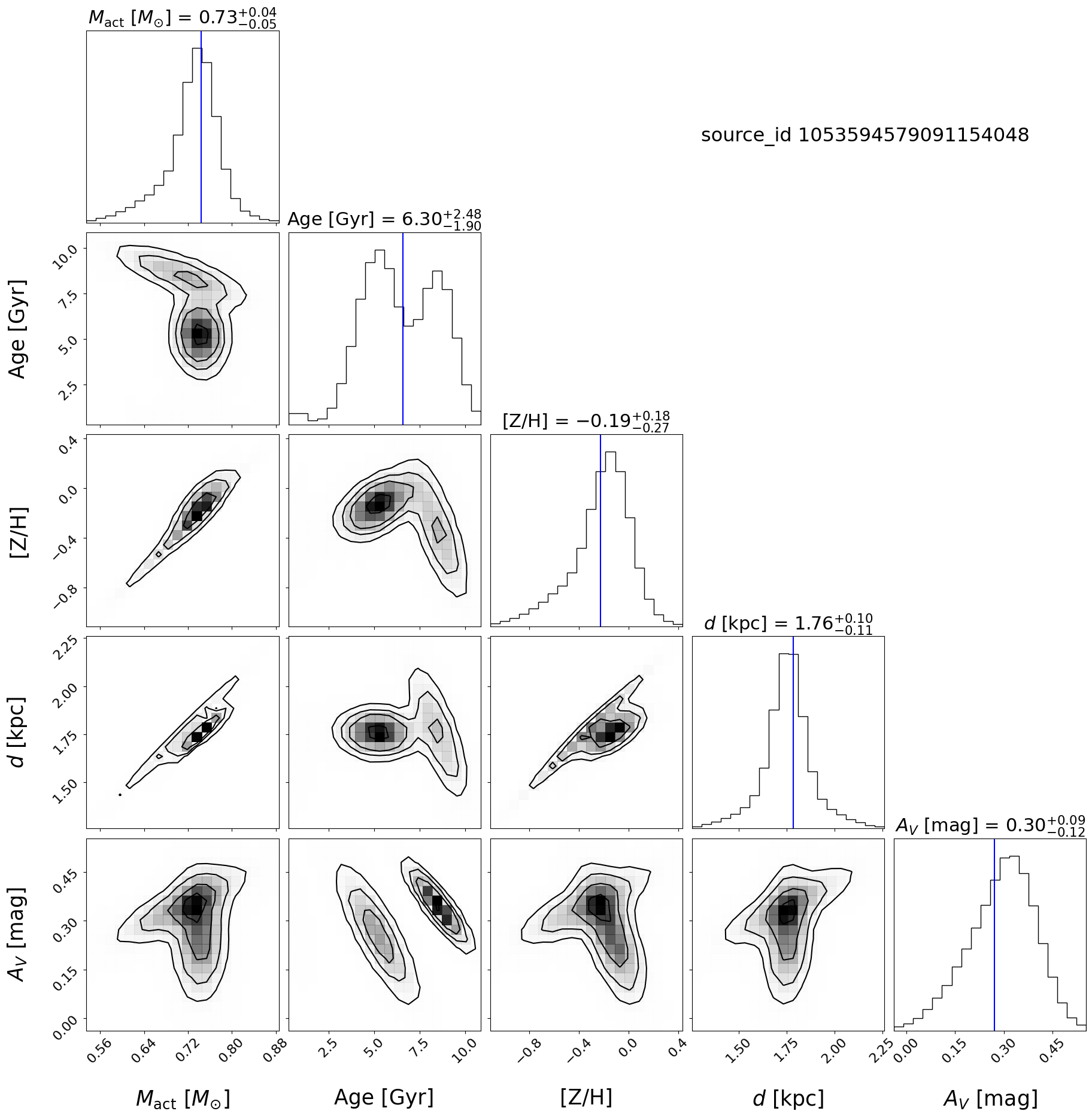}
 	\includegraphics[width=0.49\textwidth]{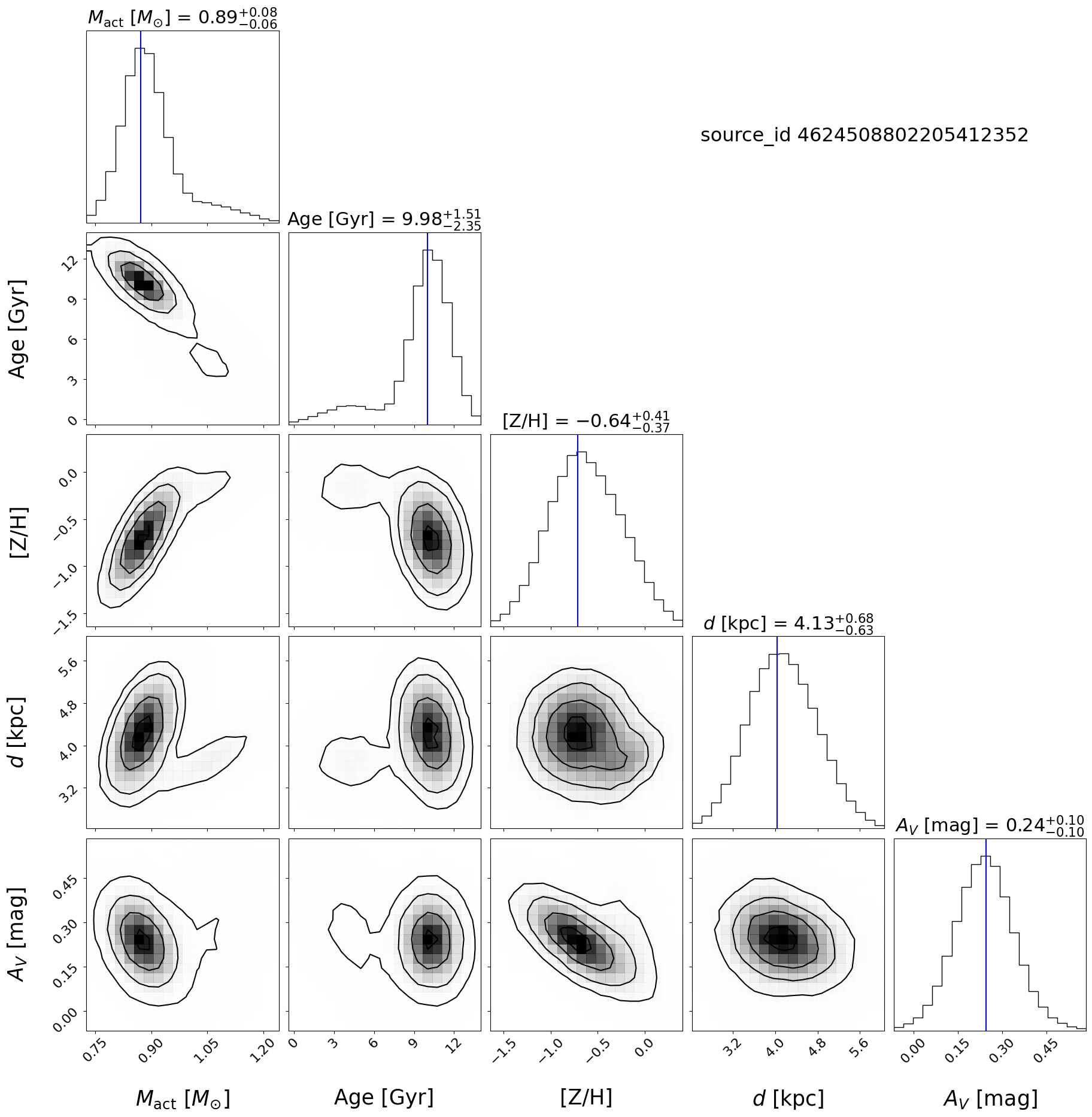}
 	\includegraphics[width=0.49\textwidth]{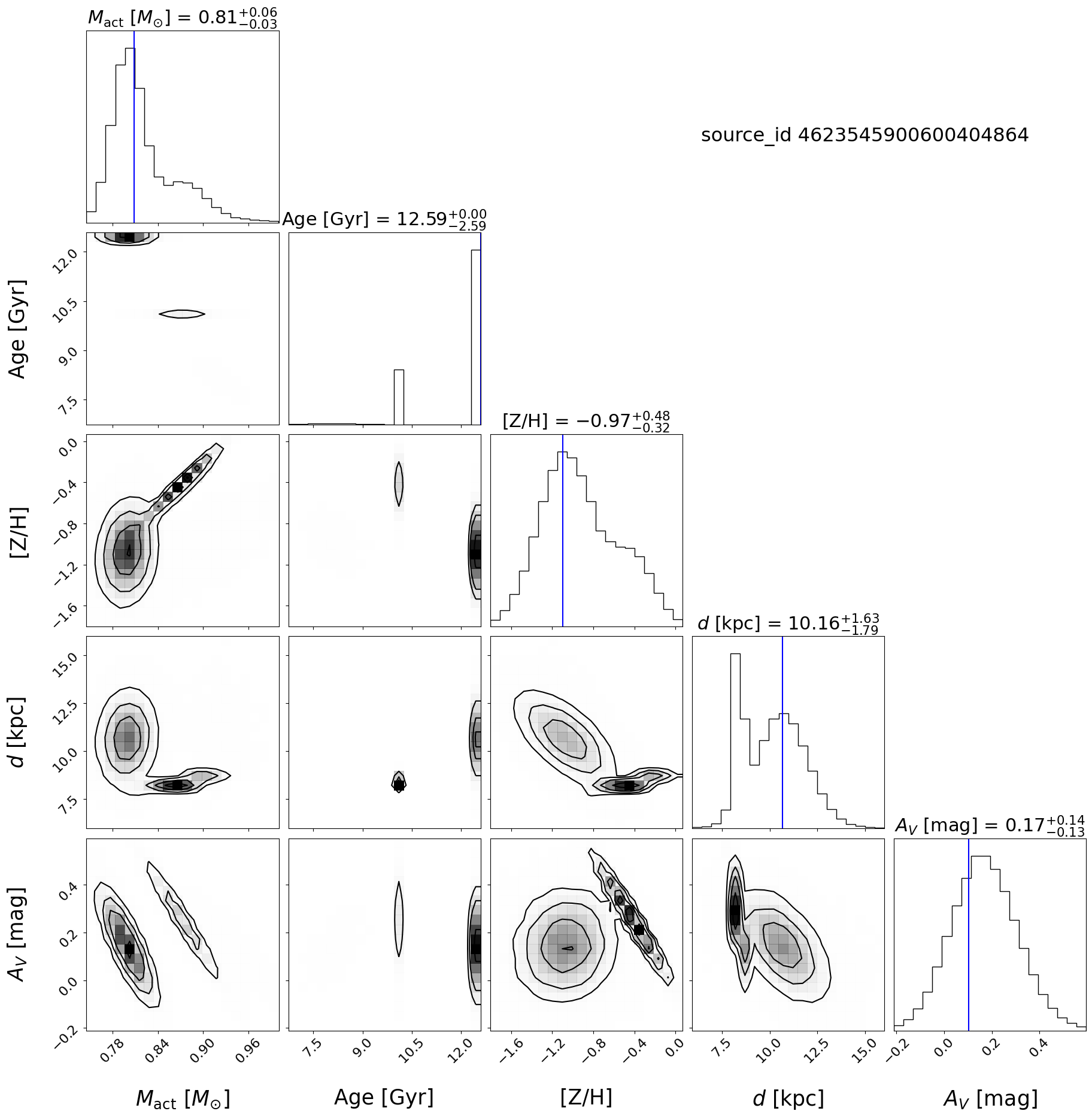}
 	\caption{{\tt StarHorse} posterior probability distributions for four example stars. In the off-diagonal sub-panels we show the two-dimensional projections of the five-dimensional posterior PDF (mass, age, metallicity, distance, and extinction) as approximated by a three-component Gaussian mixture model as black contours, while the diagonal panels show the 1D posterior approximations. The blue vertical lines show the median values directly inferred from the full posterior (available in the CDS tables and through ADQL).}
 	\label{posteriors}
 \end{figure*}

In our previous catalogues we could not publish the full posterior PDFs, since they are typically very heavy files that are not stored to disc. We since implemented an approximation of the joint posterior of the five main spectroscopic output parameters (mass, age, metallicity, distance, extinction) using the fast weighted multivariate Gaussian Mixture Model included in the {\tt python} module {\tt pomegranate} \citep{Schreiber2018}. We thus provide, along with the previously available output parameter PDF quantiles, a representation of the full posterior, stored in custom HDF5 files.

These HDF5 files can be accessed at \url{data.aip.de}. They contain, for each converged star, the weights, means, and covariances of the three Gaussian functions used to approximate the posterior. Five examples of the approximated PDFs, showing the varying complexity of the data, can be appreciated in Fig. \ref{posteriors}.

\section{Example ADQL queries}\label{examplequeries}

In this appendix we show some example ADQL queries that can be used to access the {\tt StarHorse} {\it Gaia} EDR3 results via the {\it Gaia} mirror archive at {\tt gaia.aip.de}. For example, To inspect the first 50 rows of the dataset, it is sufficient to write:

\begin{verbatim}
SELECT TOP 50 * 
FROM gaiaedr3_contrib.starhorse
\end{verbatim}

The second example shows how to access the first 50 rows of our results, cross-matched with the {\it Gaia} EDR3 catalogue, cleaned only for the first digit of the {\tt StarHorse} output flag (see Sect. \ref{outflag}), {\tt sh\_outflag}[0]$=="0"$:

\begin{verbatim}
SELECT TOP 50 g.ra, g.dec, s.*
FROM gaiaedr3.gaia_source AS g, 
     gaiaedr3_contrib.starhorse AS s 
WHERE g.source_id = s.source_id 
AND s.sh_outflag LIKE '0%%%'
\end{verbatim}

The first 50 rows of the red-clump sample shown in Fig. \ref{bar_rc_stars} can be selected using this query:

\begin{verbatim}
SELECT TOP 50 s.*
FROM gaiaedr3_contrib.starhorse AS s
WHERE s.teff50 < 5000 AND s.teff50 > 4500
AND   s.met50  < .4   AND s.met50 > -.6
AND   s.logg50 < 2.55 AND s.logg50 > 2.35
AND   abs(s.zgal) < 3
\end{verbatim}

A de-reddened CMD for a random sample can be obtained with a query like this:
\begin{verbatim}
SELECT bp_rp_index / 40 AS bp_rp, 
       g_abs_index / 10 AS g_abs, n 
FROM ( SELECT FLOOR(s.bprp0 * 40) AS bp_rp_index, 
              FLOOR(s.mg0 * 10)   AS g_abs_index, 
              COUNT(*) AS n 
       FROM gaiaedr3.gaia_source AS g, 
            gaiaedr3_contrib.starhorse AS s 
       WHERE g.source_id = s.source_id 
       AND   g.random_index < 1000000  
       GROUP BY bp_rp_index, g_abs_index ) 
AS subquery
\end{verbatim}

To retrieve a number of columns from both {\it Gaia} EDR3 and {\tt StarHorse} for a random sample of 1\,000 stars (including stars that are not in the {\tt StarHorse} catalogue and even stars with missing parallaxes), one can use this type of query:

\begin{verbatim}
SELECT g.source_id, g.ra, g.dec, g.phot_g_mean_mag, 
       g.parallax, g.parallax_error, 
       s.dist50, s.teff50, s.av50, 
       s.sh_outflag, s.sh_photoflag, s.fidelity
FROM gaiaedr3.gaia_source AS g 
LEFT OUTER JOIN gaiaedr3_contrib.starhorse AS s
ON (g.source_id=s.source_id)
WHERE g.random_index < 1000
\end{verbatim}

If one is interested in objects for which {\tt StarHorse} did not converge (e.g. white dwarfs, galaxies, stars with problematic input data), this last example query shows how to retrieve them:
\begin{verbatim}
SELECT TOP 50
       g.source_id, g.l, g.b, g.parallax,
       g.parallax_error, g.phot_g_mean_mag,
       g.phot_bp_mean_mag, g.phot_rp_mean_mag
FROM gaiaedr3.gaia_source AS g
LEFT OUTER JOIN gaiaedr3_contrib.starhorse AS s
ON (g.source_id=s.source_id)
WHERE g.phot_g_mean_mag <= 18.5
AND   g.astrometric_params_solved > 3
AND   s.source_id IS NULL
\end{verbatim}

\section{Caveats}\label{caveats}

As mentioned in Sect. \ref{caveats0}, many of the caveats present in our previous catalogue have been addressed in this work, but some important drawbacks remain and are discussed in this appendix.

\subsection{Unresolved multiple stars}\label{binaries}

Many stars, both in the field and in star clusters, come in multiple systems. {\it Gaia} is able to resolve millions of wide binaries out to significant distances \citep[e.g.][]{ElBadry2021}, but most multiple systems are still either completely or partially unresolved (resolved in the $G$ band and in astrometry, unresolved in the BP/RP photometry; see e.g. Sect. 2 in \citealt{Fabricius2021}).

A main drawback of {\tt StarHorse} and most similar codes is that they do not take into account unresolved stellar multiplicity. Especially in the case of nearly equal-mass binaries or higher-order systems on the main sequence (which are quite abundant; see e.g. \citealt{Duquennoy1991, Fuhrmann2017}), we may expect significantly biased results.
For example, our code might fit nearby main-sequence binaries (stars that are slightly brighter than predicted by single star models and for which the parallaxes are very well constrained by the {\it Gaia} EDR3 data) by moving them towards the sub-giant branch (i.e. higher effective temperatures) and to higher extinction values, so that the reddened synthetic magnitudes match the observed ones. This explains, at least in part, why our extinctions tend to be overestimated on average for nearby dwarf stars.

Properly taking into account multiplicity is beyond the scope of this work, but one way to allow for high mass-ratio binaries in the analysis would be to use data-driven stellar models \citep[e.g.][]{Anderson2018}.

\subsection{Low {\tt StarHorse} convergence in crowded fields}

In the middle panel of Fig. \ref{skydensity} we observed that {\tt StarHorse} tends to converge less for objects located close to the Galactic plane, especially towards the centre of the Galaxy and in the Magellanic Clouds. The main reason for this is that the {\it Gaia} BP/RP aperture photometry is prone to systematics in crowded regions (e.g. \citealt{Evans2018, Arenou2018, Fabricius2021, Riello2021}). Filtering the data by the astrometric fidelity and the BP/RP excess factor increases the convergence inhomogeneities on the sky (bottom panel of Fig. \ref{skydensity}), thereby rendering direct comparisons to simulations (without taking into account crowding effects in the {\it Gaia} selection function) even harder.

\subsection{Variations in the extinction law induce systematic effective temperature shifts}\label{rv_var}

\begin{figure}\centering
 	\includegraphics[width=0.49\textwidth]{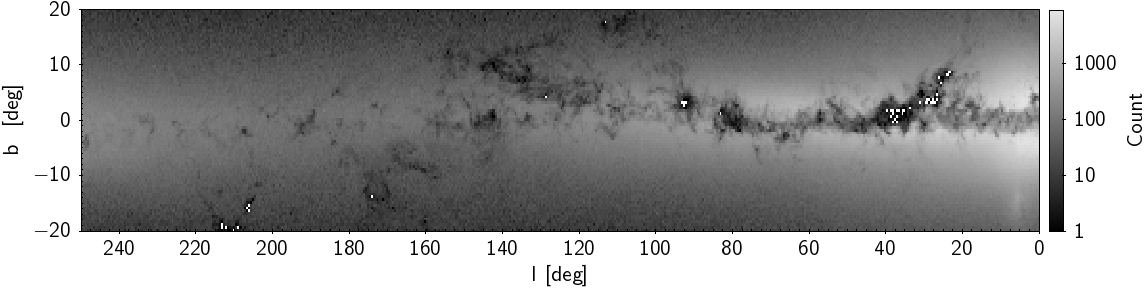}
 	\includegraphics[width=0.49\textwidth]{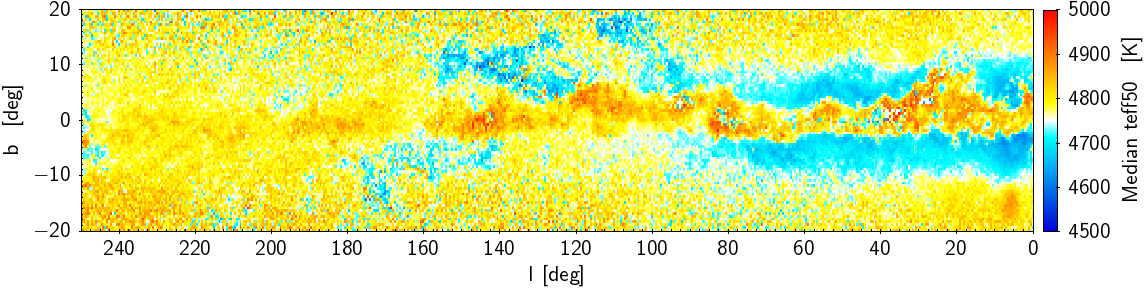}
 	\includegraphics[width=0.49\textwidth]{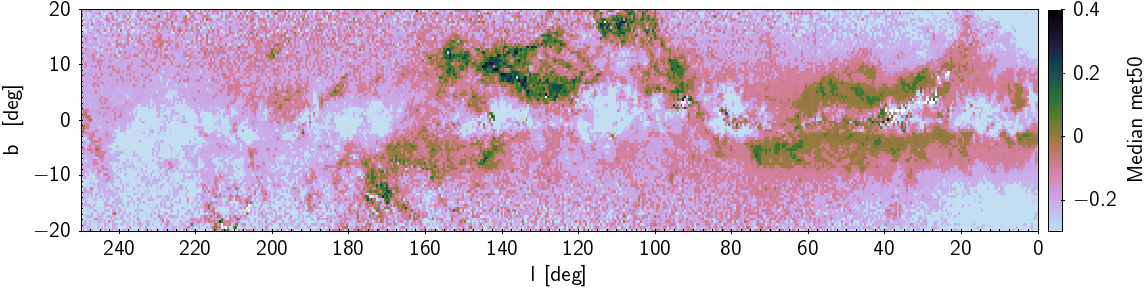}
 	\caption{Sky map of the red-clump star sample for a region close to the Galactic plane ($0<l<250$, $-20<b<20$), revealing systematics possibly related to variations in the Galactic extinction law (compare with Fig. 1 of \citealt{Schlafly2017}). The top panel is colour-coded by number density, the second panel by median effective temperature, and the bottom panel by median metallicity.}
 	\label{rc_stars_rv_variations}
\end{figure}

As mentioned already in \citetalias{Anders2019}, our results rely to some degree on the validity of the assumed extinction curve, which we fixed to the one recommended by \citet{Schlafly2016}. Figure \ref{rc_stars_rv_variations} shows sky maps for red-clump stars close to the Galactic plane, highlighting some systematic trends in both effective temperature and metallicity. While most of the systematics may possibly be explained by selection effects provoked by the complex three-dimensional dust distribution, some of the trends maight also be correlated with the highly variable total-to-selective extinction ratio ($R_V$; see the detailed study of \citealt{Schlafly2017}, for example their Fig. 1).

\subsection{Limited sky coverage of input catalogues causes slight inhomogeneities}

\begin{figure}\centering
 	\includegraphics[width=0.24\textwidth]{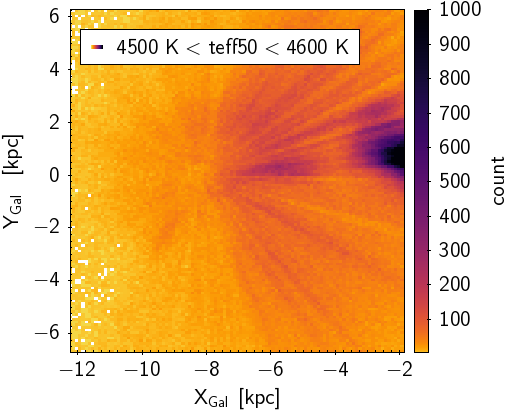}
 	\includegraphics[width=0.24\textwidth]{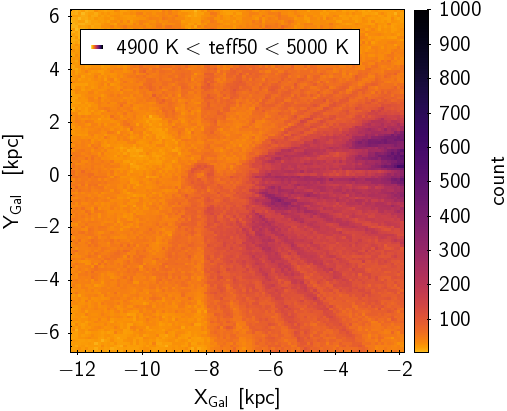}
 	\caption{Galactic distribution for two thin effective temperature slices of the disc red-clump sample ($|Z_{\rm Gal}|<3$ kpc). The left panel shows a slight underdensity in the region where Pan-STARRS1 photometry is missing, while the right panel shows an overdensity in the region where SkyMapper photometry is available.}
 	\label{rc_stars_teffbins}
\end{figure}

While the input catalogues {\it Gaia} EDR3, 2MASS, and AllWISE cover the full sky with considerable homogeneity, the sky coverage of Pan-STARRS1 and SkyMapper photometry is limited by the location of the respective telescopes, which to some degree also affects the homogeneity of the resulting {\tt StarHorse} catalogue (see e.g. the bottom panels of Fig. \ref{uncerts_vs_G}). The effect is alleviated by the similar filter system of Pan-STARRS1 and SkyMapper (and certainly less prominent than in \citetalias{Anders2019} where no SkyMapper data were used), but should nonetheless be mentioned.

Another example is given in Fig. \ref{rc_stars_teffbins}, where we show the Galactic maps for the red-clump sample shown in Fig. \ref{bar_rc_stars}, but for two narrow bins in effective temperature. Some of the underdensities in the left panel correspond to overdensities in the right panel (and vice versa), which is in part a plausible population effect, but in part also reflects the sky regions covered by Pan-STARRS1 and SkyMapper (compare to Fig. \ref{bar_rc_stars}, bottom-right panel), which cautions us about the blind use of narrow bins in effective temperature or metallicity without taking into account their uncertainties.

\subsection{Unreliable stellar parameters for massive stars}\label{obstars}

\begin{figure}\centering
 	\includegraphics[width=0.49\textwidth]{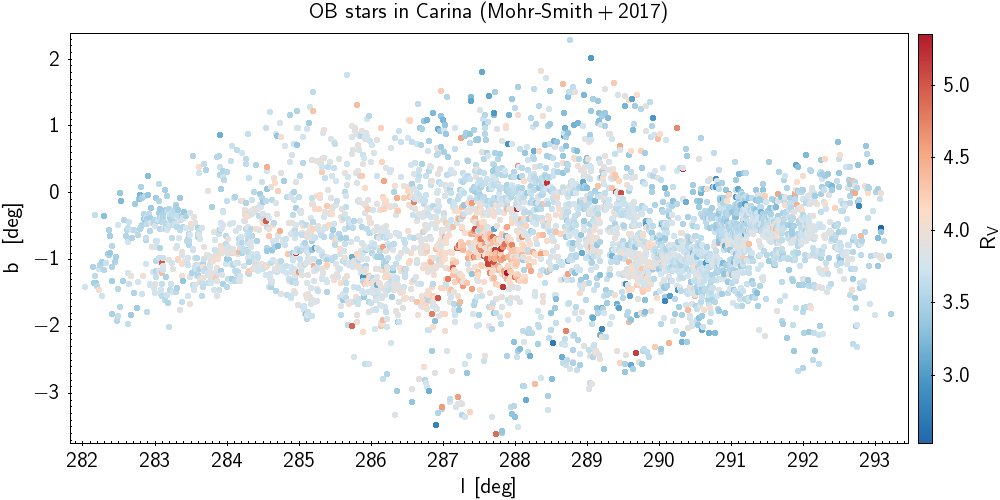}
 	\includegraphics[width=0.24\textwidth]{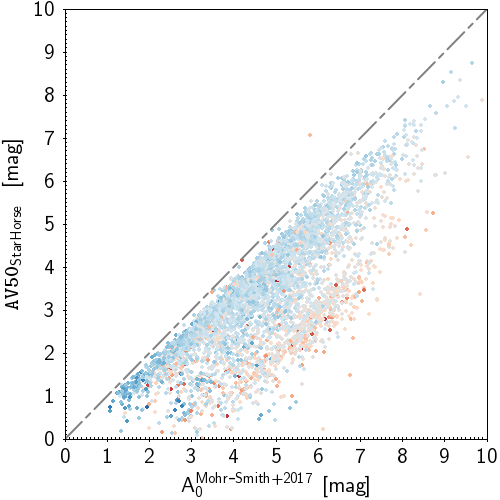}
 	\includegraphics[width=0.24\textwidth]{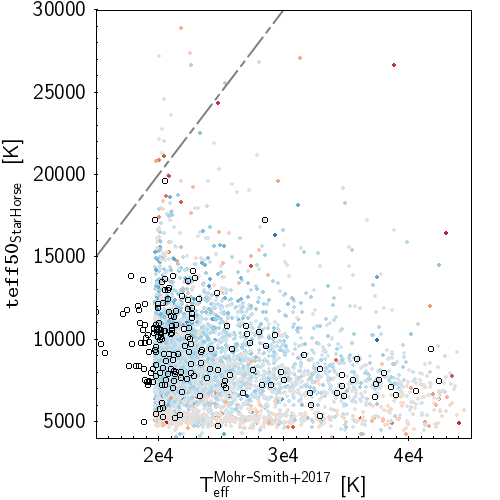}
 	\caption{Comparison to the Carina OB star sample of \citet{Mohr-Smith2017}. The top panel shows the sky distribution, while the bottom panels show one-to-one comparisons for extinction (left) and effective temperature (right). In each panel, the points are colour-coded by $R_V$, as determined by \citet{Mohr-Smith2017}. The black circles in the bottom-right panel show the spectroscopically measured effective temperatures, available for a subset of the OB stars.}
    \label{ob_stars}
\end{figure}

The OC comparisons discussed in Sect. \ref{galah} have already shown that for hot stars the present {\tt StarHorse} stellar parameters are often biased. Here we therefore investigate this statistically small but astrophysically important subset in more detail using a known sample of OB stars.

\citet{Mohr-Smith2017} selected O-B3 stars in the far Carina spiral arm using VPHAS+ data \citep{Drew2016}. Their method to detect OB stars has proven to be quite reliable (confirmed by spectroscopy for some of the targets), thanks in part to the $u$ filter (not used in our work). It also joins in 2MASS information in order to provide $T_{\rm eff}, A_0$ and $R_V$. We crossmatched this sample with the StarHorse EDR3 catalogue, resulting in $4,658$ stars with good ($\chi^2<7.82$) parameters from \citealt{Mohr-Smith2017}, and compared the effective temperatures and extinctions with the ones obtained from VPHAS+.

Figure \ref{ob_stars} shows that the {\tt StarHorse} extinctions compare relatively well to the ones of the external catalogue (modulo a small offset that also depends on $R_V$; see Sect. \ref{rv_var}). The {\tt StarHorse} effective temperatures for the O-B3 star candidates of \citet{Mohr-Smith2017}, however, are in a completely different range than estimated by those authors. While it could be argued that this photometrically selected OB star sample may still be contaminated by lower-mass field stars, the observed $T_{\rm eff}$ differences are too drastic to be explained by contamination only. We therefore caution that our results for very massive stars are very likely to be unreliable in most cases.

Similar conclusions regarding our \citetalias{Anders2019} results for OB stars have been reached by \citet{Pantaleoni2021}. The reason is that for a generic field-star approach such as {\tt StarHorse}, the initial-mass-function prior strongly suppresses hot-star solutions, since they are a very small minority among the Galactic stellar population. Especially for rare stellar populations (e.g. OB stars: \citealt{Zari2021}; or open star clusters: \citealt{Cantat-Gaudin2020, Olivares2020}), specifically tailored algorithms are therefore expected to outperform our results.

\end{document}